\documentclass[11pt,a4paper]{article}
\pdfoutput=1
\usepackage{multicol}
\usepackage[colorlinks=true, linkcolor=black!50!blue, urlcolor=blue, citecolor=blue, anchorcolor=blue]{hyperref}
\usepackage[font=small,labelfont=bf,margin=0mm,labelsep=period,tableposition=top]{caption}
\usepackage[a4paper,top=1.5cm,bottom=2cm,left=1.5cm,right=1.5cm,bindingoffset=0mm]{geometry}

\usepackage{placeins,cite}
\usepackage{graphicx}
\usepackage{soul}
\usepackage{subcaption}
\usepackage{float}
\usepackage{afterpage}
\usepackage{epsfig}
\usepackage{amssymb}
\usepackage{amsmath}
\usepackage{bm}
\usepackage{multirow}
\usepackage{url}
\usepackage{xcolor}
\usepackage{ulem}
\usepackage{url}
\usepackage{booktabs,multirow}
\usepackage{textcomp}
\usepackage{comment}
\graphicspath{{../}{./figures/}}
\newcommand{\kh}{\kappa_H}
\newcommand{\kt}{\kappa_t}
\newcommand{\kb}{\kappa_b}
\newcommand{\kc}{\kappa_c}
\newcommand{\ktau}{\kappa_\tau}
\newcommand{\kw}{\kappa_W}
\newcommand{\kz}{\kappa_Z}
\newcommand{\ka}{\kappa_\gamma}
\newcommand{\kza}{\kappa_{Z\gamma}}
\newcommand{\kg}{\kappa_g}
\newcommand{\kmu}{\kappa_\mu}
\usepackage{makecell}

\newcommand{\BrBSM}{{\rm Br}_{\rm BSM}}
\newcommand{\Brinv}{{\rm Br}_{\rm inv}}
\newcommand{\Brund}{{\rm Br}_{\rm und}}
\newcommand{\mcO}{\ensuremath{\mathcal{O}}}

\makeatletter
\def\thickhline{%
             \noalign{\ifnum0 =`}\fi\hrule \@height \thickarrayrulewidth \futurelet
             \reserved@a\@xthickhline}
\def\@xthickhline{\ifx\reserved@a\thickhline
                \vskip\doublerulesep
                \vskip -\thickarrayrulewidth
                \fi
                \ifnum0 =`{\fi}}
\makeatother
\newlength{\thickarrayrulewidth}
\usepackage{tikz}
\usetikzlibrary{positioning}
\usetikzlibrary{shapes.geometric, arrows}
\usetikzlibrary{arrows.meta}
\usepackage{varwidth}
\usepackage{xcolor}
\definecolor{mtplotlib1}{HTML}{1f77b4}
\definecolor{mtplotlib2}{HTML}{ff7f0e}
\definecolor{mtplotlib3}{HTML}{2ca02c}
\definecolor{mtplotlib4}{HTML}{d62728}

\tikzset{%
  >={Latex[width=2mm,length=2mm]},
            base/.style = {rectangle, rounded corners, draw=black,
                           minimum width=4cm, minimum height=1cm,
                           text centered}, 
            mystyle/.style={rectangle, rounded corners, draw=black,
            minimum width=12cm, minimum height=1cm,
            text centered}, 
    col0/.style = {base, fill=white!30},
    col1/.style = {base, fill=mtplotlib1!30},
    col11/.style = {mystyle, fill=mtplotlib1!30},
    col2/.style = {base, fill=mtplotlib2!30},
    col3/.style = {base, fill=mtplotlib3!30},
    col4/.style = {base, minimum width=2.5cm, fill=mtplotlib4!15,}
}

\newcommand{\hepfit}{{\sc\small HEPfit}}
\newcommand{\smefit}{{\sc\small SMEFiT}}

\newcommand{\voverL}{\,\frac{v^2}{\Lambda^2} }
\newcommand{\Tf}{T^3_f}

\newcommand{\cw}{c_W}
\newcommand{\sw}{s_W}
\newcommand{\cpwb}{c_{\varphi WB}}
\newcommand{\cpw}{c_{\varphi W}}
\newcommand{\cpb}{c_{\varphi B}}
\def\Dlr{\mathrel{\raise1.5ex\hbox{$\leftrightarrow$\kern-1em\lower1.5ex\hbox{$D$}}}}

\newcommand{\be}{\begin{equation}}
\newcommand{\ee}{\end{equation}}
\newcommand{\bea}{\begin{eqnarray}}
\newcommand{\eea}{\end{eqnarray}}
\newcommand{\bi}{\begin{itemize}}
\newcommand{\ei}{\end{itemize}}
\newcommand{\ben}{\begin{enumerate}}
\newcommand{\een}{\end{enumerate}}

\newcommand{\lc}{\left[}
\newcommand{\rc}{\right]}
\newcommand{\lp}{\left(}
\newcommand{\rp}{\right)}

\def\frac#1#2{{{#1}\over {#2}}}
\def\gsim{\mathrel{\rlap{\lower4pt\hbox{\hskip1pt$\sim$}}
    \raise1pt\hbox{$>$}}}       
\def\lsim{\mathrel{\rlap{\lower4pt\hbox{\hskip1pt$\sim$}}
    \raise1pt\hbox{$<$}}}

\newcommand{\draft}[1]{}
\newcommand{\OO}{\ensuremath{\mathcal{O}}}
\newcommand{\Op}[1]{\OO_{\sss #1}}
\newcommand{\sss}{\scriptscriptstyle}
\newcommand{\pdp}{\ensuremath{\varphi^\dagger\varphi}}
\newcommand{\ccc}[3]{c_{#2}^{#1 (#3)}}
\newcommand{\qq}[3]{\ensuremath{\mathcal{O}_{#2}^{#1 (#3)}}}
\def\beq{\begin{equation}}
\def\eeq{\end{equation}}

\numberwithin{equation}{section}
\numberwithin{figure}{section}
\numberwithin{table}{section}

\usepackage{tabularx}
\newcolumntype{C}[1]{>{\centering\arraybackslash}p{#1}}

 \def\lra#1{\overset{\text{\scriptsize$\leftrightarrow$}}{#1}}

\definecolor{darkblue}{rgb}{0.0,0,0.5}
\definecolor{darkgreen}{rgb}{0.0,0.3,0.0}
\definecolor{redish}{rgb}{0.675,0,0.2}
\definecolor{red}{rgb}{0.8,0,0}
\definecolor{green}{rgb}{0,0.6,0}
\definecolor{bluish}{rgb}{0.2,0.2,0.675}
\definecolor{mygrey}{rgb}{0.6,0.6,0.6}

\usepackage{tikz} 
\usetikzlibrary{shapes,arrows,positioning,automata,backgrounds,calc,er,patterns,arrows.meta}

\usepackage{varwidth}
\usepackage{xcolor}
\definecolor{mtplotlib1}{HTML}{1f77b4}
\definecolor{mtplotlib2}{HTML}{ff7f0e}
\definecolor{mtplotlib3}{HTML}{2ca02c}
\definecolor{mtplotlib4}{HTML}{d62728}

\tikzset{%
  >={Latex[width=2mm,length=2mm]},
            base/.style = {rectangle, rounded corners, draw=black,
                           minimum width=4cm, minimum height=1cm,
                           text centered}, 
            mystyle/.style={rectangle, rounded corners, draw=black,
            minimum width=12cm, minimum height=1cm,
            text centered}, 
    col0/.style = {base, fill=white!30},
    col1/.style = {base, fill=mtplotlib1!30},
    col11/.style = {mystyle, fill=mtplotlib1!30},
    col2/.style = {base, fill=mtplotlib2!30},
    col3/.style = {base, fill=mtplotlib3!30},
    col4/.style = {base, minimum width=2.5cm, fill=mtplotlib4!15,}
}

\usepackage{tabularx}
\usepackage{subcaption}
\newcolumntype{C}[1]{>{\centering\arraybackslash}p{#1}}

\begin{document}
\newgeometry{top=1.5cm,bottom=1.5cm,left=1.5cm,right=1.5cm,bindingoffset=0mm}

\vspace{-2.0cm}
\begin{flushright}
MPP-2026-66\\ 
CERN-TH-2026-088
\end{flushright}
\vspace{0.3cm}

\begin{center}

{\Large \bf New Physics Reach through Precision at Future Colliders:\\[0.2cm]  a Multi-Pronged Approach}\\
  \vspace{1.1cm}
  {
Tommaso~Armadillo,$^{1,2}$
Eugenia~Celada,$^{3,4}$
 Jaco~ter~Hoeve,$^{5}$
 Fabio~Maltoni,$^{1,6,7}$
 Luca~Mantani,$^8$\\[0.1cm]
 Juan~Rojo,$^{9,10}$
 Alejo~N.~Rossia,$^{11}$
 Simone~Tentori,$^{1}$
 Marion O.A. Thomas,$^{12}$ and
 Eleni~Vryonidou$^{4}$
  }\\

\vspace{1.0cm}

{\it
~$^{1}$Centre for Cosmology, Particle Physics and Phenomenology (CP3), \\Universit\'e catholique de Louvain,  Chemin du Cyclotron, 2, B-1348 Louvain-la-Neuve, Belgium\\[0.1cm]
~$^{2}$Dipartimento di Fisica “Aldo Pontremoli”, University of Milano and \\ INFN, Sezione di Milano, Milano, I-20133, Italy\\[0.1cm]
~$^{3}$Department of Physics and Astronomy, University of Manchester,\\ Oxford Road, Manchester M13 9PL, UK\\[0.1cm]
~$^{4}$Department of Physics, University of Cyprus,\\  Panepistimiou Street 1, Aglantzia, 
CY-1678 Nicosia, Cyprus\\[0.1cm]
~$^{5}$The Higgs Centre for Theoretical Physics, University of Edinburgh,\\[0.1cm]
JCMB, KB, Mayfield Rd, Edinburgh EH9 3FD, Scotland\\[0.1cm]
~$^{6}$Dipartimento di Fisica e Astronomia, Università di Bologna, and INFN, Sezione di Bologna,\\ via Irnerio 46, 40126 Bologna, Italy\\[0.1cm]
~$^{7}$Theoretical Physics Department, CERN, 1211 Geneva 23, Switzerland\\[0.1cm]
~$^{8}$Instituto de F\'isica Corpuscular (IFIC), Universidad de Valencia-CSIC, E-46980 Valencia, Spain\\[0.1cm]
~$^{9}$Nikhef Theory Group, Science Park 105, 1098 XG Amsterdam, The Netherlands\\[0.1cm]
~$^{10}$Department of Physics and Astronomy, Vrije Universiteit Amsterdam, \\NL-1081 HV Amsterdam, The Netherlands\\[0.1cm]
~$^{11}$ Dipartimento di Fisica e Astronomia ``G. Galilei'', Universit\`a di Padova, and Istituto Nazionale di Fisica Nucleare, Sezione di Padova, Via F. Marzolo 8, I-35131, Padova, Italy \\[0.1cm]
~$^{12}$ Max-Planck-Institut für Physik, Boltzmannstrasse 8, 85748 Garching, Germany
}

\vspace{0.7cm}

\vspace{1.0cm}

{\bf \large Abstract}

\end{center}

We present projections for the sensitivity of future high-energy colliders to new physics through precision measurements of the Standard Model (SM) interactions, focusing on near-term electron-positron facilities: FCC-ee, LEP3, and the Linear Collider Facility. 
We interpret these projections in three complementary frameworks: Higgs coupling modifiers, effective Higgs and electroweak couplings, and global SMEFT fits. The SMEFT analysis includes renormalisation-group evolution, linear/quadratic contributions, and NLO corrections to EFT cross sections where available. 
By matching the EFT to UV-complete models, we also quantify the sensitivity of future colliders to representative benchmark scenarios, including composite Higgs models and single-particle SM extensions.
In parallel, we release an updated  version of the open-source SMEFiT framework, enabling the results presented here to be fully reproduced, extended, and customised.

\clearpage

\tableofcontents
\vfill
\section{Introduction}

Comparing the physics potential of future collider projects has become one of the central tasks of the particle physics community as it plans for the post-HL-LHC era. 
This broad effort has been shaped by successive community exercises, from the 2020 update of the European Strategy for Particle Physics~\cite{deBlas:2022ofj} to the Snowmass 2022 process\cite{Butler:2023glv}, and most recently by the conclusions of the European Strategy Group~\cite{CERN-ESU-2025-002} for the 2026 Update. 
In this context, a number of future collider options have been put forward and assessed, including the Future Circular Collider (FCC-ee), the integrated FCC programme with the FCC-ee followed by a hadron collider operating at $\sqrt{s}\sim 84$ TeV (FCC-hh), a Linear Collider Facility (LCF)~\cite{LinearCollider:2025lya,LinearColliderVision:2025hlt} providing linear electron-positron collisions based on different acceleration technologies and reaching up to $\sqrt{s}=1$ TeV, an electron-positron Higgs factory operating in the LHC tunnel (LEP3)~\cite{Anastopoulos:2025jyh}, the Large Hadron electron Collider (LHeC)~\cite{LHeC:2020van,Ahmadova:2025vzd,LHeCStudyGroup:2012zhm}, a stand-alone lower-energy variant of the FCC-hh~\cite{Cavaliere:2025ujf}, a Circular Electron Positron Collider (CEPC)~\cite{CEPCPhysicsStudyGroup:2022uwl}, and a muon collider (MuCol)~\cite{InternationalMuonCollider:2025sys}, among others.

Assessing the physics reach, and in particular the New Physics (NP) reach, of future colliders is a highly non-trivial endeavour. Different approaches can be pursued, each characterised by its own assumptions, advantages, and intrinsic limitations. One strategy is based on direct searches for new particles, usually formulated within explicit beyond the Standard Model (BSM) scenarios; however, this approach is complicated by the large multiplicity of candidate models and by the broad parameter space that each of them typically entails. A complementary strategy relies instead on indirect probes, where precision measurements of Standard Model (SM) observables are used to constrain the low-energy effects of heavy degrees of freedom and thereby gain information on the possible structure of ultraviolet physics. Within this broader programme, the reach of future colliders can be quantified using different benchmark scenarios and interpretation frameworks.

In the case of NP at high scales, complementary descriptions have been adopted: modified Higgs couplings (the so-called $\kappa$-framework) based on a constant rescaling of SM couplings; the global SM Effective Field Theory (SMEFT) analysis~\cite{Brivio:2017vri,Isidori:2023pyp}, in which possible departures from the SM are parametrised in terms of higher-dimensional operators; and the effective couplings formalism~\cite{Barklow:2017suo}, based on pseudo-observables that encode distortions in Higgs and electroweak interactions and can be written in terms of a subset of SMEFT Wilson coefficients. 
Compared to the $\kappa$-framework, an important advantage of SMEFT is that it can be matched consistently onto specific UV-complete scenarios. 
This makes it especially well suited for comparing the reach of indirect EFT-based probes with projections for direct searches for NP at high-energy frontier colliders.

In this work we present a set of new developments and tools that enable a multi-pronged approach to constraining, and potentially discovering, NP through precision measurements at future colliders.
Building on the global SMEFT fitting programme, our framework connects bottom-up and top-down strategies within a common analysis environment. Precision measurements can be interpreted in a model-independent way in terms of modified couplings and SMEFT Wilson coefficients, and can also be matched consistently onto simplified or UV-complete scenarios, thereby allowing their implications for specific classes of BSM models to be assessed. 
This provides a flexible platform to compare complementary interpretations of indirect collider sensitivity and to relate them to explicit realisations of physics beyond the SM. 
More generally, the framework being modular and automated could be further extended, including through new analysis strategies, including those based on AI techniques, see {\it e.g.}, Refs.~\cite{Hirsch:2025qya,Agrawal:2026lvg}.
In this context, we present projections for a representative set of future particle colliders, quantifying their physics reach in terms of modified Higgs couplings, SMEFT operators, and the parameters of UV-complete models, all obtained within the open-source {\sc\small SMEFiT} framework~\cite{Hartland:2019bjb,vanBeek:2019evb,Ethier:2021bye,Ethier:2021ydt,Giani:2023gfq,Celada:2024mcf,terHoeve:2023pvs,terHoeve:2025gey,terHoeve:2025omu}.

Our analysis systematically includes higher-order effects in the SMEFT, such as Renormalisation Group Evolution (RGE), assesses the impact of quadratic EFT corrections to the input cross sections, and automates the matching of general UV-complete models onto the SMEFT at tree level and one loop, presenting results for representative benchmark scenarios. We also devote particular attention to the role of theory uncertainties, both in clarifying the limitations of current projections and in identifying the theoretical advances required to strengthen indirect constraints on the NP scale. Altogether, these developments extend  and complement the methodology followed in the analysis presented in the ESPPU26 Physics Briefing Book~\cite{deBlas:2025gyz} and provide a framework not only to examine its assumptions and challenges critically, but also to explore concrete paths towards more powerful approaches to the search of BSM physics.

Through these projections, we quantify the reach of FCC-ee, LCF (for centre-of-mass energies up to $\sqrt{s}=1$~TeV), and LEP3 in the parameter space of the SMEFT and of UV-complete models. 
In addition, for the $\kappa$-framework analysis we also consider projections for the MuCol at $\sqrt{s}=3$ and $10$~TeV, the FCC-hh at $\sqrt{s}=84$~TeV, and the LHeC. 
Our results address several questions that are central to future collider studies, including the reach on the Higgs self-coupling and the comparison between indirect sensitivity to benchmark UV scenarios, such as composite Higgs models, and the projections for direct searches.

A central feature of this work is that our results are fully reproducible and can be readily extended and customised with the public version of the {\sc\small SMEFiT} framework. 
In particular, users can add new future collider options or consider different combinations of those already implemented, {\it e.g.}, to investigate alternative running scenarios such as the descoped FCC-ee, by adapting the current data and theory files as well as the associated run cards. It is also straightforward to include new observables in the future collider projections and quantify their impact relative to the baseline dataset. 
Our framework thus provides an open, reproducible, and customisable platform for current and future collider projections, accessible to the wider community and readily adaptable to specific physics objectives, collider scenarios, and analysis needs, thereby helping to fully exploit the full potential of future colliders.

The structure of the paper is as follows. In Sect.~\ref{sec:interpretation}, we review the three interpretation formalisms adopted in this work: the modified couplings, effective couplings, and global SMEFT fit frameworks. In Sect.~\ref{sec:future_colliders_description}, we describe the main features of the future colliders considered and the corresponding inputs used in our analysis. Results for the $\kappa$-framework, the effective coupling formalism, and the global SMEFT fit are presented in Sects.~\ref{sec:kappa_results}, \ref{sec:effective_couplings_results}, and \ref{sec:global_smeft_fits_results}, respectively. In Sect.~\ref{sec:uv_benchmarks}, we study the reach of future colliders in the parameter space of benchmark UV models matched onto the SMEFT. Finally, in Sect.~\ref{sec:summary}, we summarise our findings, outline possible follow-up studies, and describe how to reproduce our results.

Additional technical information and dedicated benchmarks are provided in various appendices.
First, App.~\ref{app:operator_basis} summarises the definitions of the SMEFT operator basis used in this work. 
App.~\ref{app:effective-derivation} includes details on the effective coupling dictionary.
App.~\ref{app:kappa-validation} compares our results obtained in the kappa and SMEFT frameworks with the {\sc\small HepFit} ones presented for Snowmass 2022 study.
Subsequently, App.~\ref{app:pbb-bench} carries out a similar comparison with the ESPPU26 Physics Briefing Book (PBB)~\cite{deBlas:2025gyz} projections for individual EFT operators. 
Finally, App.~\ref{app:numerics_and_tables} collects inputs and supplementary numerical results from the kappa, effective and SMEFT analyses.

\section{Interpretation frameworks}
\label{sec:interpretation}

Here we present the interpretation frameworks that we consider in this work to assess the reach of future particle colliders.
We start by describing the coupling modifier framework, {\it i.e.} the $\kappa$-framework (Sect.~\ref{subsec:kappa_formalism}), 
then we review the SMEFT formalism (Sect.~\ref{sec:global_smeft_fits})
and finally the effective couplings approach (Sect.~\ref{sec:effective_couplings}). 
\subsection{Coupling modifier analysis}
\label{subsec:kappa_formalism}

The coupling modifier framework~\cite{CMS:2018uag,ATLAS:2019nkf,deBlas:2019rxi}, more commonly known as the $\kappa$-framework, is based on the characterisation of Higgs properties in terms
of multiplicative coupling strength modifier parameters $\kappa_{i}$. While the $\kappa$-framework is theoretically less general than the SMEFT, it remains an important and widely used benchmark for Higgs phenomenology. Its usefulness lies in the fact that it provides a simple and experimentally intuitive parametrisation of possible departures from the SM in terms of rescaled production and decay couplings, thereby ignoring shape modifications, yet closely matching the language in which many Higgs measurements are reported.  As such, the $\kappa$-framework continues to play a central role in the comparison of projected Higgs sensitivities across present and future collider facilities, despite the fact that it probes only a limited subset of possible NP effects and cannot match the full consistency and model-independent scope of the SMEFT. 

A further practical consideration is that the resulting fit is intrinsically non-linear, since the observables depend on products and ratios of coupling modifiers, in particular through the total Higgs width. 
This motivates a dedicated implementation of the $\kappa$-framework within {\sc\small SMEFiT}. 

In the $\kappa$-framework a given Higgs production and decay process is decomposed as follows:
\be
\label{eq:signal_strenght}
\lp \sigma \cdot {\rm Br}\rp ( i \to H \to f) = \frac{\sigma_i \cdot \Gamma_f}{\Gamma_{\rm H}} \, ,
\ee
where $i$ labels the particles involved in the production mode, $f$ the ones in the final state, and $\Gamma_{\rm H}$ is the total Higgs width.
In  terms of the coupling modifiers to the SM predictions, the  signal strengths in Eq.~(\ref{eq:signal_strenght}) become
\be
\lp \sigma \cdot {\rm Br}\rp ( i \to H \to f) = \frac{\lp \sigma_i^{(\rm SM)} \kappa_i^2 \rp \cdot \lp \Gamma_f^{(\rm SM)} \kappa_f^2 \rp}{\Gamma^{\rm (SM)}_{\rm H}\kappa_H^2} \, ,
\ee
or in terms of signal strengths normalised to the SM predictions, 
\be
\mu_i^f \equiv \frac{\lp \sigma \cdot {\rm Br}\rp ( i \to H \to f)}{\lp \sigma \cdot {\rm Br}\rp ( i \to H \to f)_{\rm SM}} \, ,
\ee
one obtains the following compact expression
\be
\label{eq:signal_strenght_kappas}
\mu_i^f =  \frac{\kappa_i^2 \kappa_f^2}{\kappa_H^2} \, ,
\ee
in terms of the three relevant coupling modifiers: production, decay, and total width.

In Eq.~(\ref{eq:signal_strenght_kappas}), the modifier of the total Higgs width is defined in terms of all coupling modifiers weighted by the values of the SM partial decay widths, namely
\be
\kappa_H^2 \equiv \sum_j \kappa_j^2 \frac{\Gamma_j^{(\rm SM)}}{\Gamma^{\rm (SM)}_{\rm H} } \, ,
\ee
where the sum over $j$ runs over the possible final states in which the Higgs boson may decay.
Within this modified coupling formalism, one can choose to resolve loop-induced processes either in production or decay or else leave them unresolved and include an effective coupling modifier. 
Here, for the sake of comparison with previous studies, we adopt the latter and hence leave loop-induced processes unresolved.

In this modified coupling framework, projections for Higgs production and decay measurements at future colliders are interpreted in terms of the ten parameters defined in Table~\ref{tab:kappadef}. Regarding $\kappa_t$, we note that since in the $\kappa$-framework-fit we include neither off-shell probes nor single-top processes, $\kappa_t$ cannot be accessed  below the $t\bar{t}H$ kinematic threshold. 
For Higgs production via VBF we follow~\cite{ATLAS:2019nkf} using the parametrisation of the production cross-section
\bea
   \frac{\sigma({\rm VBF})}{ \sigma_{\rm SM}({\rm VBF})}&=& 0.73\,\kw^2+0.27\,\kz^2 \, ,
\eea
in terms of the coupling modifiers.

\begin{table}[t]
\renewcommand{\arraystretch}{1.3}
    \begin{center}
\begin{tabular}{ c|c }
\toprule
$\qquad \kappa_i\qquad $ & Relevant processes\\
\midrule
$\kappa_W$ & $H\to WW$ decays, vector-boson fusion (VBF), $VH$ production  \\
$\kappa_Z$ & $H\to ZZ$ decays, VBF, $VH$ production \\
$\kappa_c$& $H\to c\bar{c}$ decays\\
$\kappa_b$ & $H\to b\bar{b}$ decays \\
$\kappa_t$& $t\bar{t}H$ and $tH$ production, off-shell $t\bar{t}$ kinematical distributions~\cite{Martini:2021uey,Maltoni:2024wyh} \\
$\kappa_{\tau}$ & $H\to \tau^+\tau^-$ decays \\
$\kappa_{\mu}$ & $H\to \mu^+\mu^-$ decays\\
$\kappa_g$ & Higgs production in gluon fusion, $H\to gg$ decays\\
$\kappa_{\gamma}$ & $H\to \gamma \gamma$ decays \\
$\kappa_{\gamma Z}$&  $H\to \gamma Z$ decays \\
\bottomrule
\end{tabular}
\end{center}
    \caption{ Production and decay processes relevant to constrain each of the coupling modifier $\kappa_i$ parameters.    }
      \label{tab:kappadef}
\end{table}

%
In the simplest incarnation of the coupling modifier framework, no invisible or undetected decays of the Higgs boson are considered: this is denoted as the `kappa-0' framework, and the ten fitted parameters (when data or projections are available) are those described in Table~\ref{tab:kappadef}, namely
\be
\kappa_W, \kappa_Z, \kappa_c,\kappa_b, \kappa_t, \kappa_\tau, \kappa_\mu, \kappa_g, \kappa_\gamma, \kappa_{\gamma Z} \, .
\ee
If one or more $\kappa$-parameters do not contribute to any theoretical prediction, either because the center-of-mass energy is insufficient to access a given final state (e.g., $t\bar{t}H$ for $\kappa_t$), or because a decay channel is too rare or experimentally challenging to isolate (e.g., $H \rightarrow Z\gamma$ or $H \rightarrow c\bar{c}$ for $\kappa_{Z\gamma}, \kappa_c$), their values are fixed to the Standard Model expectation, $\kappa_p = 1$, in the fit described in \ref{sec:kappa_results}.
Under these assumptions, the modifier of the total Higgs boson decay width is:
\bea
\kh^2&=&0.577\kb^2+0.063\ktau^2+0.0291\kc^2+0.215\kw^2+0.0264\kz^2 \\
&+&0.0857\kg^2+0.00228\ka^2+0.00154\kza^2+0.000219\kmu^2\,. \nonumber
\eea

Together with the kappa-0 framework, in which no additional Higgs boson decays are assumed,  we  also present our results in the so-called `kappa-3' framework, where decays  into new light/exotic-states are considered.
In presence of new physics affecting the Higgs decays, the SM Higgs width can be written as
\begin{equation}
\Gamma_{H}= \frac{\kappa_H^2 }{1-\BrBSM} \Gamma_{H,{\rm SM}}=\bar{\kappa}_H^2\Gamma_{H,{\rm SM}} \, ,
\label{eq:gammaH_kappa3}
\end{equation}
where $\BrBSM$  indicates the branching ratio for Higgs bosons decaying via BSM interactions.
In turn, this branching ratio can be separated into $\Brinv$, containing Higgs decays into BSM particles escaping detection and therefore leading to missing momentum, because of their feebly or non-existing interactions, and $\Brund$, containing possibly undetected decays of the Higgs to final states that can not be distinguished by other SM Higgs decays.\footnote{A paradigmatic example is any scalar extension of the SM featuring a lighter-than-Higgs (pseudo)scalar $S$ decaying into light hadronic SM particles.}
The decay channel $H\rightarrow SS\rightarrow bb\bar{b}\bar{b}$ would for example contribute to Br$_{\rm und}$. 
In the kappa-3 framework, the total Higgs width coupling modifier $\bar{\kappa}_H$ in Eq.~(\ref{eq:gammaH_kappa3}) is therefore expressed as
\begin{equation}
\bar{\kappa}_H^2=\frac{\kappa_H^2}{1-\Brinv-\Brund} \, .
\end{equation}
The invisible branching ratio $\Brinv$ can be constrained both at future and present colliders by targeting missing  momentum/energy signatures in Higgs decays.
By construction, no direct measurement can constrain $\Brund$, however its value can be inferred by measuring the total Higgs width $\Gamma_H$ in a simultaneous fit together with the coupling modifiers. 
Future lepton colliders play a crucial role in this regard by measuring the $\ell^+\ell^-\rightarrow ZH$ inclusive cross-section with the $Z$-recoil method, which enables a rather model-independent constraint on the total Higgs boson width.

For colliders where there is no path to constrain the total width $\Gamma_H$ independently from a global fit, such as proton-proton or lepton-proton colliders, the kappa-3 fit becomes degenerate due to the unconstrained $\Brund$.
To bypass this problem, it is customary to impose an additional condition to the coupling modifier for vector bosons, namely
\begin{equation}
\label{eq:kvcond}
\kappa_V\le 1, \qquad {\rm for}\quad V =W,Z.
\end{equation}
The $\kappa_V$ bound in Eq.~(\ref{eq:kvcond}) is motivated by the fact that UV-complete models which increase the coupling between Higgs bosons and vector bosons as compared to the SM are very constrained~\cite{Falkowski:2012vh}. 
Therefore, in the kappa-3 framework, the fitted parameters are:
\be
\kappa_W, \kappa_Z, \kappa_c,\kappa_b, \kappa_t, \kappa_\tau, \kappa_\mu, \kappa_g, \kappa_\gamma, \kappa_{\gamma Z}, {\rm Br}_{\rm inv}, {\rm Br}_{\rm und} \, ,
\ee
hence extending kappa-0 with the Higgs branching ratio into invisible, ${\rm Br}_{\rm inv}$ and undetected ${\rm Br}_{\rm und}$ final states, with the additional constraint of Eq.~(\ref{eq:kvcond}) for hadronic and lepton-hadron colliders.

In Sect.~\ref{sec:kappa_results} we present projections for future colliders in different variants of the modified couplings framework: kappa-0, kappa-3, and two restricted versions of kappa-3 assuming the universality of Higgs couplings. 
We present results both without imposing the constraint of Eq.~(\ref{eq:kvcond}), in which case only lepton colliders can be considered, and with this additional constraint included, which allows all colliders to be compared on a consistent footing.

As mentioned above, to carry out the analyses of future projections in these two variants of the $\kappa$-framework, {\sc\small SMEFiT} has been extended with new functionalities which enable an arbitrary functional dependence of the fitted observables on the parameters of interest (in this case the coupling modifiers $\kappa_{\rm i}$), to be defined by the user through the run card. 
Examples of run cards reproducing the $\kappa$-framework fits of this paper are found in the {\sc\small SMEFiT} repository, see Sect.~\ref{sec:summary} for more details. 
\subsection{The SMEFT framework}
\label{sec:global_smeft_fits}

The SMEFT~\cite{Buchmuller:1985jz,Grzadkowski:2010es, Brivio:2017vri,Isidori:2023pyp,Aebischer:2025qhh} provides a powerful and flexible framework to interpret high-energy physics data while reducing model assumptions.
By systematically extending the SM with higher-dimensional operators that encode the low-energy effects of heavy new particles, the SMEFT captures a broad class of potential deviations using a consistent quantum field–theoretic approach and can be matched to a plethora of UV completions to the SM.
This framework enables measurements from collider experiments and low-energy observables to be combined coherently within a unified parametrisation of new physics, searching for potential deviations with respect to the SM  while remaining reasonably agnostic about the underlying UV theory.
See~\cite{Brivio:2019ius,Aoude:2020dwv,Bruggisser:2022rhb,Grunwald:2023nli,Allwicher:2023shc,Bartocci:2023nvp,Ellis:2020unq,Garosi:2023yxg,deBlas:2025xhe,Mantani:2025bqu,Hirsch:2025qya} for recent SMEFT interpretations of LHC processes, often complemented with precision observables from LEP and eventually also with low-energy and flavour data.

The settings for the global SMEFT analyses presented in this work follow closely recent {\sc\small SMEFiT} studies~\cite{Celada:2024mcf,terHoeve:2025gey,terHoeve:2025omu} with a number of extensions.
Here we focus on describing additions or modifications with respect to these previous studies, and point the reader to those references for further details. 

\paragraph{Operator basis.}
In our recent study of the Higgs self-coupling~\cite{terHoeve:2025omu}, we extended the {\sc\small SMEFiT} operator basis of Ref.~\cite{Celada:2024mcf} by including two-lepton-two-quark operators involving heavy quark fields. In the present work, we further enlarge this basis to include the corresponding operators with light quark fields.
At the LHC, the two-lepton-two-quark operators can be constrained from Drell-Yan (DY) production, which as discussed in Sect.~\ref{sec:future_colliders_description} is now part of the input LHC dataset.\footnote{
Two-lepton-two-quark operators can also be constrained from flavour observables, whose inclusion is left for future work. }
Additionally, these operators can be directly constrained at future leptonic colliders, both for light and for heavy quark fields, in the latter case provided data at or above the $t\bar{t}$ threshold becomes available.
Their inclusion is actually required for the interpretation of $e^+e^- \to t \bar{t}$ measurements at lepton colliders above the top-quark pair threshold, including for indirect constraints on the Higgs self-coupling.
Therefore, the relevant two-lepton-two-quark operators added to our fitting basis are associated to the following Wilson coefficients:
\begin{equation}
\label{eq:2L2Qoperators}
c_{Q\ell}^{(-)}, \, c_{Q\ell}^{(3)}, \, c_{Q e}, \, c_{t\ell}, \, c_{t e} ,\, 
c_{q\ell}^{(-)}, \, c_{q\ell}^{(3)},\, c_{q e}, \,c_{\ell u},\, c_{\ell d},\, c_{e u},\, c_{e d} \,  .
\end{equation}
In addition, we now also include the four-lepton operators that are constrained both from the EWPOs as well as from difermion production at electron-positron colliders. In particular, we consider the Wilson coefficients $c_{\ell \ell}$, $c_{\ell\ell}'$, $c_{\ell e}$ and $c_{ee}$. We also include the muon Yukawa operator with coefficient $c_{\mu\varphi}$. A last difference with respect to the operator basis used in~\cite{Celada:2024mcf} is the inclusion of the coefficient $c_\varphi$, already present in \cite{terHoeve:2025gey,terHoeve:2025omu}. This purely-Higgs Wilson coefficient enters in the Higgs trilinear coupling, and is therefore of utmost interest at future particle colliders. 

The baseline flavour symmetries assumed in our basis are 
\be
U(2)_{q_L}\times U(2)_{u_R}\times U(3)_{d_R} \,,
\ee
in the quark sector and
\be
\label{eq:smeft_predictions_flav_lepton}
(U(1)_{\ell}\times U(1)_e)^3 \,,
\ee
in the lepton sector. While these symmetries are adopted for all the {\sc\small SMEFiT} predictions, at the level of the fits presented in Sect.~\ref{sec:global_smeft_fits_results} we enforce an additional symmetry in the lepton sector,
\be
U(2)_{q_L}\times U(2)_{u_R}\times U(3)_{d_R}\times U(3)_{\ell}\times U(3)_e \, .
\label{eq:flavour_symmetry}
\ee 
The latter choice is motivated by practical considerations of both operator counting and sensitivity. 
Retaining the lepton-sector flavour assumptions used for the {\sc\small SMEFiT} predictions, Eq.~(\ref{eq:smeft_predictions_flav_lepton}), would yield a 
fitting basis of approximately 130 operators. A significant fraction of these would enter the 
observables of the present analysis only at one-loop level, meaning their constraints would be 
driven primarily by running effects rather than by direct tree-level sensitivity. This is, for instance, the case of the four-lepton operators involving second- and third-generation leptons. Fitting the more general flavour assumptions
of Eq.~(\ref{eq:smeft_predictions_flav_lepton}) would therefore result in poorly determined coefficients and quasi-flat directions. 
Adopting the extended lepton-flavour 
universal symmetry of Eq.~\eqref{eq:flavour_symmetry} instead reduces the operator count to a 
manageable level and ensures that every coefficient in the fitting basis enters at least one 
observable at tree level, guaranteeing meaningful sensitivity throughout.

In total, we consider $n_{\rm op}=61$ independent dimension-six operators in the fitting basis used in 
Sect.~\ref{sec:global_smeft_fits_results}, with the possibility to relax the flavour assumption 
of Eq.~\eqref{eq:flavour_symmetry} in future work.
For completeness, the definition of the complete operator basis is reviewed in App.~\ref{app:operator_basis}.
Furthermore, note that in this work we absorb factors of $\Lambda^{-2}$ (the EFT cut-off) in a redefinition of the Wilson coefficients, such that the SMEFT Lagrangian truncated at dimension-six reads
\be
\mathcal{L}_{\rm SMEFT} = \mathcal{L}_{\rm SM} + \sum_{i=1}^{n_{\rm op}} c_i \mathcal{O}_i^{(6)} \, ,
\ee
with $\mathcal{O}_i^{(6)}$ the $d=6$ Warsaw-basis operators, and hence the coefficients $c_i$ will be dimensionful.
\subsection{The effective coupling formalism}
\label{sec:effective_couplings}

Future colliders will enable precise measurements of the Higgs and electroweak boson couplings to SM particles, allowing potential NP effects to be identified through deviations from their predicted values. In the effective coupling formalism\cite{Barklow:2017awn} one defines effective couplings as pseudo-observables, in the SMEFT context these can be expressed in terms of Wilson coefficients. 

The details of the derivation of the dictionary between the effective couplings and the Warsaw basis are discussed in App.~\ref{app:effective-derivation}.
Here we limit ourselves to stating the definition of the effective Higgs couplings $g^{\rm (eff)}_{HX}$ to the particle(s) $X$ in terms of partial widths, namely  
\be
\label{eq:gHeff}
    g^{\rm (eff)}_{HX} \equiv \sqrt{\frac{\Gamma_{H \to X}}{\Gamma^{\rm (SM)}_{H \to X}}} \, .
\ee
The expression of $g^{\rm (eff)}_{HX}$ in terms of the SMEFT coefficients then follows from the corresponding one for $\Gamma_{H \to X}$. 
Note that the analogous pseudo-observable for the top quark  is not available due to kinematical reasons. However, it is still possible to define an effective coupling $g_{Htt}$ entering processes such as $ttH$ and $tH$ production.  

In analogy with Eq.~(\ref{eq:gHeff}), one can define the effective couplings of electroweak gauge bosons in terms of SMEFT coefficients using pseudo-observables.
Specifically, the left- and right-handed effective couplings of the $Z$ boson to fermions, $g_{fL}$ and $g_{fR}$, are defined in terms of $Z$-pole observables,
\begin{align}
    \Gamma_{Z\to f \bar{f}} &\equiv \frac{G_F m_Z^3}{3 \sqrt{2} \pi} N_C^f \lp |g_{fL}|^2 + |g_{fR}|^2\rp  \, , \\
    A_f &\equiv \frac{|g_{fL}|^2 - |g_{fR}|^2}{|g_{fL}|^2 + |g_{fR}|^2} \, ,\label{eq:gHeff2}
\end{align}
with $\Gamma_{Z\to f \bar{f}}$ being the $Z$-boson decay width into a $f\bar{f}$ pair, $A_f$ the forward-backward asymmetry, and $N_C^f$ the number of colors for the fermion $f$. 

Following the approach used in the ESPPU20~\cite{deBlas:2019rxi,EuropeanStrategyforParticlePhysicsPreparatoryGroup:2019qin}, we also consider here the anomalous Triple Gauge Couplings (aTGCs), namely $\delta g_{1,Z}$, $\delta \kappa_{\gamma}$, and $\lambda_Z$, defined from the SMEFT correction to the triple gauge interaction Lagrangian~\cite{Brivio:2017bnu}
\begin{equation}
\label{aTGClagrangian}
	-\frac{\mathcal L_{\text{TGC}}}{g_{VWW}} = 
	i g_{1,V} (W_{\mu \nu}^+W^{- \,\mu}V^{\nu} - W_{\mu}^+ V_{\nu} W^{- \,\mu \nu}) +
	i \kappa_V W_{\mu}^+W_{\nu}^- V^{\mu \nu} - \frac{\lambda_V}{{m_W}^2} V^{\mu \nu} W_{\nu}^{+\rho} W_{\rho \mu}^-
	\, ,
\end{equation}
where $V=\gamma,Z$, $W_{\mu \nu}^{\pm} = \partial_{\mu}W_{\nu}^{\pm} - \partial_{\nu}W_{\mu}^{\pm}$ and analogously for $V_{\mu \nu}$. The SM couplings are defined as $g_{\gamma WW}=e$, $g_{ZWW}=e \,\cw/\sw$, while $\kappa_{V} = 1+\delta \kappa_{V}$, $g_{1,V} = 1+\delta g_{1,V}$, $\lambda_{V} =\delta \lambda_{V}$ are the SMEFT corrections.

\paragraph{Analysis overview.}
All in all, for the implementation of the effective couplings formalism adopted in this work, we consider a total of 26 independent couplings, namely the 10 Higgs effective couplings
\be
\label{eq:eff_couplings_1}
g_{Htt},\,g_{Hbb},\, g_{Hcc},\, g_{H\mu\mu},\, g_{H\tau\tau},\, g_{HZZ}, \,g_{HWW}, \,g_{H\gamma\gamma}, \,g_{HZ\gamma},\, g_{Hgg} \, ,
\ee
the 13 effective couplings of weak vector bosons to fermions
\be
\label{eq:eff_couplings_2}
g_{We}, \,g_{W\mu}, \,g_{W\tau},
g_{eL}, \,g_{eR}, \,g_{\mu L}, \,g_{\mu R}, \,
g_{\tau L}, \, g_{\tau R}, \, g_{u L}, \, g_{uR}, \, g_{dL}, \, g_{dR} \, ,
\ee
and the three anomalous Triple Gauge Couplings
\be
\label{eq:eff_couplings_3}
g_{1Z}, \, \kappa_\gamma, \,\lambda_Z \, .
\ee
As discussed in App. \ref{app:effective-derivation}, a dictionary can be derived to connect these effective couplings to the Warsaw operator coefficients and hence the {\sc\small SMEFiT} operator coefficients. %
The relevant operators are the following purely bosonic operators
\be
 \mathcal{O}_{W}, \mathcal{O}_{\varphi G}, \mathcal{O}_{\varphi D},\mathcal{O}_{\varphi \Box},  \mathcal{O}_{\varphi W},\mathcal{O}_{\varphi B}, \mathcal{O}_{\varphi WB} \, ,
\ee
the two-fermion current operators
\begin{gather}
    \nonumber \mathcal{O}_{\varphi \ell_1},\mathcal{O}_{\varphi \ell_2},\mathcal{O}_{\varphi \ell_3},\mathcal{O}^{(1)}_{\varphi q},\mathcal{O}^{(1)}_{\varphi Q} \, ,\\[6pt]
    \nonumber \mathcal{O}^{(3)}_{\varphi \ell_1},\mathcal{O}^{(3)}_{\varphi \ell_2},\mathcal{O}^{(3)}_{\varphi \ell_3},\mathcal{O}^{(3)}_{\varphi q},\mathcal{O}^{(3)}_{\varphi Q}\, , \\[6pt]
    \mathcal{O}_{\varphi e},\mathcal{O}_{\varphi \mu},\mathcal{O}_{\varphi \tau},\mathcal{O}_{\varphi d},\mathcal{O}_{\varphi u}\, ,
\end{gather}
the Higgs-Yukawa operators
\be
\mathcal{O}_{t\varphi},\mathcal{O}_{b\varphi},\mathcal{O}_{c\varphi},\mathcal{O}_{\tau\varphi},\mathcal{O}_{\mu\varphi},
\ee
and the four fermions operator 
\be
\mathcal{O}_{\ell\ell},\ee
entering the EW parameter shifts, see Eq.~\eqref{eq:deltavev}. The dictionary is given by Eqs.~(\ref{eq:eff_couplings_gLf})--(\ref{eq:eff_couplings_gRf}) for $g_{R,f} $ and $g_{L,f}$, Eq.~(\ref{eq:deltagWl}) for $g_{W,\ell}$, Eqs.~(\ref{eq:gHff})--(\ref{eq:gHZa}) and
(\ref{eq:gHww})--(\ref{eq:gHZZ}) for the Higgs effective couplings, and Eqs.~(\ref{eq:atgc1})--(\ref{eq:atgc3}) for the anomalous triple gauge couplings.

Within the {\sc\small SMEFiT} framework, fits in the effective coupling framework are performed by restricting the global SMEFT analysis to the subset of Wilson coefficients entering the definitions of the effective couplings in Eqs.~(\ref{eq:eff_couplings_1})--(\ref{eq:eff_couplings_3}). All other SMEFT operators are set to zero at the fit level in our implementation. Subsequently  we apply the dictionary of App.~\ref{app:effective-derivation} to translate the results in terms of the effective couplings. 

Furthermore, when presenting results in this framework, RGE effects are always switched off, since their inclusion would generate theoretical inconsistencies in the definition of effective couplings.
Therefore, Higgs and electroweak effective couplings defined in this manner are scale independent. As discussed in App.~\ref{app:kappa-validation}, we compared our implementation of the effective couplings with that of {\sc\small HepFit} and found general agreement.

\section{Experimental input and future collider projections}
\label{sec:future_colliders_description}

This section summarises the experimental projections for future colliders based on the inputs submitted to the ESPPU26, adopting in all cases their assumed running scenarios and their estimates of statistical and systematic uncertainties.

We group future colliders in two categories, discussed in turn.
The first category consists of the colliders which will be interpreted in the 
three formalisms presented in the previous section,
namely the hadron collider HL-LHC and the electron-positron colliders LEP3, FCC-ee, and LCF.
The second category consists of those colliders which are studied exclusively in the modified couplings formalism of Sect.~\ref{subsec:kappa_formalism}, namely FCC-hh, MuCol, and LHeC. 
Note that the projections for the electron-positron colliders are also considered for the kappa-framework fits.
Table~\ref{tab:FutureCollider_runs} summarizes the running scenarios assumed for the projections considered in this work.
For each collider, we indicate the geometry, the number of interaction points (IPs), the beam polarisation, the centre of mass energy, and the integrated luminosity.

\begin{table}[t]
  \centering
  \footnotesize
   \renewcommand{\arraystretch}{1.65}
  \begin{tabularx}{\linewidth}{|X|c|c|c|c|c|}
   \toprule
Collider &Initial state& Geometry, \# IPs & Beam Polarisation $(e^-,e^+)$& Energy $(\sqrt{s})$  &  Luminosity ($\mathcal{L}_{\text{int}}$)  \\
\midrule
     \midrule
\multirow{3}{*}{LEP3} &\multirow{3}{*}{$e^+e^-$} &\multirow{3}{*}{Circular, 2 IPs}  &\multirow{3}{*}{Unpolarised}  & 91.2 GeV & 48 $\text{ab}^{-1}$ \\
     & & & & 160 GeV & 5.6 $\text{ab}^{-1}$    \\
     & & & & 230 GeV & 2.304  $
     \text{ab}^{-1}$      \\      
\midrule
\multirow{4}{*}{FCC-ee} &\multirow{4}{*}{$e^+e^-$} &\multirow{4}{*}{Circular, 4 IPs}  &\multirow{4}{*}{Unpolarised}  & 91.2 GeV & 205 $\text{ab}^{-1}$ \\
     & & &  & 161 GeV & 19.2 $\text{ab}^{-1}$    \\
     & & &  & 240 GeV & 10.8 $\text{ab}^{-1}$      \\
     & & & & 365 GeV & 3.12 $\text{ab}^{-1}$      \\
\midrule
     \multirow{5}{*}{LCF} &\multirow{5}{*}{$e^+e^-$}
     &\multirow{5}{*}{Linear, 1 IP}
     &\multirow{4}{*}{$(\mp 80\%, \pm 30 \%)$}  & 91.2 GeV & 0.1 $\text{ab}^{-1}$ \\
     & &  &  & 250 GeV & 3 $\text{ab}^{-1}$     \\
     & & &  & 350 GeV &   0.2 $\text{ab}^{-1}$    \\
     & & &  & 550 GeV &  4 (8) $\text{ab}^{-1}$     \\
     & & & $(\mp 80\%, \pm 20 \%)$   & 1000 GeV &  8 (0) $\text{ab}^{-1}$     \\
     \midrule
    \multirow{1}{*}{FCC-hh (*)} &\multirow{1}{*}{$pp$}  &  \multirow{1}{*}{Circular, 4 IP} & \multirow{1}{*}{unpolarised}  & 84 TeV & 30 $\text{ab}^{-1}$ \\
\midrule
\multirow{1}{*}{MuCol3 (*)} &\multirow{1}{*}{$\mu^+\mu^-$}  &  \multirow{1}{*}{Circular, 2 IP} & \multirow{1}{*}{unpolarised}  & 3 TeV & 1 $\text{ab}^{-1}$ \\
\midrule
\multirow{1}{*}{MuCol10 (*)} &\multirow{1}{*}{$\mu^+\mu^-$}  &  \multirow{1}{*}{Circular, 2 IP} & \multirow{1}{*}{unpolarised}  & 10 TeV & 10 $\text{ab}^{-1}$ \\
\midrule
\multirow{1}{*}{LHeC (*)} &\multirow{1}{*}{$e^-\,p$} &  \multirow{1}{*}{Hybrid, 1 IP} & \multirow{1}{*}{unpolarised}  & 1.2 TeV & 1 $\text{ab}^{-1}$ \\
\bottomrule
    \end{tabularx}
    \vspace{0.2cm}
  \caption{The assumed running scenarios for the LEP3, FCC-ee, LCF, FCC-hh, MuCol and LHeC colliders.
  Colliders marked with (*) are only considered for  the  kappa framework analyses.
  In the case of LCF, the luminosities correspond to the LCF1000 programme, while those in parentheses correspond to the LCF550 scenario.
\label{tab:FutureCollider_runs}}
\end{table}

Before turning to the future collider projections we discuss the updates to the (HL-)LHC dataset, in comparison with previous {\sc\small SMEFiT} studies,  that serves as the baseline for our analysis, 

\subsection{Updates in the (HL-)LHC dataset}
\label{subsec:updates_hllhc}

We describe here the updates in the LHC and HL-LHC datasets, highlighting differences with our previous studies~\cite{Celada:2024mcf,terHoeve:2025gey,terHoeve:2025omu}.

\paragraph{LHC.}
The LHC dataset used in this work follows the one adopted in~\cite{terHoeve:2025omu,terHoeve:2025gey} with two improvements.
First, the inclusion of neutral-current Drell-Yan production, specifically of the CMS $Z'$ dilepton search of Ref.~\cite{CMS:2021ctt} at $\sqrt{s}=13$ TeV.
This dataset is particularly relevant to constrain four-fermion operators composed of two light quark and two leptonic fields listed in Eq.~(\ref{eq:2L2Qoperators}) by means of energy-growing effects~\cite{Greljo:2021kvv,Greljo:2022jac}.
Second, the ATLAS measurement of $Z$ production in vector boson fusion (VBF) at 13 TeV~\cite{ATLAS:2020nzk}, which provides independent constraints on the modifications of the triple gauge couplings.

The CMS $Z'$ dilepton search of~\cite{CMS:2021ctt}, targeting the high invariant-mass region of the dilepton pair, can be recast as a measurement of neutral-current DY production as follows.
This analysis, being based on the full Run-2 statistics, extends the dilepton invariant-mass reach up to 6 TeV~\cite{Maguire:2017ypu}. 
In our study, we retain the $n_b=79$ bins with $m_{\ell\ell}\ge 500$ GeV.
For each bin $i$, the analysis reports the observed dilepton yield, $n_{i,\ell\ell}$, as well as the corresponding SM expectation $b_{i,\text{SM}}$, which contains contributions from DY, top, and diboson production. 
To isolate the DY signal, $n_{i}$, we subtract the non-DY background (top and diboson) from the observed yield,
\begin{equation}
    n_{i} = n_{i,\ell\ell}- b_i,
\end{equation}
where the non-DY background, $b_i$, is obtained as
\begin{equation}
    b_i=b_{i,\text{SM}} - a_i \; d_{i,{\rm SM}} \, .
\end{equation}
Here $d_{i,\text{SM}}$ denotes our prediction for the DY event yields based on {\sc\small MadGraph5\_aMC@NLO}~\cite{Alwall:2014hca}, and  $a_i$ represents the experimental acceptance and selection efficiency.
Theoretical predictions for the DY differential distributions in the SMEFT, $d_{i,\text{SMEFT}}$, have been computed using {\sc\small MadGraph5\_aMC@NLO} together with {\sc\small SMEFTsim}~\cite{Brivio:2020onw} and {\sc\small SMEFT@NLO}~\cite{Degrande:2020evl}, the latter by means of a dedicated version extended to two-lepton-two-quark (2$\ell$2Q) operators, for the LO and NLO QCD calculations respectively. 
Cross-checks have been performed between {\sc\small SMEFTsim} and this new implementation of 2$\ell$2Q operators in {\sc\small SMEFT@NLO}, finding agreement.
This way, we obtain the expected number of events in the $i$-th bin as a function of the Wilson coefficients as $y_i=a_i\;d_{i,\text{SMEFT}}$.
We furthermore assume that background processes entering this dataset are described by the SM.

Since we focus on the tails of the distributions, where a small number of events is observed, for this dataset the likelihood is non-Gaussian and given by the product of Poisson probabilities per bin:
\begin{equation}
\label{eq:DY_likelihood}
    \mathcal{L}(n_i,y_i)=\prod_{i=1}^m \frac{{y_i}^{n_i}}{n_i!}e^{-y_i} \, .
\end{equation}
The best-fit values for the Wilson coefficients are then determined by minimising the difference between the experimental and theoretical likelihoods,
\be
    D=-2\left(\log\mathcal{L}(n_i,y_i)-\log\mathcal{L}(n_i,n_i)\right)
    =2\sum_{i=1}^m \left( y_i - n_i + n_i \log\frac{n_i}{y_i}\right) \, ,
\ee
through the external likelihood module of {\sc\small SMEFiT}. 
We note that the Poissonian likelihood of Eq.~(\ref{eq:DY_likelihood}) introduces non-linear effects even in the $\mathcal{O}\lp \Lambda^{-2}\rp$ fits, and hence one may find non-Gaussian posteriors in linear EFT fits which include this DY dataset.
In practice, deviations from Gaussianity are small. 

Concerning the $Zjj$ production in VBF from ATLAS~\cite{ATLAS:2020nzk} at 13 TeV, here we include the $n_{\rm dat}=12$ bins of the absolute differential distribution in $\Delta \phi_{jj}$, namely the difference in azimuthal angle between the two tagged VBF jets of the event.
This distribution is known to provide sensitivity to modifications in the triple-gauge coupling $c_{WWW}$~\cite{Ellis:2020unq,Celada:2024cxw}.
For this dataset, theoretical SMEFT predictions have been computed at LO in {\sc\small MadGraph5\_aMC@NLO} interfaced with {\sc\small SMEFT@NLO}. We adopt the SM prediction obtained with {\sc\small Herwig7+Vbfnlo}, including theoretical errors, also given in Ref.~\cite{ATLAS:2020nzk}.

\paragraph{HL-LHC.}
Our strategy for generating HL-LHC projections from the extrapolation of existing LHC measurements follows the one adopted in~\cite{Celada:2024mcf}.
This strategy is based on Level-0 (L0) projections of Run II datasets with systematic uncertainties reduced by a factor 1/2 and statistical uncertainties rescaled according to the projected luminosity of 3 ab$^{-1}$. 

In addition to these extrapolated projections, we include (following~\cite{terHoeve:2025omu}) the dedicated HL-LHC projections  for Higgs pair production jointly from ATLAS and CMS ~\cite{ATLAS:2025wdq,ATLAS:2024voi,ATLAS:2025bsu,Collaboration:2928096,ATLAS:2025eii}.
As in~\cite{terHoeve:2025gey}, we  also include dedicated projections for the $m_{t\bar{t}}$ distribution in top-quark pair production, this time with higher statistics in the theoretical predictions to tame spurious fluctuations in the previous calculation. 
Furthermore, as compared to our previous studies, we also include dedicated projections for single Higgs production in the $\mu^+\mu^-$ and $Z\gamma$ channels following the new ESPPU26 projections of~\cite{ATLAS:2025eii}.

\subsection{Projections for lepton colliders}
Next we describe the projections for future lepton collider observables used in this work.
The full dataset, including theoretical SM and SMEFT predictions, and the projected experimental and theoretical uncertainties, are publicly available in the {\sc\small SMEFiT} database,
\begin{center}\url{https://github.com/LHCfitNikhef/smefit_database}
\end{center}
see also Sect.~\ref{sec:summary}, to which we refer the reader for more detailed information.

\paragraph{LEP3.}
LEP3 is a proposed electron-positron collider operating as an electroweak and Higgs factory in the LHC tunnel.
Here we follow~\cite{LEP3-ESPPU} and assume three different runs at three different $\sqrt{s}$ and two IPs for the operation of LEP3.
The $Z$-pole run at 91.2 GeV would run for 5 years at an integrated luminosity per year and per IP of 4.8 $\mathrm{ab}^{-1}$, resulting in a total luminosity of 48  $\mathrm{ab}^{-1}$. 
The $WW$ threshold run at 160 GeV would last 4 years, collecting a total of 5.6 $\mathrm{ab}^{-1}$.
The Higgs factory $ZH$ run at 230 GeV would collect a total of 2.304 $\mathrm{ab}^{-1}$.

Statistical uncertainties for LEP3 observables are obtained by rescaling the corresponding ones from the FCC-ee by the luminosity ratio $\sqrt{\mathcal{L}_{\rm LEP3}/\mathcal{L}_{\rm FCC-ee}}$, while the systematic uncertainties are kept the same as for FCC-ee (though we note that dedicated detector concepts and the corresponding simulations are missing for LEP3).
In terms of input observables, for LEP3 we consider the same processes as for FCC-ee, with the exception of those belonging to the $\sqrt{s}=365$ GeV run which are not accessible at LEP3.
We further neglect the effect of changing the centre-of-mass energy from $240$~GeV down to $230$~GeV, which makes our LEP3 projections somewhat optimistic.

\paragraph{FCC-ee.}
In the integrated FCC programme~\cite{FCC:2025lpp,FCC:2025uan}, an  electron-positron collider (FCC-ee) operating in a new 91 km tunnel would be followed by a hadron-hadron collider (FCC-hh) with centre of mass energies between 70 TeV and 120 TeV, depending on the chosen magnet technology.
The FCC-ee inputs used in this analysis are described in~\cite{Celada:2024mcf} and have been updated following the Feasibility Report~\cite{FCC:2025lpp,FCC:2025uan} studies submitted to the ESPPU26, with  projected experimental uncertainties taken from~\cite{Selvaggi:2025kmd,blondel_2025_sq5pm-c8334}. 

The $Z$-pole measurements considered for the FCC-ee are the following: the electroweak coupling constant $\alpha_{\rm EW}(m_Z)$, the total and partial $W$ decay widths $\Gamma_W$, the EWPOs, namely the total $Z$ width $\Gamma_Z$, the total cross section into hadrons $\sigma^{0}_{\rm had}$, the ratios to difermion final states $R_e,R_\mu,R_\tau$, $R_b$, $R_c$, and the corresponding forward-backward asymmetries $A_e, A_\mu, A_\tau,A_b$, $A_c$.

For the data-taking periods at $\sqrt{s}=240$ GeV and $\sqrt{s}=365$ GeV, we consider projections for single Higgs production, difermion production, and the total cross section into hadrons $\sigma^{0}_{\rm had}$. Specifically, for single Higgs production, we consider the $HZ$ channel (both inclusive cross section and its decays to the $b\bar{b}$, $c\bar{c}$, $gg$, $WW$, $ZZ$, $\gamma Z$, $\gamma\gamma$ $\tau^+\tau^-$ and $\mu^+\mu^-$ final states) and the VBF channel (only hadronic Higgs decays at 240 GeV, and all channels except $\mu^+\mu^-$ at 365 GeV). For difermion production, we include both the forward-backward asymmetries ($A_e, A_\mu, A_\tau,A_b$, $A_c$) and the associated partial ratios ($R_\mu$, $R_\tau$, $R_b$, $R_c$), except for the $e^+e^-$ final state which is included as total production rate (in addition to $A_e$),

Additional constraints provided by the FCC-ee projections arise from $W^+W^-$ production (total rate, branching fractions, and normalised optimal observables for the fully- and semi-leptonic final states) at 161 GeV, 240 GeV, and 365 GeV.
We also consider optimal observables for $t\bar{t}$ production at $\sqrt{s}=365$ GeV, where a 10\% selection efficiency to reconstruct the $t\bar{t}$ system is assumed.
Note that optimal observables in $t\bar{t}$ production are defined such that they also include the inclusive cross-section (total rate), differently from their counterparts in $W^+W^-$ production which are normalised and hence are only sensitive to shape distortions.

\paragraph{LCF.}
Several possible realisations of a linear electron-positron collider, from the International Linear Collider (ILC) in Japan to the Compact Linear Collider (CLIC) at CERN, have been proposed for consideration. 
Here we adopt as a benchmark the settings of the Linear Collider Facility (LCF) proposed in~\cite{LinearColliderVision:2025hlt}.
As indicated in Table~\ref{tab:FutureCollider_runs}, the LCF would run at five different energies, from a Giga-$Z$ run at 91.2 GeV and Higgs and top-quark factory runs at 250 GeV and 350 GeV, to the high-energy runs at 550 GeV and 1 TeV. 
An alternative running scenario is also being considered, where the 1 TeV run is replaced by a longer 550 GeV run collecting 8 $\mathrm{ab}^{-1}$ of luminosity.
In this work, these two scenarios are labelled as LCF550 and LCF1000.
In Table~\ref{tab:FutureCollider_runs} we indicate in brackets the luminosities corresponding to LCF550.

A unique feature of linear colliders is the possibility of polarising the lepton beams, which provides additional sensitivity to the electroweak couplings. 
The polarised $e^-$ and $e^+$ beams of the LCF with polarisations $P=(P_-,P_+)$ contain the following fractions of right and left-handed particles~\cite{ILCInternationalDevelopmentTeam:2022izu}:
\begin{equation}
    f_R(e^{\pm}) = \frac{1+P_{\pm}}{2}\, , \quad \quad f_L(e^{\pm})= \frac{1-P_{\pm}}{2}\, ,
\end{equation}
{\it e.g.} a beam with polarisation $P=(+80 \%, -20\%)$ results in handedness fractions of
\begin{equation}
   f_R(e^-) = 90\%,\quad  f_L(e^-) = 10\%,\quad f_R(e^+) = 40\%, \quad f_L(e^+) = 60\% \, ,
\end{equation}
and hence collisions dominated by the $e^-_R + e^+_L$ initial state. 
Since right- and left-handed fermions are charged differently under the  $SU(2)_L\otimes U(1)_Y$ gauge group, varying polarisations probe different aspects of the electroweak interactions, enhancing sensitivity to BSM deviations in electroweak couplings.

In terms of experimental input, for the LCF projections considered in this work we include the following observables:
\begin{itemize}
    \item $\alpha_{\rm EW}(m_Z)$ and EWPOs, measured at $\sqrt{s}=250$ GeV via radiative return.
    EWPOs are also measured in the Giga-$Z$ run at 91 GeV.
    \item Single Higgs production in the $ZH$ channel,  measured both inclusively and in  different decay channels at $\sqrt{s}=250$, 350 and 550 GeV.
    
    \item Single Higgs production in the VBF channel, accessible in the different final states at 350, 550 and 1000 GeV.
    For the run at $\sqrt{s}=250$ GeV, only the $b \bar b$ final state is available.
    
    \item The total rate and forward-backward asymmetries in difermion production, measured at all energies.
    
    \item $W^+ W^-$ optimal observables, included for all energies.
    
    \item $t \bar t$ optimal observables, included for all runs with $\sqrt{s}\ge 350$ GeV, with a 10\% selection efficiency assumed up to 550 GeV and then a 6\% efficiency for the 1000 GeV run.
    
    \item The high-energy runs
    at $\sqrt{s}=550$ GeV and 1000 GeV give access to $t\bar{t}H$ and double Higgs production, providing direct sensitivity to the top Yukawa coupling and to the Higgs self-coupling,  respectively.
    Here we include the total inclusive cross sections for $t\bar{t}H$ and $HH$ production at these two energies, for the latter considering both VBF ($W^+W^-$ fusion) and associated $ZHH$ production.
\end{itemize}

\subsection{Projections for hadron-hadron, lepton-hadron, and muon colliders}

Next we summarise the main features of the future colliders whose projections are used exclusively for the kappa framework analysis: the FCC-hh, the MuCol, and the LHeC.

\paragraph{FCC-hh.}
For the FCC-hh projections we use the ESPPU2026 projections for inclusive cross-sections presented in \cite{Selvaggi:2025kmd} and based on a center-of-mass energy of $\sqrt{s}=84$ TeV.
We note that the ESPPU26 inputs for the FCC-hh also contain differential distributions for processes such as diboson, top-quark pair, and Higgs production: these are not considered in this work and will be studied in follow-up analyses where FCC-hh projections are added to the global SMEFT fit.

\paragraph{Muon Collider.}
As in the case of the FCC-hh, here we consider only MuCol projections for the kappa framework studies, with their inclusion in the global SMEFT analysis left for future work. 
The MuCol input for the kappa analysis is based on inclusive cross-sections for Higgs boson production and decay~\cite{deBlas:2022ofj,Forslund:2022xjq}.
The running scenarios assumed are $\sqrt{s}=3$ TeV with $\mathcal{L}=1~ {\rm ab}^{-1}$ and $\sqrt{s}=10$ TeV with $\mathcal{L}=10~{\rm ab}^{-1}$.

\paragraph{LHeC.}
For the LHeC inputs to the kappa fit, we restrict ourselves to those of the ESPPU19 analysis~\cite{deBlas:2019rxi}. A more detailed and updated study, incorporating the impact of reduced LHeC uncertainties on HL-LHC measurements, has recently been presented in \cite{Ahmadova:2025vzd}.
These LHeC projections assume a total integrated luminosity of $\mathcal{L}_{\rm int}=1$ ab$^{-1}$ for a centre-of-mass energy of electron-proton collisions of $\sqrt{s}=1.2$ TeV.\\

\noindent
App.~\ref{app:kappa-additional} collects additional information on the experimental inputs for the kappa framework analysis.
Furthermore, for the kappa framework benchmarks presented in App.~\ref{subsec:kappa_framework_bench}, for consistency we adopt exactly the same experimental inputs as in the corresponding  Snowmass 2022 {\sc\small HepFit} analysis, which differ from those listed in Table~\ref{tab:future_colliders_kappafits}.

\subsection{Theoretical calculations and uncertainties}
\label{subsec:th_errors}

Here we summarise our setup to obtain theoretical predictions and describe how theoretical uncertainties on the SM calculations are accounted for. 

\paragraph{Theoretical calculations.}
We account for NLO QCD corrections in the SMEFT for most LHC and HL-LHC processes considered~\cite{Celada:2024mcf}.
Following~\cite{terHoeve:2025gey}, one-loop RGE effects in both the strong and electroweak couplings are included for all operators in our fitting basis.
NLO electroweak corrections to $ZH$ production at lepton colliders are also included~\cite{Asteriadis:2024xts,Asteriadis:2024xuk}.\footnote{NLO electroweak corrections to other processes relevant for future collider studies have been recently made available, such as to EWPOs~\cite{Biekotter:2025nln} and Higgs branching ratios~\cite{Bellafronte:2025jbk}.
Their implementation is left for future work. } 
Recently, the complete two-loop RGE anomalous dimension matrix for SMEFT operators has been 
computed~\cite{Born:2026xkr}. 
A companion paper~\cite{Mantani:2026xyz} studies their impact for the same HL-LHC and  FCC-ee datasets considered in this work, finding the overall effect to be small.
Therefore, inclusion of two-loop RGE effects does not 
significantly affect the results presented here, particularly in the quadratic fits, while some 
residual sensitivity is found in the linear fits, most notably for $c_{\varphi G}$, $c_{t\varphi}$, 
and the four-heavy-quark operators. 
We refer the reader to~\cite{Mantani:2026xyz} for a 
detailed discussion.

\paragraph{Theoretical uncertainties.}
In this work, we consider multiple sources of theoretical uncertainties on the SM calculations following the discussions in the ESPPU2026 PBB~\cite{deBlas:2025gyz}. 
First, we include SM theory uncertainties due to missing higher-order calculations on the following lepton collider observables:  Higgs production and decay, $Z$-pole EWPOs, $\alpha_{\rm EW}$, $W$-boson width and branching ratios.
Then, the extraction of $Z$-pole EWPOs from experimental data requires theory input, for example for background subtraction, leading to another source of theoretical uncertainties. 
Finally, we include for the $Z$-pole EWPOs and for the $\alpha_{\rm EW}$ extraction the effects of parametric uncertainties arising from the propagation of experimental uncertainties on the measurements of input parameters. 

The first two sources of theoretical uncertainties (missing higher orders and theory inputs entering the EWPO measurements) are implemented in this work following the four possible scenarios as discussed in the ESPPU2026 PBB~\cite{deBlas:2025gyz}:

\begin{itemize}
    \item \textit{``Current''}: theoretical uncertainties on SM processes at future lepton colliders are assumed to be of the same size as current estimates. 
    
    \item \textit{``Conservative''}: theoretical uncertainties on lepton collider processes are assumed to improve by a moderate amount, based on the expansion of existing higher-order computational techniques.
    
    \item \textit{``Aggressive''}: this scenario assumes a substantial improvement of  theoretical uncertainties, requiring breakthroughs in higher-order calculation methods.
    
    \item \textit{``Ideal''}: we assume all  SM theory uncertainties at future colliders are sub-dominant compared to experimental precision and can be neglected.

\end{itemize}

These four scenarios for future theoretical uncertainties are considered in the effective coupling formalism and the global SMEFT analysis of Sect.~\ref{sec:effective_couplings_results} and Sect.~\ref{sec:global_smeft_fits_results} respectively. 
The $\kappa$-framework fits in Sect.~\ref{sec:kappa_results} use instead the theoretical uncertainties of Ref.~\cite{deBlas:2022ofj}.
We note that in previous {\sc\small SMEFiT} analyses~\cite{Celada:2024mcf,terHoeve:2025omu}, only the ``current" and ``ideal" theory scenarios were considered, and uncertainties due to $Z$-pole EWPOs extraction were neglected.

Regarding the parametric uncertainties, they were assumed to be uncorrelated in previous {\sc\small SMEFiT} analyses, while here we include their full theory covariance matrix. 
This is done using the code presented in~\cite{Mildner:2024wbl} for LEP and future colliders by implementing the projected uncertainties on input parameters.
In the $m_W$ scheme used here, the input parameters relevant for these parametric uncertainties are $\{m_W, m_Z, G_F, m_H, m_t, \alpha_s\}$, for which we assume the expected precision of their determination at the corresponding future collider~\cite{deBlas:2025gyz}.

As already explained in~\cite{Celada:2024mcf}, in this work we neglect theory uncertainties on the EFT predictions as we already include NLO QCD corrections to the EFT cross-sections in (HL-)LHC processes, as well as NLO electroweak corrections to $ZH$ production for future lepton colliders. 
In addition, we ensure that the Monte Carlo statistical uncertainties on the EFT predictions are kept below $1 \%$.

\section{Coupling modifiers analysis}
\label{sec:kappa_results}

Here we present the results of fits to future collider projections based on the coupling modifiers formalism described in Sect.~\ref{subsec:kappa_formalism}.
First of all, we present the kappa-0 framework results, treating each future collider (including the HL-LHC) separately.
We then consider the kappa-3 framework, which also supports Higgs decays into invisible and undetected final states, when each collider is added to a HL-LHC baseline.
Finally, we show results for kappa-3 variants in terms of universal models consisting of one or two overall coupling modifiers. 
For reference purposes, numerical tables containing the results presented in this section are collected in App.~\ref{app:kappa-additional}. Within the $\kappa$-framework, we include only the experimental correlations available for the HL-LHC, in order to place all the future collider on an equal footing. This contrasts with the SMEFT fits, where we exploit the full set of available experimental inputs, such as the correlations among Higgs hadronic decay channels at FCC \cite{del_vecchio_2025_3jjdh-6fz97}.

\begin{figure}[t]
    \centering
\includegraphics[width=0.49\linewidth]{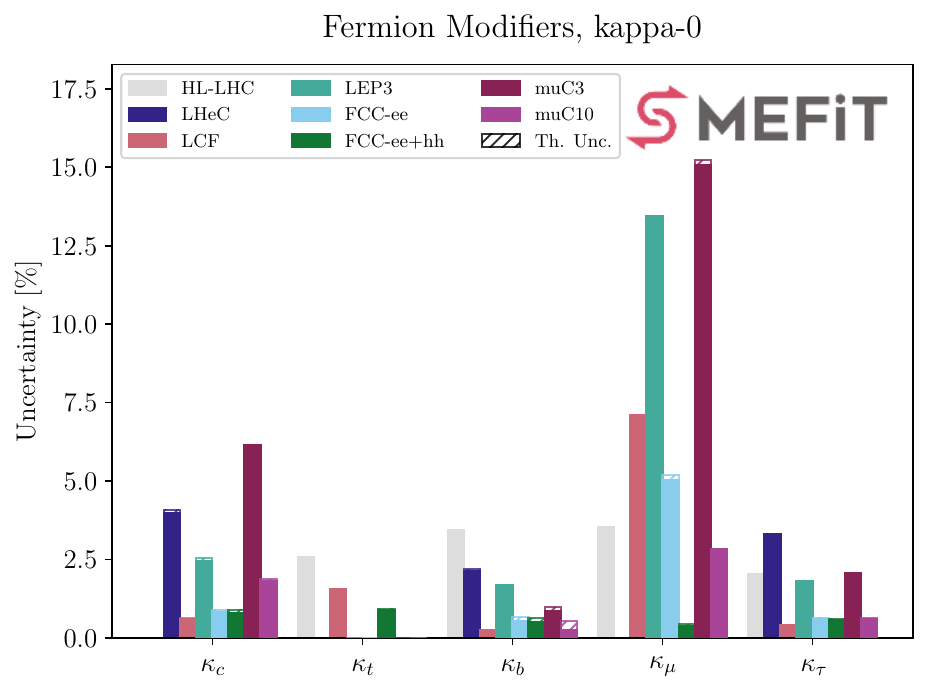} \includegraphics[width=0.49\linewidth]{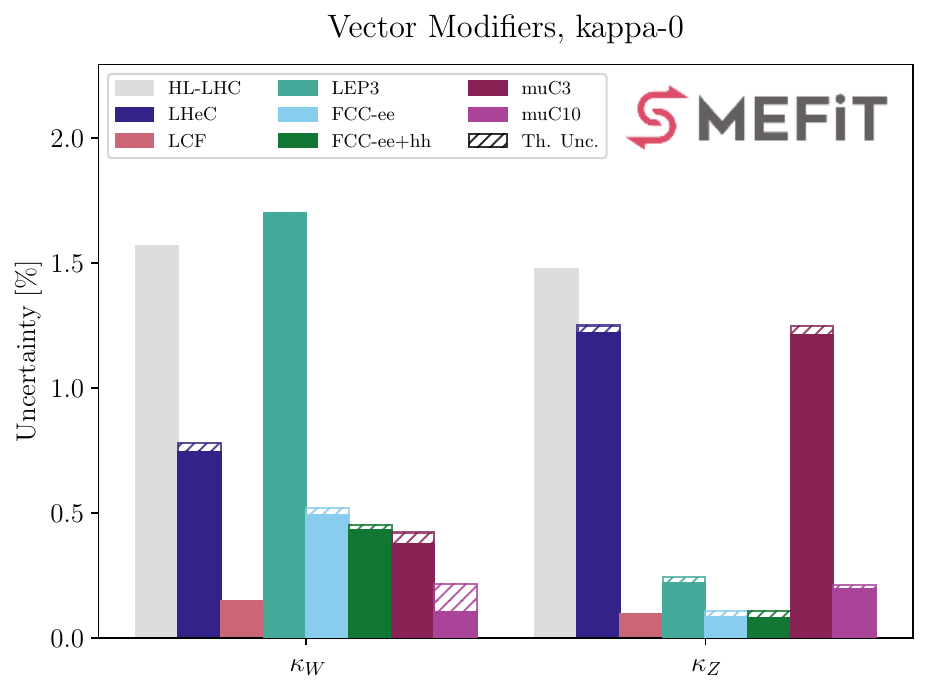}
\includegraphics[width=0.49\linewidth]{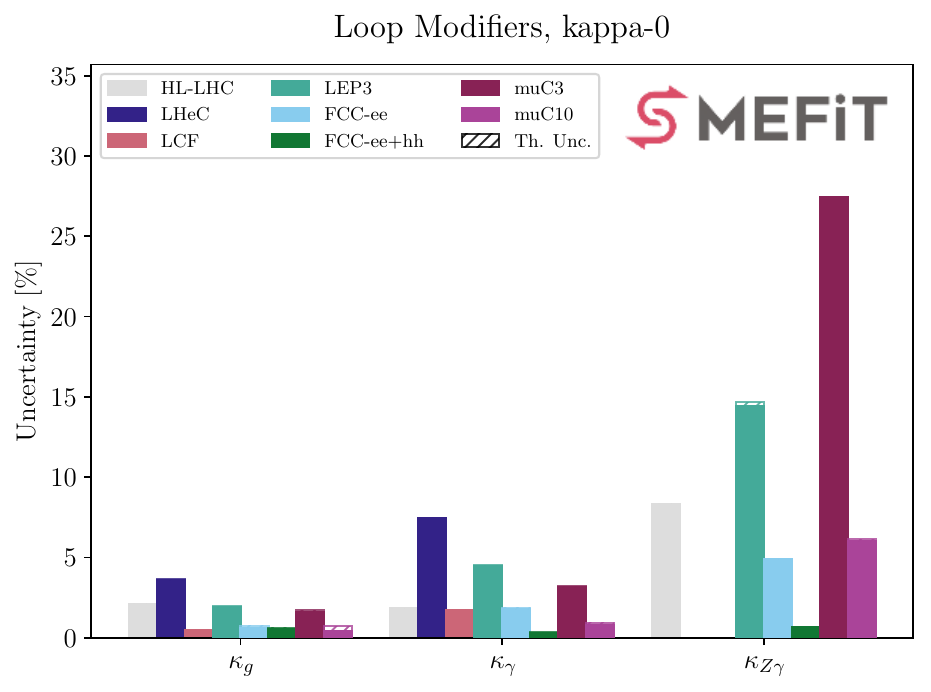}
    \caption{The projected percentage uncertainties (from the half-width of the 68\% C.I.) for the measurement of the coupling modifiers $\kappa_i$ for fermions, gauge bosons, and loop-induced processes respectively, in the kappa-0 framework.
    Each collider is treated separately, and in particular no common HL-LHC baseline is assumed. 
    The dashed bars indicate the sensitivity degradation once  theory uncertainties in the SM predictions are accounted for.
    For $\kappa_t$, $\kappa_\mu$, and $\kappa_{Z\gamma}$, missing bars indicate those colliders which cannot constrain the coupling modifiers due to lack of sensitivity.
    }
    \label{fig:kappa0_histo}
\end{figure}

\paragraph{Kappa-0 framework fits.}
Fig.~\ref{fig:kappa0_histo} displays the projected percentage uncertainty (as half the width of the 68\% C.I.\footnote{In this work, we compute the Credible Intervals (C.I.) as Highest Density Intervals (H.D.I.).})  for the measurement of the coupling modifiers $\kappa_i$ in the kappa-0 framework. 
We show separately the coupling modifiers for fermions ($\kappa_c,\kappa_b,\kappa_t,\kappa_\mu,\kappa_\tau$), for gauge bosons ($\kappa_W,\kappa_Z$), and for loop-induced processes ($\kappa_g,\kappa_\gamma,\kappa_{Z\gamma}$). 
Each collider is treated separately, without a common HL-LHC baseline being assumed. 
In this comparison, results for the following colliders are shown: HL-LHC, FCC-ee, FCC-ee together with FCC-hh, LCF (all $\sqrt{s}$ runs up to 1 TeV included), LEP3, LHeC, and the muon collider (both at $\sqrt{s}=3$ and 10 TeV).
The dashed bars indicate  the degradation in sensitivity as a consequence of the projected  parametric and intrinsic theory uncertainties considered in~\cite{deBlas:2019rxi} for the leptonic colliders. 

The results of Fig.~\ref{fig:kappa0_histo} indicate which of the proposed colliders could improve the coupling modifiers in the kappa-0 framework as compared to the HL-LHC expectations, and among these which exhibits the best precision.  
For instance, for $\kappa_W$ the best reach is achieved by LCF and MuCol10, with FCC (both ee and ee+hh) following closely. FCC-hh projections are however just a subset of the possible measurements, so this results could be improved with an extend input (something along this line was proposed in the context of the ESPPU 2019 in \cite{Mangano:2681378}).
For Higgs decays with small branching fractions, such as $\kappa_{Z\gamma}$ and $\kappa_\mu$, the integrated FCC programme leads to the highest sensitivity.
Note also that not all colliders can meaningfully constrain all coupling modifiers, for instance $\kappa_t$ requires a high energy collider such as LCF1000 or FCC-hh.
In terms of improvements as compared to the HL-LHC, electroweak couplings ($\kappa_W$, $\kappa_Z$), $\kappa_g$, and $\kappa_b,\kappa_\tau$ will all significantly benefit from future lepton collider constraints.

\paragraph{Kappa-3 framework fits.}
As discussed in Sect.~\ref{subsec:kappa_formalism}, the kappa framework can be extended to account for additional final states, which are either invisible or which cannot be resolved by a given collider, the kappa-3 framework.
In the following, whenever the constraint Eq.~(\ref{eq:kvcond}) is imposed to stabilise the fit, we  indicate it with a (*).
In this case, the uncertainty on the $\kappa_V$ modifier is defined as
\begin{equation}\label{eq:err_kv}
 \delta\kappa_V = 1 - \lp {\rm 68\%\,\, C.I.  \,on}\,\, \kappa_V  \rp  \, .
\end{equation} 
To ensure a consistent comparison, whenever this constraint is applied to a subset of colliders, it will also be applied to the lepton colliders.

\begin{figure}[htbp]
    \centering
\includegraphics[width=0.43\linewidth]{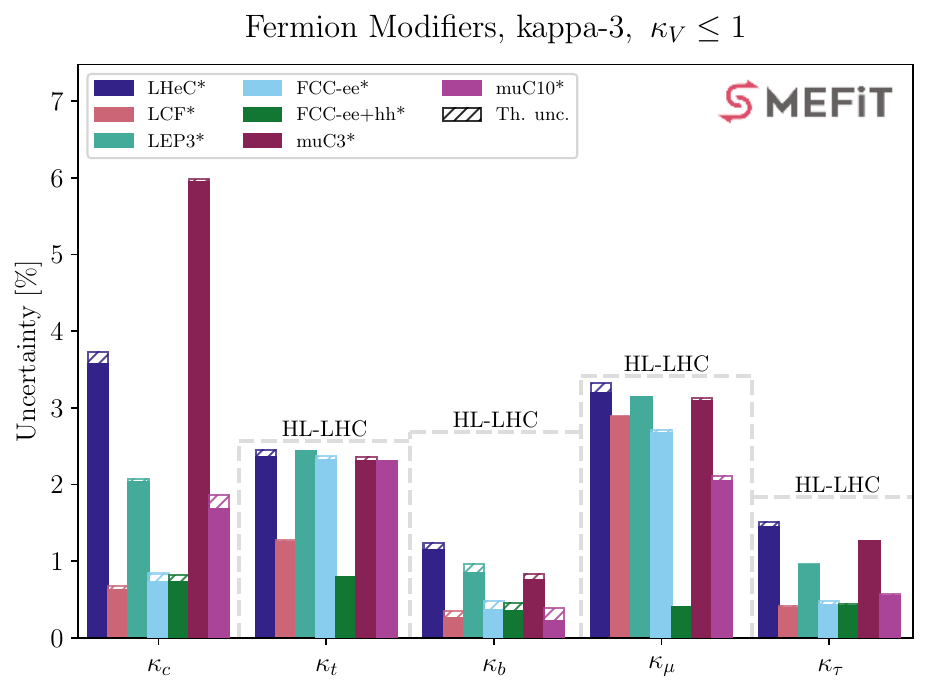}
\includegraphics[width=0.43\linewidth]{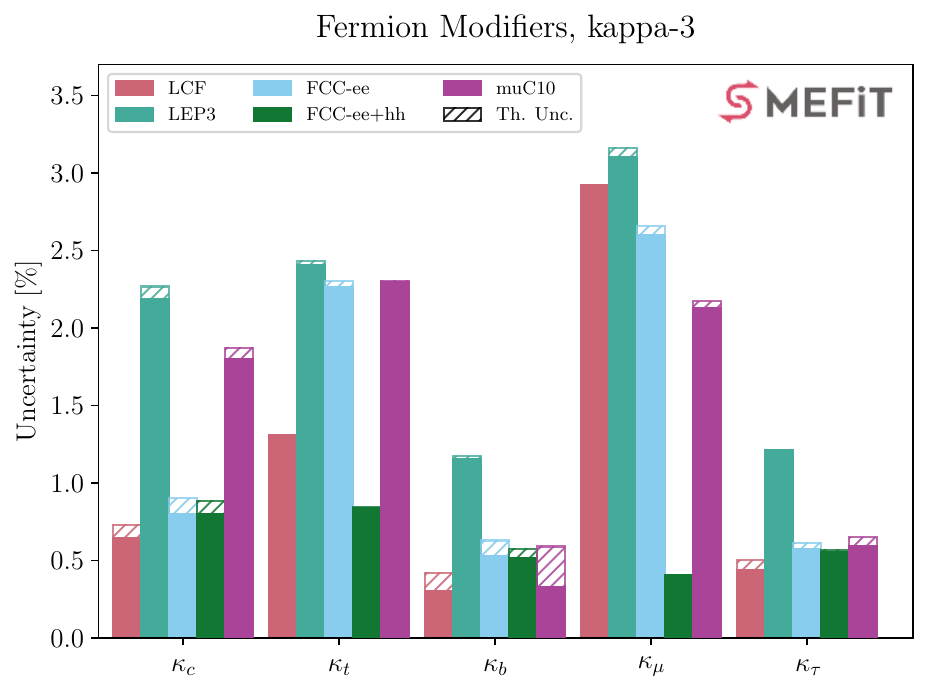}
\includegraphics[width=0.43\linewidth]{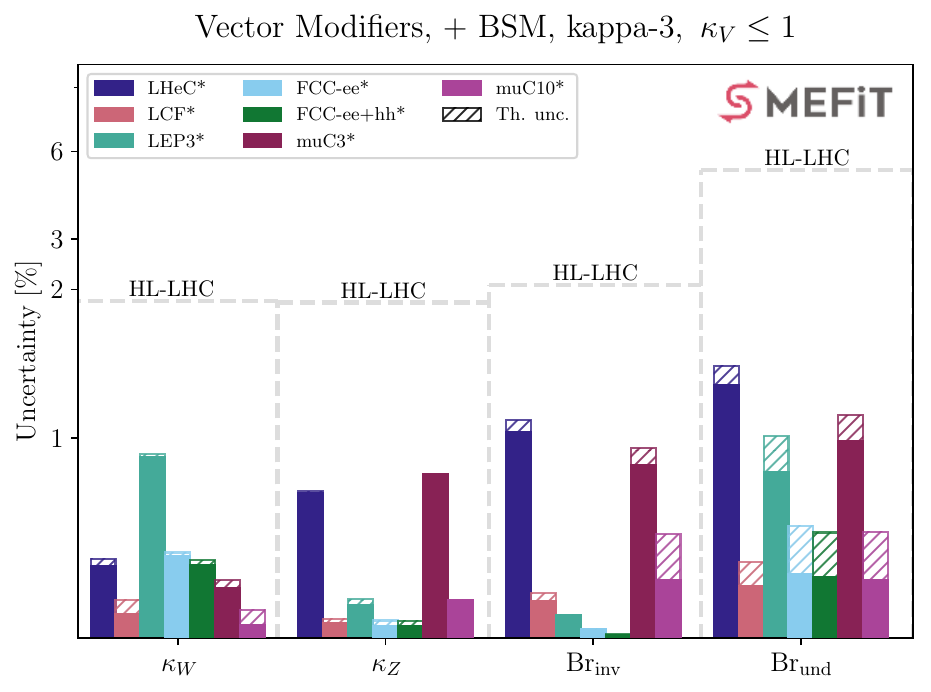}
\includegraphics[width=0.43\linewidth]{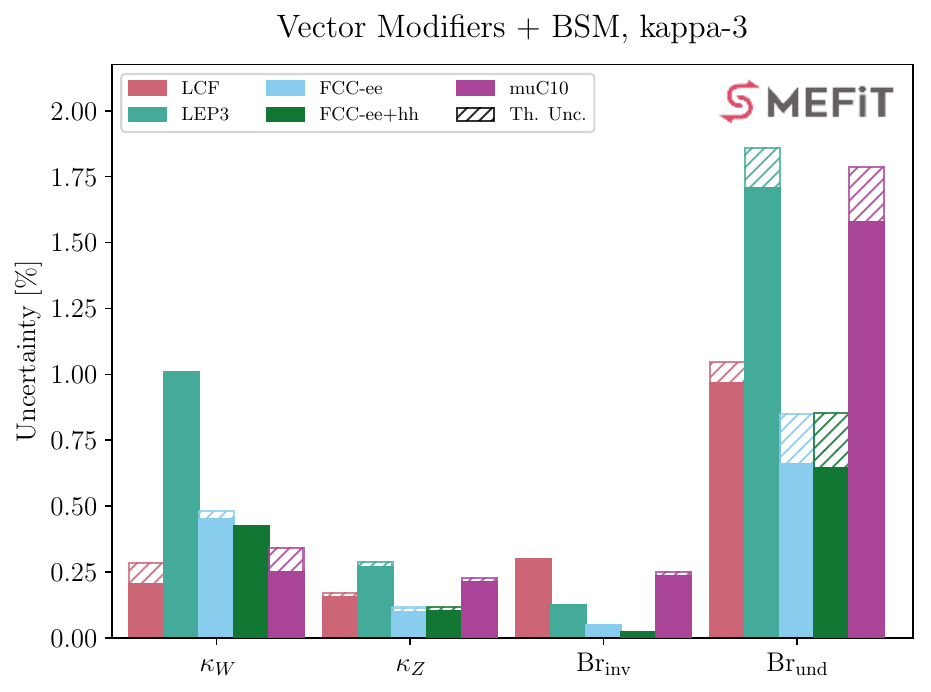}
\includegraphics[width=0.43\linewidth]{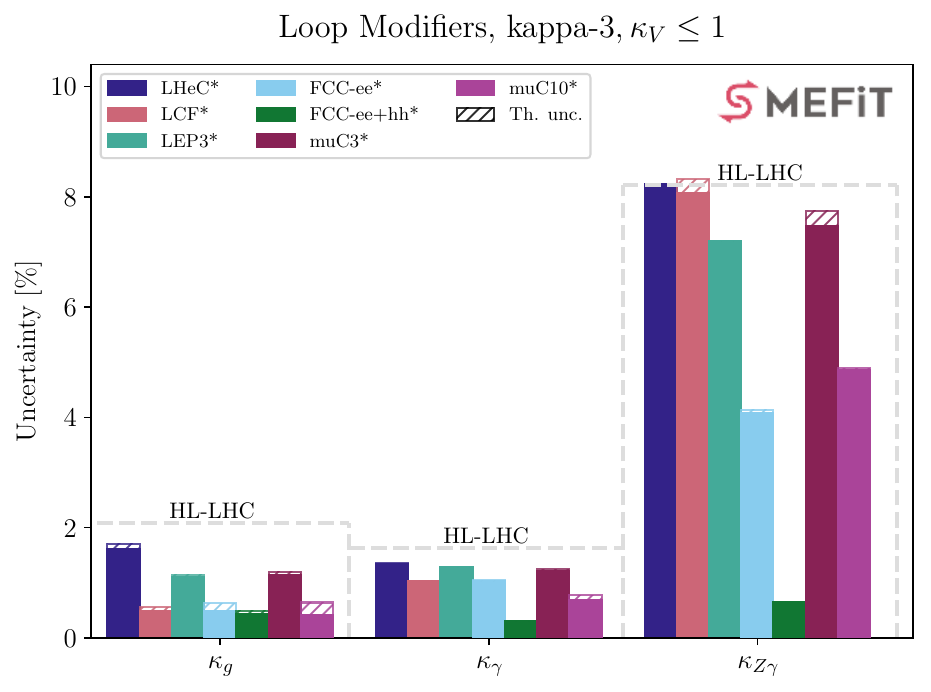}
\includegraphics[width=0.43\linewidth]{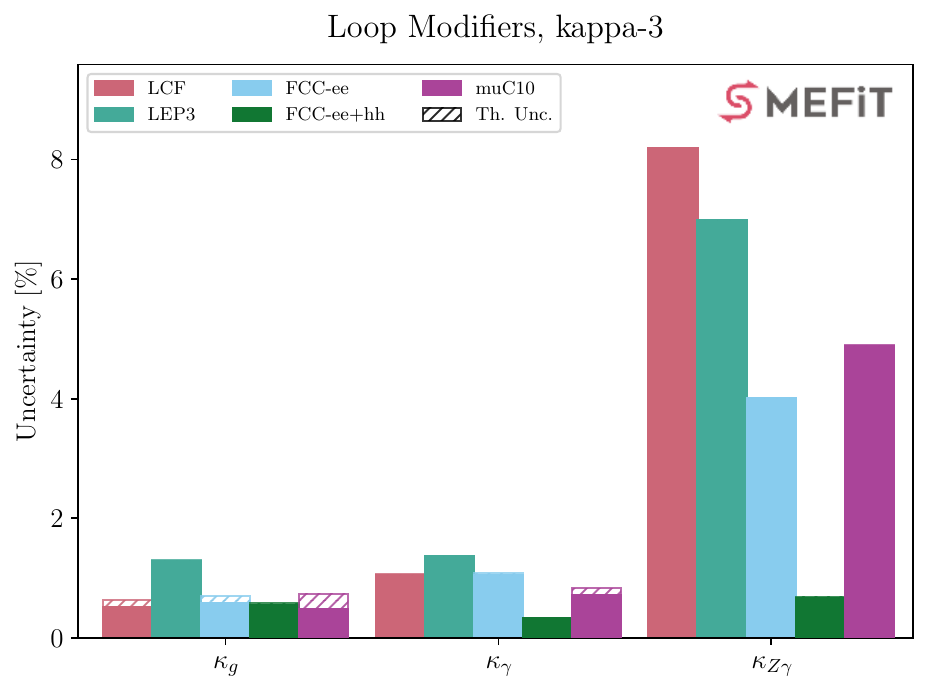}
        \caption{Same as Fig.~\ref{fig:kappa0_histo} for the fit results in the kappa-3 framework.
        In the left panels, we impose the condition $\kappa_V\le 1$ to stabilise the fit (indicated by a star next to a collider name), while in the right panels $\kappa_V$ is left unrestricted and hence only the colliders which can measure $\Gamma_H$ in a model-independent manner are kept.
        In these kappa-3  fits, a common baseline consistent on HL-LHC projections is assumed for all colliders.
        For reference, the left panels indicate the expectation for the coupling modifiers at the end of the HL-LHC.
    }
\label{fig:kappa3_histo_comparison}
\end{figure}

Fig.~\ref{fig:kappa3_histo_comparison} displays a similar comparison as in Fig.~\ref{fig:kappa0_histo} now for the results in the kappa-3 framework.
In the left panels, we impose the conditions in Eq.~(\ref{eq:kvcond}) to stabilise the fit, while in the right panels $\kappa_V$ is unrestricted.
Therefore, the right panels display results only for the colliders which can directly measure $\Gamma_H$ in a model-independent manner: FCC-ee, LEP3, LCF, and MuCol10.
As opposed to what is done in the kappa-0 fits of Fig.~\ref{fig:kappa0_histo}, now a common baseline consistent with HL-LHC projections is assumed for all future colliders.
For reference, the left panels indicate the expectation for the coupling modifiers at the end of the HL-LHC. 

From the results of the kappa-3 fits of Fig.~\ref{fig:kappa3_histo_comparison}, we can compare the reach of future colliders for the measurement of $\Brinv$ and $\Brund$, which is one of their core scientific goals.
For instance, from the unrestricted kappa-3 fit, we find that the integrated FCC programme has the best sensitivity to $\Brinv$, achieving 0.1\% precision, while for instance LCF would reach 0.3\% precision. 
Concerning $\Brund$, once theory uncertainties are accounted for, FCC-ee achieves the best precision (0.8\%), with LCF following closely.
Concerning the coupling modifiers $\kappa_i$, similar observations as compared to the kappa-0 fits can be drawn.
In particular, FCC-ee leads in general to the best precision on the coupling modifiers, with the possible exceptions of $\kappa_c$, $\kappa_b$, and $\kappa_W$, where LCF performs somewhat better.
We emphasise that the outcome of these fits may vary strongly should one make other choices for the collider scenarios, \textit{e.g.}~FCC-ee followed by MuCol would lead to a better sensitivity than a stand-alone MuCol.

\paragraph{Universal coupling modifier scenario.}
In different UV-complete scenarios, such as the Minimal Scalar Extension of the SM~\cite{OConnell:2006rsp}, all Higgs couplings are rescaled by a common factor with respect to their SM values. 
Motivated by these scenarios, we consider now a variant of the kappa-3 framework with a universal coupling modifier $\kappa_{\rm univ}$.
Furthermore, these models often contain new light scalars to which the Higgs boson can decay, resulting in non-zero invisible and undetected branching fractions.
Therefore, this simplified variant contains three free parameters: the universal modifier $\kappa_{\rm univ}$ and the branching ratios $\Brinv$ and $\Brund$.

\begin{figure}[t]
    \centering
\includegraphics[width=0.48\linewidth]{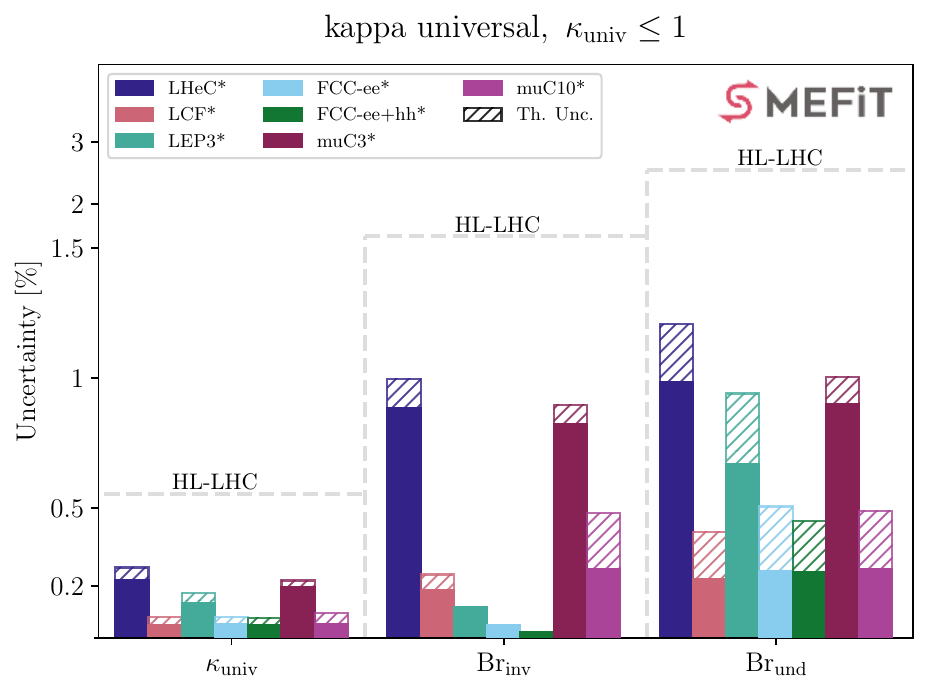}
\includegraphics[width=0.48\linewidth]{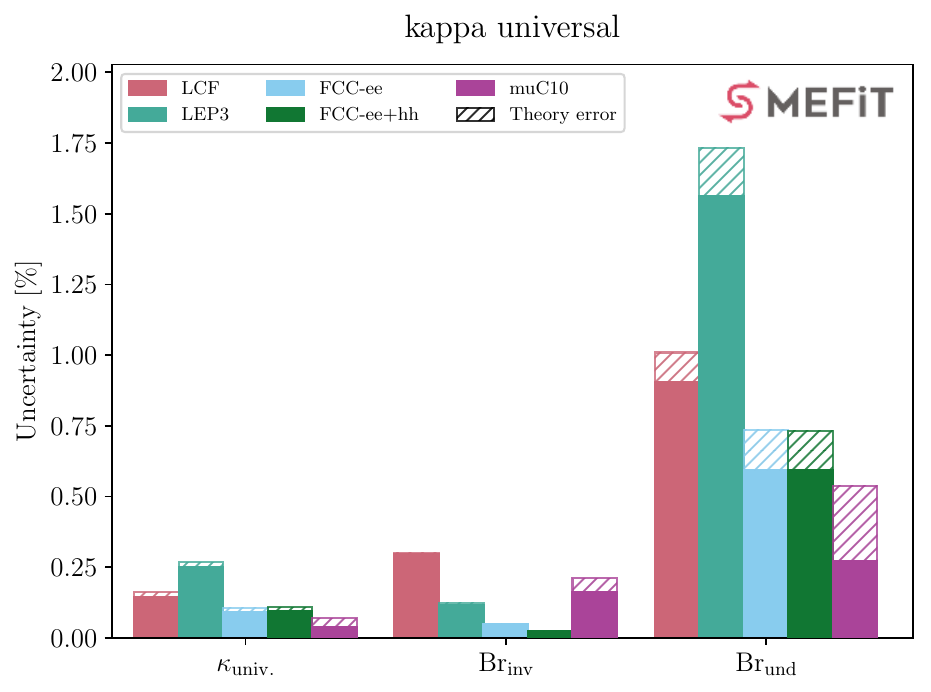}
    \caption{Same as Fig.~\ref{fig:kappa3_histo_comparison}
in the universal coupling modifier scenario.    
These variants of the kappa-3 fits contain three free parameters: the universal coupling modifier $\kappa_{\rm univ}$ and the branching ratios $\Brinv$ and $\Brund$. 
In both cases, projections for the indicated colliders are added on top of the HL-LHC baseline.
    \label{fig:kappa-universal}
    }
\end{figure}

 Fig.~\ref{fig:kappa-universal} presents, in the same format of Fig.~\ref{fig:kappa3_histo_comparison}, the results for the kappa-3 fits in the universal coupling modifier scenario, where projections for the indicated colliders are added on top of a common HL-LHC baseline for all colliders.
The conclusions for $\Brinv$ and $\Brund$ are the same as in the standard kappa-3 fits, highlighting the limited correlation between these parameters and the coupling modifiers.
Concerning the results for the universal coupling modifier $\kappa_{\rm univ}$, this time MuCol leads to the strongest bounds (0.1\% precision), with FCC-ee and LCF following closely.

\begin{figure}[htb]
    \centering
\includegraphics[width=0.48\linewidth]{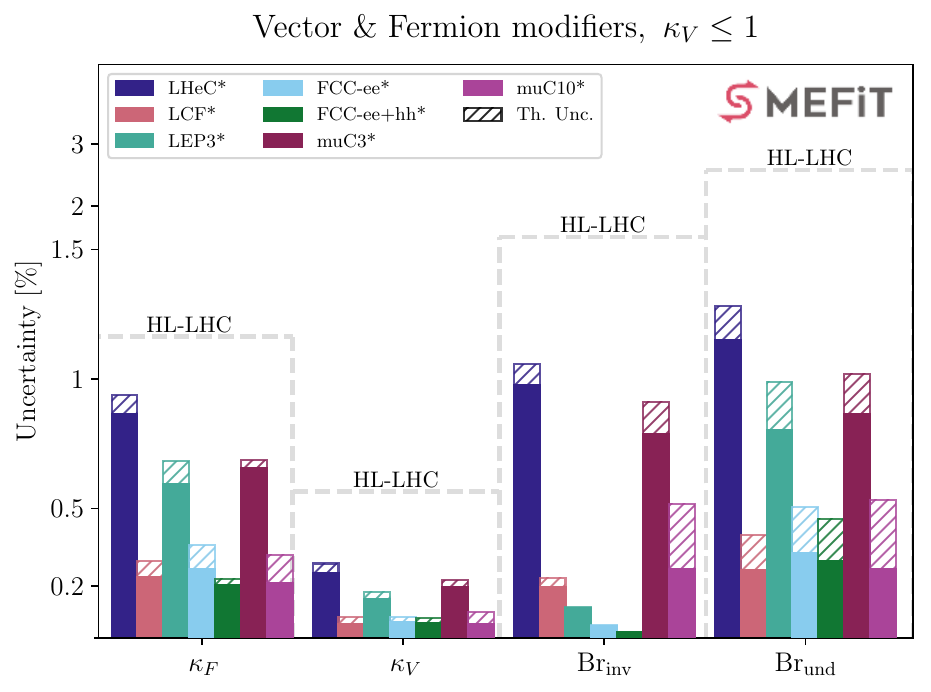}
\includegraphics[width=0.48\linewidth]{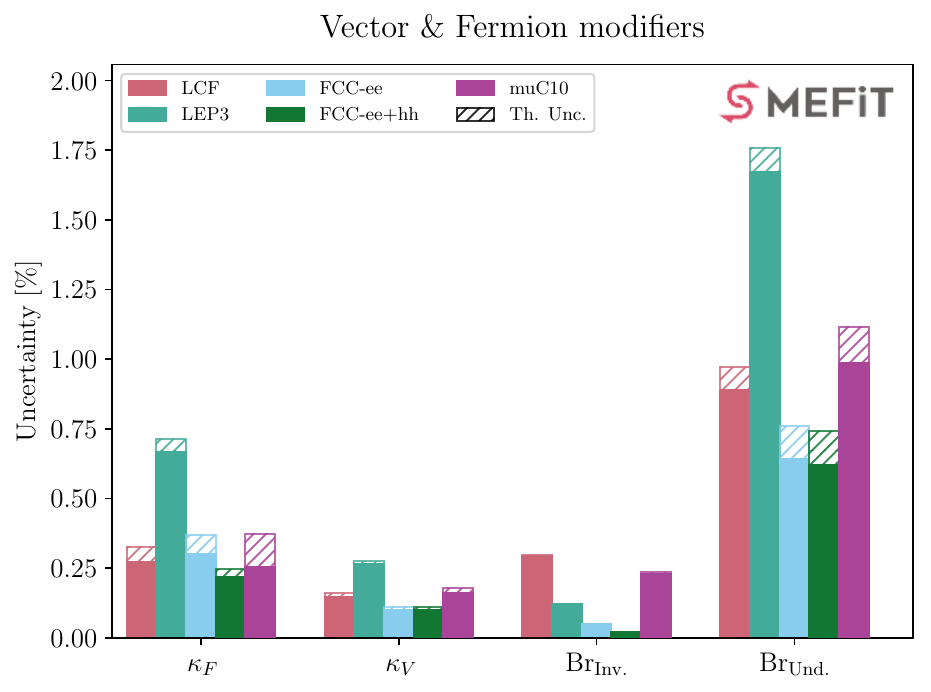}
    \caption{Same as Fig.~\ref{fig:kappa3_histo_comparison} for the kappa-3 variant with universal couplings to fermions, $\kappa_F$, and to vector bosons, $\kappa_V$.}
    \label{fig:kappa-FvsV_histo}
\end{figure}

    \begin{figure}[htb]
    \centering
\includegraphics[width=0.48\linewidth]{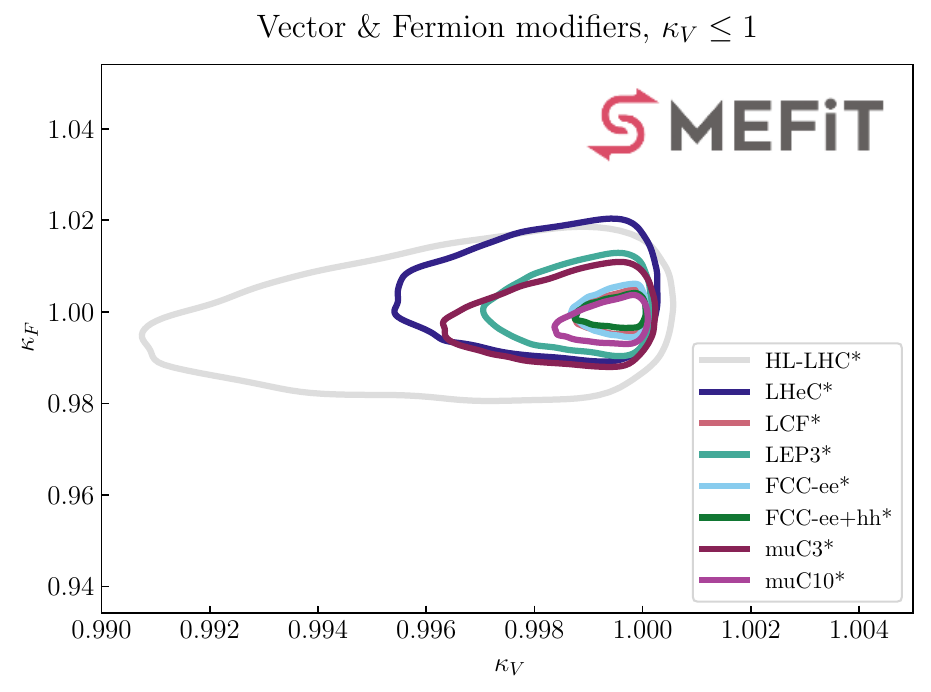}\includegraphics[width=0.48\linewidth]{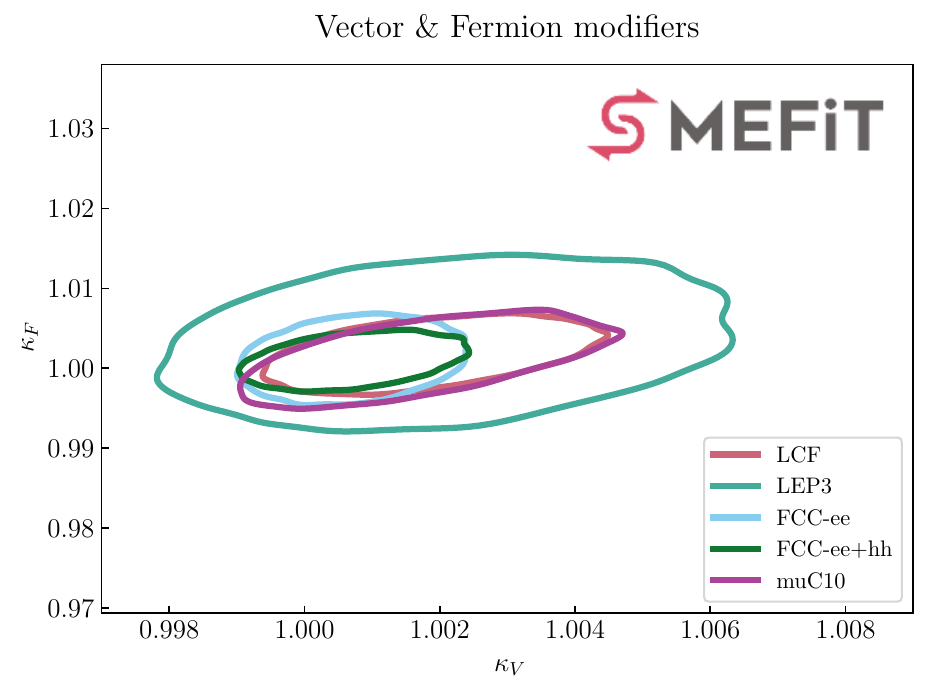}
    \caption{Same as  Fig.~\ref{fig:kappa-FvsV_histo} for the 68\% C.L. contours in the
    $(\kappa_F,\kappa_V)$ plane.
}
    \label{fig:kfvskv_contours}
\end{figure}

\paragraph{Universal vector and fermion modifiers.}
Finally, we consider another variant of the kappa-3 fits, where we assume a common coupling modifier for all fermions and another one for the vector bosons.
As in the case of the universal coupling modifier, this scenario is motivated by popular UV-complete models, such as the Type I Two-Higgs Doublet Model (2HDM)~\cite{Branco:2011iw}, in which the Higgs couplings to fermions are rescaled by a common factor $\kappa_F$ and those to vector bosons by a different factor $\kappa_V$.

The settings of this variant of the kappa-3 fits follow closely the previous one, with the only difference that now instead of $\kappa_{\rm univ}$ we have $\kappa_V$ and $\kappa_F$ as free parameters in the fit, for a total of 4 independent fitted parameters.
Furthermore, given that now $\kappa_F\ne\kappa_V$, it is necessary to resolve the loop substructure in order to express the loop-induced processes modifiers $\kappa_g$, $\kappa_\gamma$, and $\kappa_{Z\gamma}$ in terms of the two coupling modifiers.
For this we use the following expressions:
\begin{align}
\kg^2(\kt^2,\kb^2,\kt\kb)&=\kappa_F^2 \, ,\\
\ka^2(\kt^2,\kw^2,\kt\kw)&=1.59\,\kappa_V^2+0.07\,\kappa_F^2-0.66\,\kappa_F\kappa_V\,,\\
\kza^2(\kw^2,\kt\kw)&=1.12\,\kappa_V^2-0.12\,\kappa_F\kappa_V \, ,
\end{align}
where the $(\kappa)$ notation indicates how the unresolved kappa modifiers $\kappa_g$, $\kappa_\gamma$, and $\kappa_{Z\gamma}$ are changed once the loop is resolved.
Note that when $\kappa_F=\kappa_V$ we recover the previous universal scenario in which $\kappa_g=\kappa_\gamma=\kappa_{Z\gamma}=\kappa_{\rm univ}$, the common coupling modifier.

Fig.~\ref{fig:kappa-FvsV_histo} displays the same comparison as in Fig.~\ref{fig:kappa-universal} now for this scenario with separate universal couplings for fermions and for vector bosons. 
A similar picture as in the other variants of the kappa-3 fits arises.
In the unconstrained fits, $\kappa_F$ can be measured at the 0.25\% level at the integrated FCC, with other lepton colliders (with the exception of LEP3) exhibiting a similar performance.
For the vector boson coupling modifier $\kappa_V$, 0.1\% precision would be reached by FCC-ee, followed closely by LCF and MuCol.
To complement the results of Fig.~\ref{fig:kappa-FvsV_histo}, we also show in Fig.~\ref{fig:kfvskv_contours} the 68\% C.I. contours obtained through a Kernel Density Estimate (KDE) in the $(\kappa_V,\kappa_F)$ plane, quantifying the correlation between these two universal coupling modifiers. 
From this comparison it is apparent that the correlation between $\kappa_V$ and $\kappa_F$ is small, indicating that these two modifiers are being independently constrained.

\section{Effective Higgs and electroweak couplings}
\label{sec:effective_couplings_results}

In this section we present results for the SMEFT fits based on the effective couplings formalism of Sect.~\ref{sec:effective_couplings}.
In our implementation of this formalism, we fit a restricted subset of the full basis of SMEFT operators listed in Sect.~\ref{sec:global_smeft_fits} in a manner that allows the consistent definition of pseudo-observables for the Higgs and electroweak effective couplings. 
In the following, results are obtained for $\mathcal{O}\lp \Lambda^{-2}\rp$ EFT fits without RGE effects, as demanded by the theoretical consistency of the effective coupling dictionary presented in Sect.~\ref{sec:effective_couplings}.
We note that our fitting methodology is different from the {\sc\small HepFit} analysis entering the ESPPU2026 PBB~\cite{deBlas:2025gyz}, which instead recasts the global SMEFT fit outcome, including RGE and in some case NLO expansion effects, in the language of the effective couplings.
Additional details on our effective coupling analysis are provided in App.~\ref{app:effective_coupling_benchmark}, where in particular we benchmark our setup with the Snowmass 2022 analysis from {\sc\small HepFit}.

For the three future $e^+e^-$ colliders, namely LEP3, FCC-ee, and LCF, we indicate the expected uncertainties in the determination of the Higgs and electroweak couplings in four scenarios for the theory uncertainties, following the ESPPU2026 PBB~\cite{deBlas:2025gyz} convention: current, conservative, aggressive, and no theoretical uncertainties at all (denoted as the `ideal' scenario). 
The corresponding numerical results for these bounds on the effective couplings are collected in App.~\ref{app:effectivecouplings_supplementary}.

\begin{figure}[t]
    \centering
\includegraphics[width=0.49\linewidth]{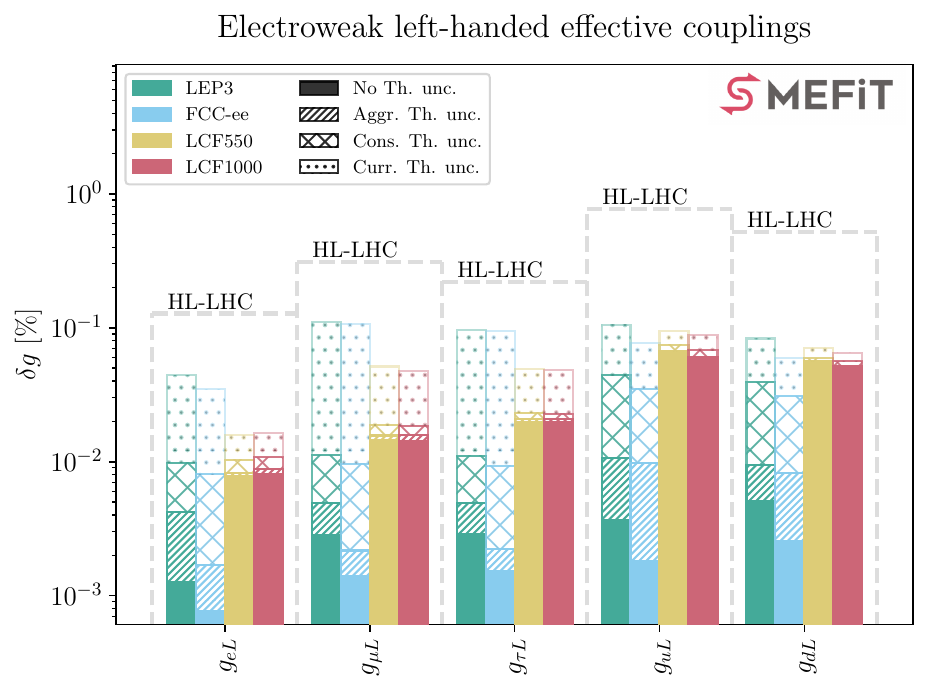}
\includegraphics[width=0.49\linewidth]{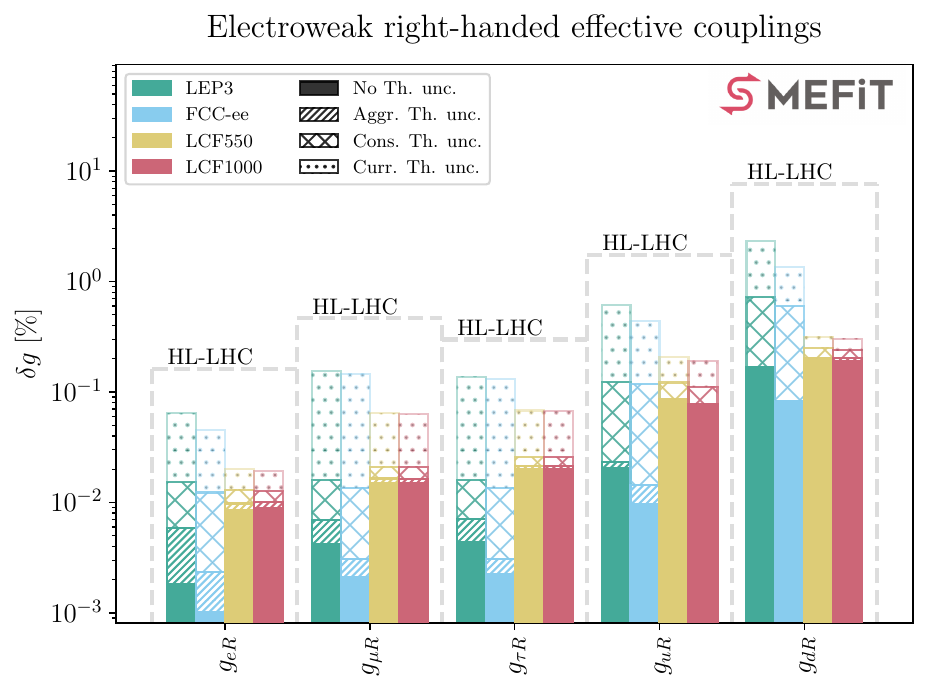}
\includegraphics[width=0.5\linewidth]{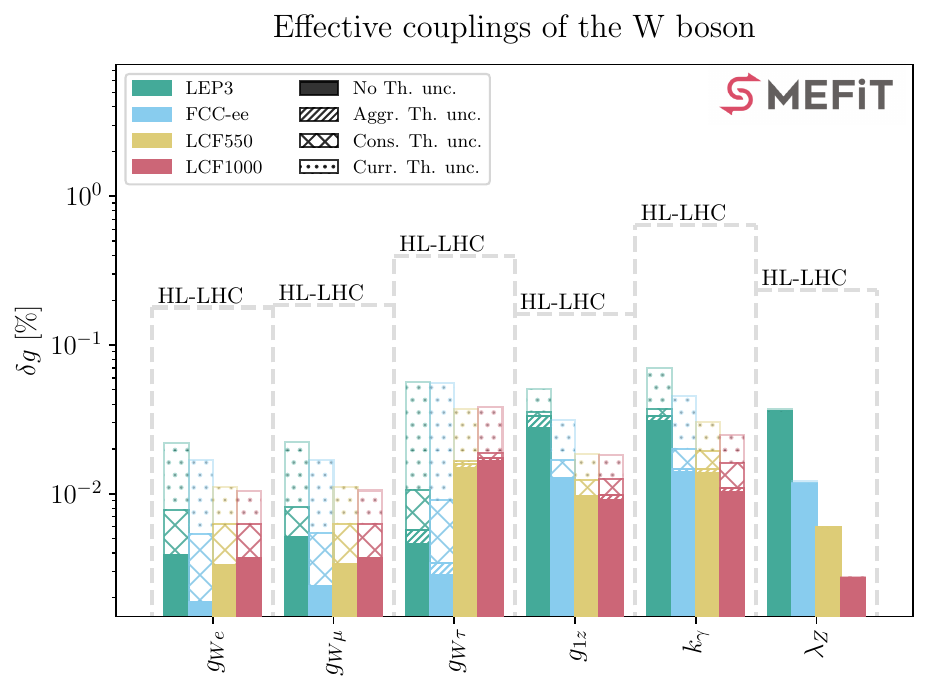}
    \caption{The projected relative uncertainties (at the 68\% C.I.) for the determination of the electroweak effective couplings defined in Sect.~\ref{sec:effective_couplings}: the left- and right-handed couplings to the $Z$-boson (left and right upper panels, respectively) and the $W$-boson effective couplings (bottom panel).
    The empty dashed bars correspond to the HL-LHC projections, taken as baseline. 
    For each of the proposed future colliders, we indicate the projected uncertainties in four scenarios for the theoretical uncertainties following the PBB analysis~\cite{deBlas:2025gyz} (see also Sect.~\ref{subsec:th_errors}).
    See Fig.~\ref{fig:eff_couplings_results_Higgs} for the corresponding results in the case of the Higgs effective couplings.
    }
\label{fig:eff_couplings_results_EW}
\end{figure}

\begin{figure}[t]
    \centering
\includegraphics[width=0.49\linewidth]{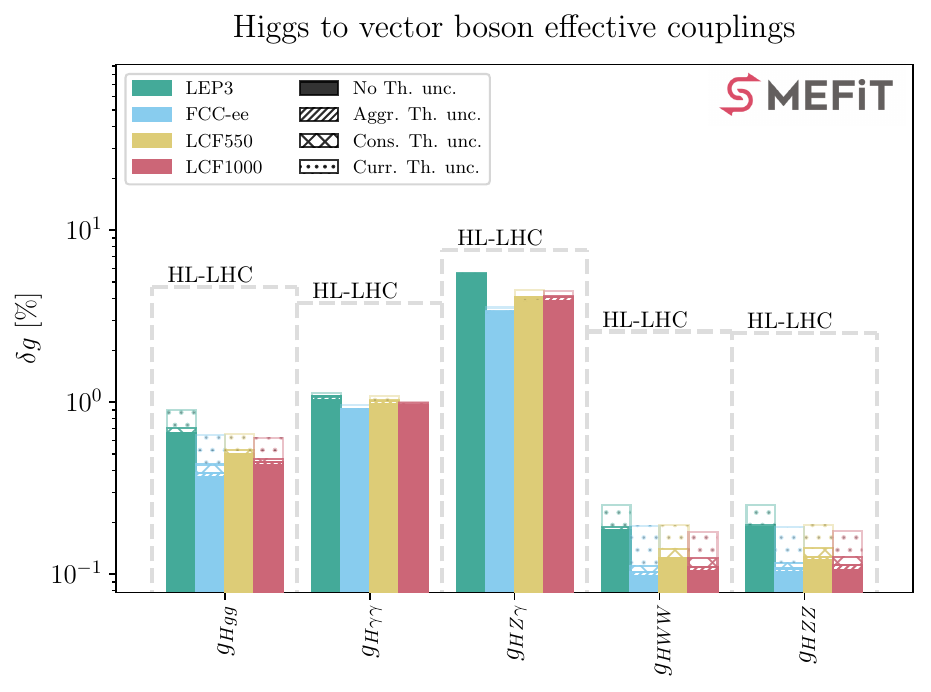}
\includegraphics[width=0.49\linewidth]{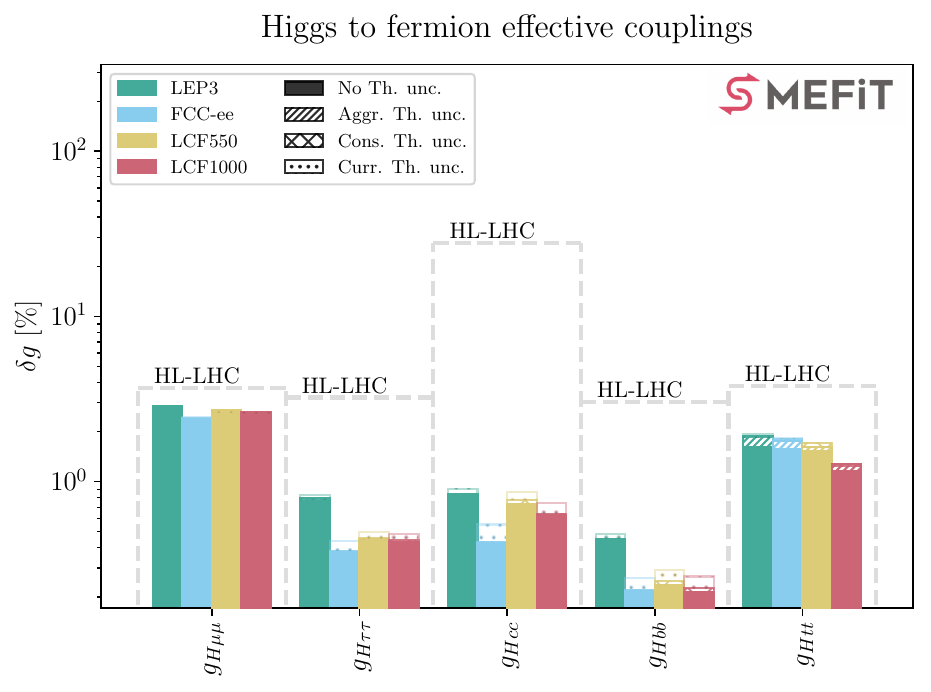}
    \caption{Same as Fig.~\ref{fig:eff_couplings_results_EW} for the Higgs effective couplings to vector bosons (left) and to fermions (right panel).
\label{fig:eff_couplings_results_Higgs}
}
\end{figure}

Fig.~\ref{fig:eff_couplings_results_EW} displays the  projected relative uncertainties, at the 68\% C.I., for the determination of the electroweak effective couplings: left- and right-handed couplings to the $Z$-boson, and the $W$ boson couplings.
Then Fig.~\ref{fig:eff_couplings_results_Higgs} provides a similar comparison now for the Higgs effective couplings to vector bosons and to fermions, following the definitions of Sect.~\ref{sec:effective_couplings}.
In both cases, the empty dashed bars correspond to the HL-LHC expectations for these effective couplings, which are taken here as baseline. 
For all future colliders, significant improvements as compared to HL-LHC are observed, typically by an order of magnitude or better, especially for the Higgs-electroweak couplings $g_{HWW}$
and $g_{HZZ}$ and for couplings poorly constrained at the HL-LHC such as the Higgs-charm coupling $g_{Hcc}$.

One interesting feature of the results in Fig.~\ref{fig:eff_couplings_results_EW} is related to the impact of theoretical uncertainties.
Differences in the projected sensitivity between the current and the aggressive scenarios can be as large as a factor 50 depending on the collider and the effective coupling considered.
For instance, the left-handed $Z$-boson coupling to electrons $g_{eL}$ can be measured at the FCC-ee with 0.002\% precision in the aggressive theory errors scenario, a bound which is degraded to almost 0.03\% assuming current theoretical uncertainties. 
It is also noteworthy how the hierarchies between the sensitivity achieved in different colliders can be markedly modified by theoretical uncertainties, due to their rebalancing of the relative impact of the Tera-$Z$ run with respect to the higher-energy runs.  

Considering in the following the aggressive theory uncertainties scenario as baseline, we find that FCC-ee provides the best overall sensitivity on the effective couplings of the $W$ and $Z$ bosons, thanks to the higher luminosities of the $Z$-pole and $WW$ runs.
As compared to LCF, for example, FCC-ee leads to more stringent bounds by an order of magnitude for most of the left- and right-handed effective couplings. 
One exception is provided by the anomalous triple gauge couplings $\kappa_\gamma,g_{1z},$ and $\lambda_Z$, for which LCF is projected to offer the best sensitivity due to the possibility of accessing higher centre-of-mass energies.

In contrast with the case of the electroweak effective couplings, for the Higgs effective couplings shown in Fig.~\ref{fig:eff_couplings_results_Higgs} the impact of theoretical uncertainties is more moderate.
This is explained by the fact that for Higgs production the experimental accuracy target is not too different from current theory uncertainties. 
For most effective couplings, again major improvements as compared to HL-LHC are observed, in particular for $g_{HWW}$, $g_{HZZ}$, $g_{Hgg}$, and the effective couplings to the charm and bottom quarks.
Effective couplings related to rare Higgs decays such as $H\to Z\gamma$ and $H\to \mu^+\mu^-$ will only be mildly improved at future electron-positron colliders, and only a higher-energy collider such as the FCC-hh or the MuCol can produce a sufficiently large sample of Higgs bosons to meaningfully extend the post-HL-LHC results. 

In terms of the relative impact of the future colliders being considered, again in most cases FCC-ee offers the best overall sensitivity for the Higgs effective couplings, although differences with other colliders such as LCF are more moderate than in the case of the electroweak couplings shown in Fig.~\ref{fig:eff_couplings_results_EW}, and LEP3 is found to have the worse reach. 
In the case of LCF projections, we find that the highest energy run only improves the sensitivity of the previous lower energy runs by a small amount, as expected since the $\sqrt{s}=1$ TeV run does not bring new constraints on the Higgs effective couplings considered here, with the only exception of $g_{Htt}$ through the Yukawa coupling. 

From this effective coupling analysis, we conclude that FCC-ee offers the best overall sensitivity, and that reaching its ultimate precision demands an aggressive programme to reduce theory uncertainties in the SM predictions.
\section{Global SMEFT analysis}
\label{sec:global_smeft_fits_results}

In this section we present results for the global SMEFT fit at the three 
$e^{+}e^{-}$
colliders listed in Table~\ref{tab:FutureCollider_runs}: LEP3, FCC-ee, and LCF. 
For LCF, we consider both the LCF550  (without the 1 TeV run, but with higher luminosity at $\sqrt{s}=550$ GeV) and the LCF1000 variants separately.
While comparing the projections from these three colliders among them, we also quantify the impact  of quadratic EFT corrections, how the sensitivity to SMEFT operators varies in the different scenarios on the theoretical uncertainties associated to the SM predictions, and revisit our study of~\cite{terHoeve:2025omu} for the Higgs self-coupling at the HL-LHC and future electron-positron colliders.

In this section, we show results for the $n_{\rm op}=61$ Wilson coefficients that compose our fitting basis (see also App.~\ref{app:operator_basis}) evaluated at the reference scale $\mu_0=10$ TeV. 
The one exception is our analysis of the Higgs self-coupling in Sect.~\ref{subsec:higgs_self_coupling}, where we instead adopt $\mu_0=250$ GeV as reference scale. 
This choice corresponds to the characteristic energy scale of double-Higgs production, and represents a natural reference scale for the physical definition of the Higgs trilinear coupling.
Furthermore, we present our results both in global fits, where all the $n_{\rm op}$ Wilson coefficients are allowed to float simultaneously and then one marginalises for each of them, as well as individual fits, where only the corresponding operator at $\mu_0=10$ TeV is assumed to be non-zero (and where additional operators may be generated at the data scale through RGEs).

First, we present in Sect.~\ref{subsec:results_baseline_linear} results for the baseline fits, carried out at the linear level in the EFT expansion and account for theoretical uncertainties in the aggressive scenario.
Then we study in Sect.~\ref{subsec:quadratic_corrections} the impact of quadratic EFT corrections in the global fit.
Subsequently, Sect.~\ref{subsec:theory_errors_impact}  presents results for the four scenarios for the theoretical uncertainties discussed in Sect.~\ref{subsec:th_errors}.
Finally, we compare in Sect.~\ref{subsec:higgs_self_coupling} the reach of future lepton colliders on the Higgs self-coupling modifier $\delta \kappa_3$.
In addition, App.~\ref{app:extra_smeft} collects supplementary results and comparisons complementing these global SMEFT fit studies.

\subsection{Baseline results}
\label{subsec:results_baseline_linear}

For the presentation of the results in this section, we cluster the Wilson coefficients into different families: those associated with four-heavy-quark (4H), two-light-two-heavy quark (2L2H), purely bosonic (B), two-lepton-two-light-quark (2$\ell$2q), two-fermion (2FB),  two-lepton-two-heavy-quark (2$\ell$2Q), and four-lepton (4$\ell$) operators, see also the discussion in App.~\ref{app:operator_basis}.
We start by presenting our baseline results for the global SMEFT fit, which are obtained with settings aligned whenever possible with those of the ESPPU26 PBB (see App.~\ref{app:pbb-bench} for selected comparisons at the level of individual fits). 
In particular, here we consider as baseline linear EFT calculations and the aggressive scenario for theoretical uncertainties at lepton colliders, and subsequently explore the impact of varying these settings. 

\begin{figure}[htbp]
    \centering
\includegraphics[width=\linewidth]{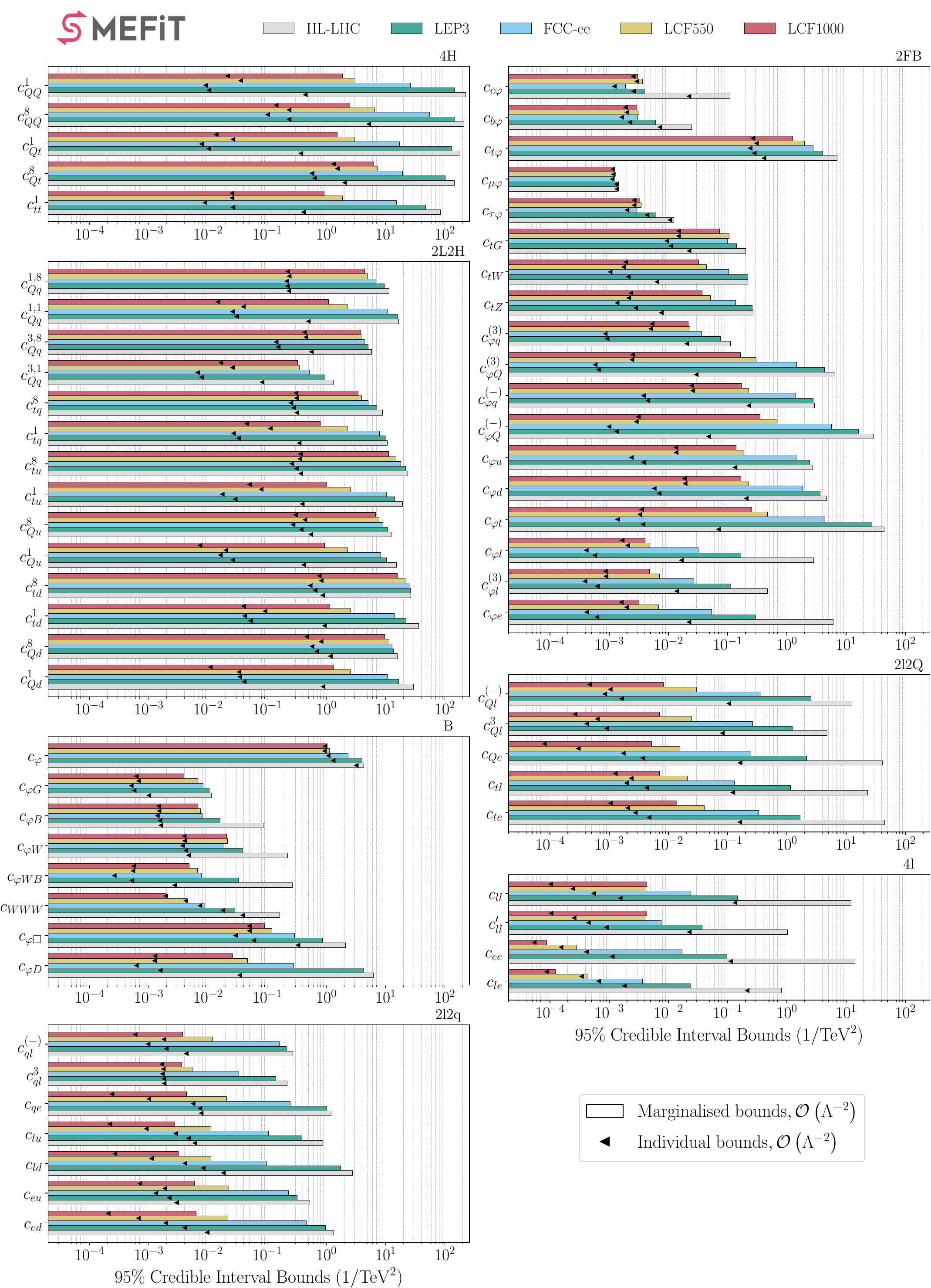}
    \caption{Results (length of the 95\% C.I. bounds) of individual (triangles) and global marginalised (filled bars) fits to the projections for different future colliders for the $n_{\rm op}=61$ Wilson coefficients that compose our basis, displayed at the reference scale of $\mu_0=10$ TeV. 
    These fits are carried out at the linear level,
    $\mathcal{O}\lp \Lambda^{-2}\rp$, in the EFT expansion, account for RGE effects, and assume the aggressive scenario for theory uncertainties.
    }
    \label{fig:lin_ind_global}
\end{figure}

To begin with, Fig.~\ref{fig:lin_ind_global} displays the 95\% C.I. bounds obtained for the Wilson coefficients in the SMEFT analysis, comparing the outcome of individual and global marginalised fits for the HL-LHC, LEP3, FCC-ee, LCF550, and LCF1000 projections.
Note that since we absorb factors of $\Lambda^{-2}$ in our definition of the Wilson coefficients $c_i$ (see the discussion in Sect.~\ref{sec:global_smeft_fits}), these are dimensionful and have units of TeV$^{-2}$.
These fits are carried out at $\mathcal{O}\lp \Lambda^{-2}\rp$ in the EFT expansion; results accounting for quadratic corrections are presented in Sect.~\ref{subsec:quadratic_corrections}.
One-loop RGE effects are included, and theoretical uncertainties are included in the `aggressive' scenario; the impact of varying the latter is quantified in Sect.~\ref{subsec:theory_errors_impact}.
In all cases, the bounds on the Wilson coefficients are provided at the reference scale of $\mu_0=10$ TeV.

\begin{figure}[htbp]
    \centering
\includegraphics[width=\linewidth]{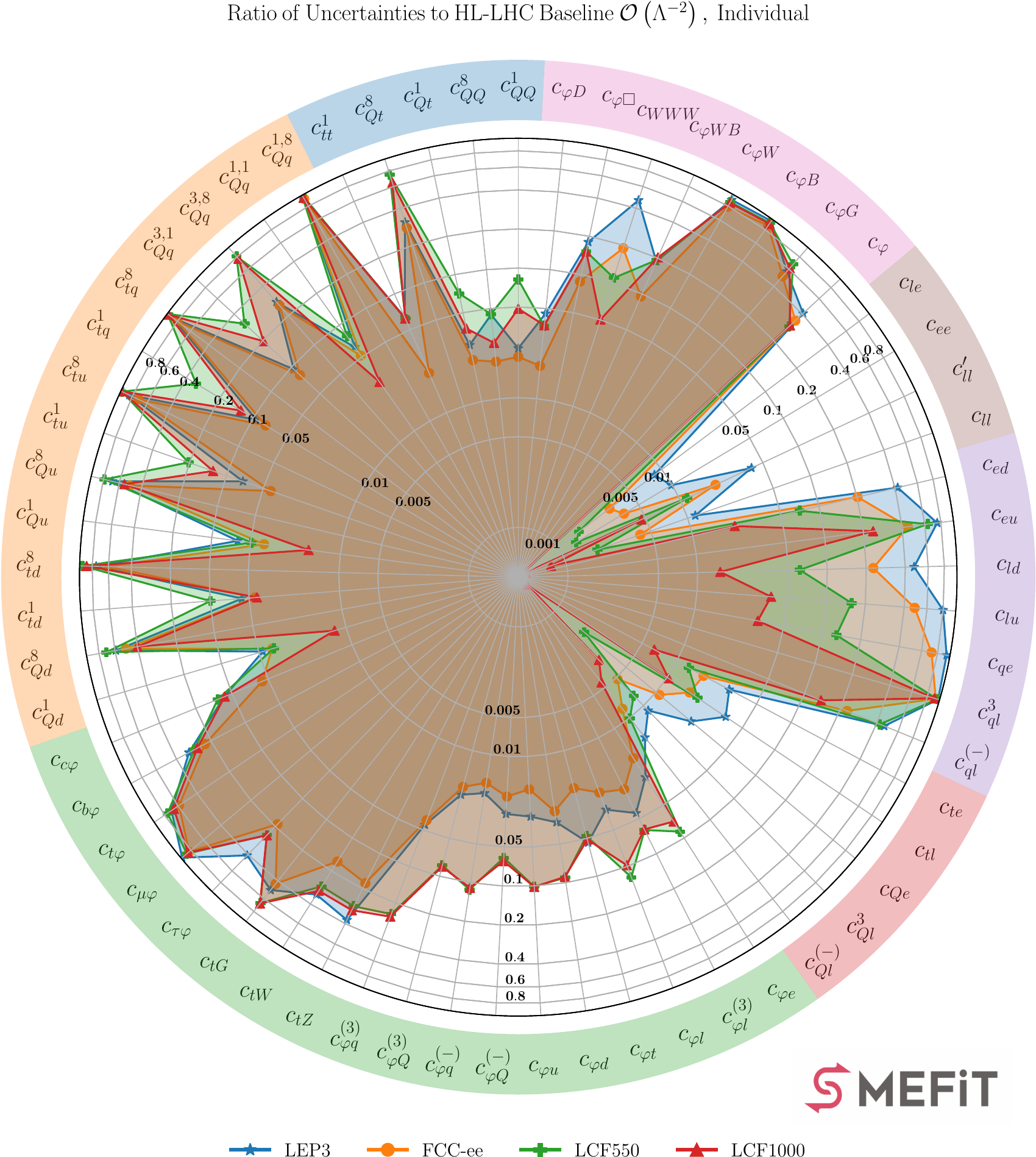}
    \caption{Same as Fig.~\ref{fig:lin_ind_global} now with the projected uncertainties in the individual bounds on the Wilson coefficients in a linear SMEFT fit with aggressive theory uncertainties displayed as ratios to the HL-LHC expectations,
    Eq.~(\ref{eq:RatioC}).
    We compare LEP3, FCC-ee, LCF550, 
    and LCF1000.
    Note the logarithmic scale of the radial axis.
    The Wilson coefficients are clustered in terms of the categories defined in Table~\ref{tab:operators_summary}. 
    \label{fig:spider_lin_ind_aggressive}
}
\end{figure}

\begin{figure}[htbp]
    \centering
\includegraphics[width=\linewidth]{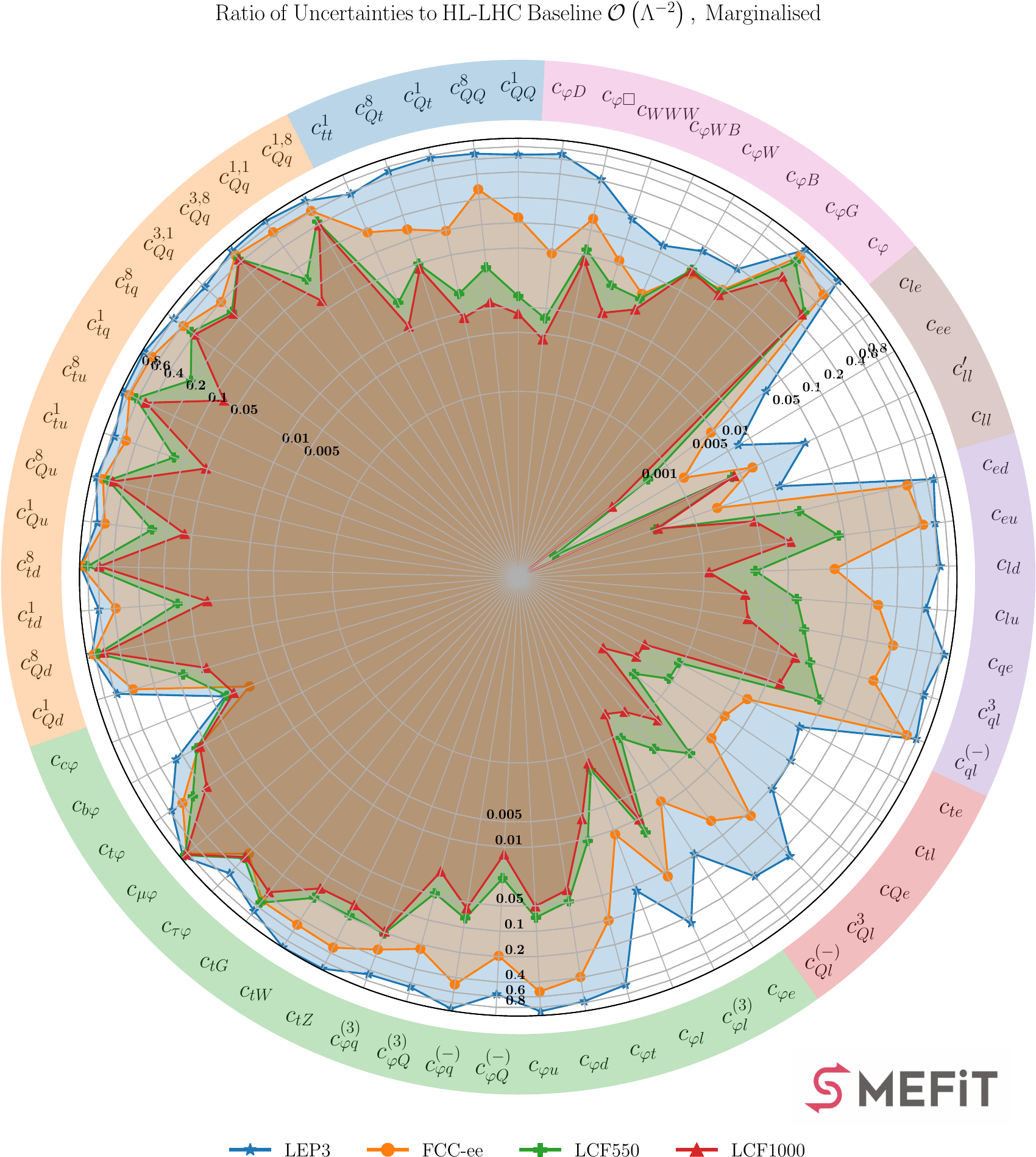}
    \caption{
Same as Fig.~\ref{fig:spider_lin_ind_aggressive} for the global marginalised bounds.
\label{fig:spider_lin_global_aggressive}
}
\end{figure}

\begin{figure}[htbp]
    \centering
\includegraphics[width=\linewidth]{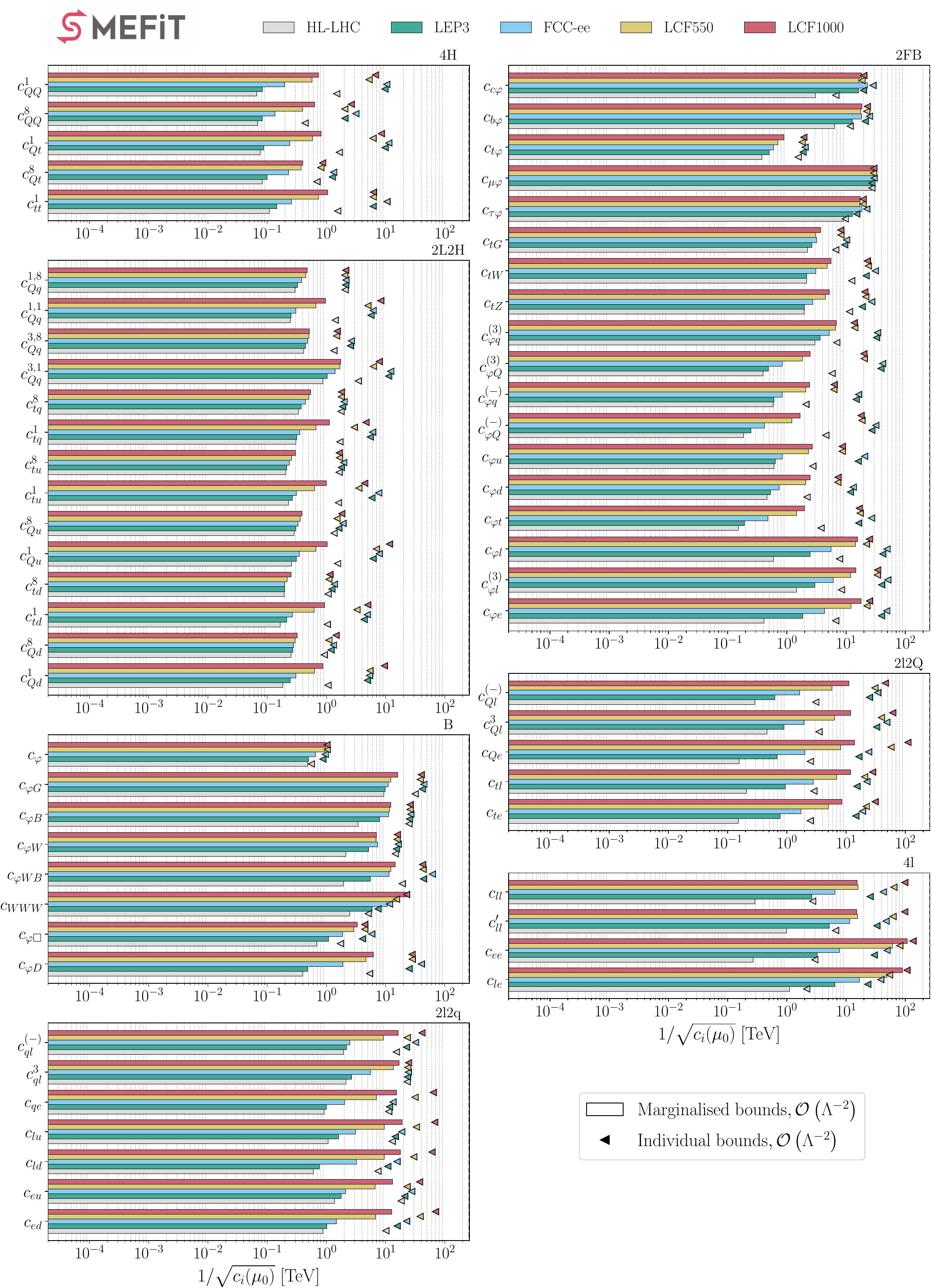}
    \caption{Same as Fig.~\ref{fig:lin_ind_global} for the lower bounds on the mass scale $\Lambda \equiv c_i^{-1/2}(\mu_0)$, which can be interpreted as the characteristic NP mass scale being probed by the SMEFT analysis assuming Wilson coefficients of order unity.
    Results are evaluated at
    the reference energy scale of $\mu_0=10$ TeV.  
\label{fig:lin_ind_global_lambda}
}
\end{figure}

The same information contained in Fig.~\ref{fig:lin_ind_global} for the constraints of the different Wilson coefficients can be presented with alternative visualisations to facilitate its perusal.
To this end, Figs.~\ref{fig:spider_lin_ind_aggressive} and~\ref{fig:spider_lin_global_aggressive} display the same bounds as in Fig.~\ref{fig:lin_ind_global}, now in a spider-plot format, with the projected uncertainties in the bounds on the Wilson coefficients  displayed as ratios to the HL-LHC expectations, for individual and global marginalised fits respectively.
Specifically, we plot the metric $R_{\delta c_i}$, 
defined as the ratio between the magnitude of the 95\% C.I. interval for a given EFT coefficient $c_i$ obtained in a future collider fit, to that of the same quantity in the HL-LHC fit:
\be
\label{eq:RatioC}
R_{\delta c_i} = \frac{ \lc c_i^{\rm min}, c_i^{\rm max} \rc^{95\%~{\rm CI}}~({\rm HL\textnormal{-}LHC+\text{future collider}})}{
 \lc c_i^{\rm min}, c_i^{\rm max} \rc^{95\%~{\rm CI}}~({\rm HL\textnormal{-}LHC})
} \, , \qquad i=1,\ldots, n_{\rm eft} \, .
\ee
Therefore, in Figs.~\ref{fig:spider_lin_ind_aggressive} and~\ref{fig:spider_lin_global_aggressive}, for a given  coefficient $c_i$, the smaller the value of $R_{\delta c_i}$ the more significant the impact of the new data.
Note also the logarithmic scale of the radial axis, and that the Wilson coefficients are clustered in terms of the categories defined in Table~\ref{tab:operators_summary}.
Furthermore, Fig.~\ref{fig:lin_ind_global_lambda} shows the same results as in  Fig.~\ref{fig:lin_ind_global} presented as bounds on $\Lambda\equiv 1/\sqrt{c_i(\mu_0)}$, which can be interpreted as the mass scale being probed by the SMEFT analysis for Wilson coefficients of order unity at the reference scale $\mu_0=10$ TeV.  

\begin{figure}[htbp]
    \centering
    \includegraphics[height=24cm, keepaspectratio]{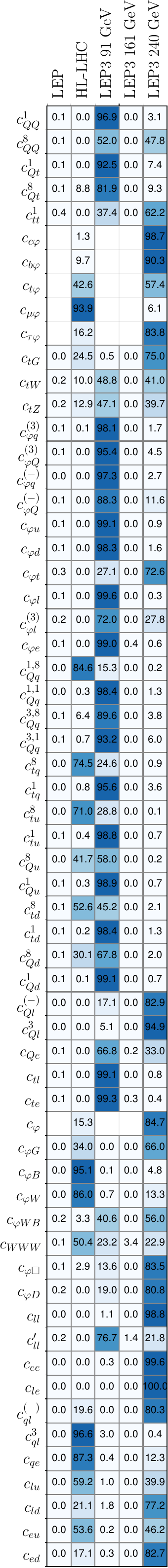}
    \includegraphics[height=24.2cm, keepaspectratio]{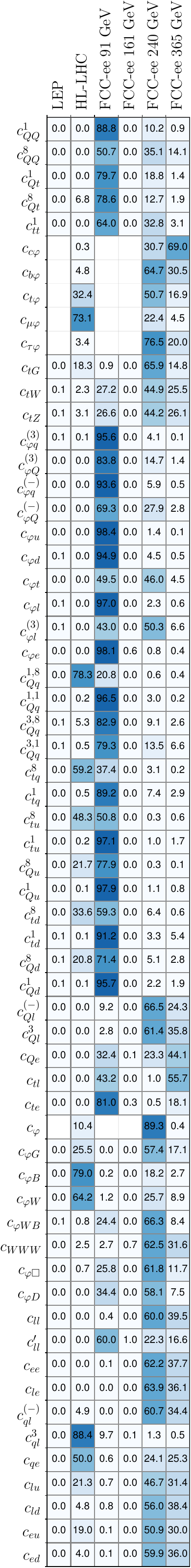}
    \includegraphics[height=24.5cm, keepaspectratio]{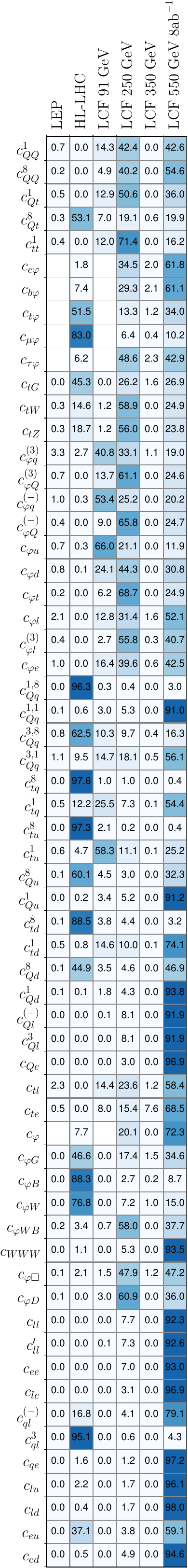}
    \includegraphics[height=24.5cm, keepaspectratio]{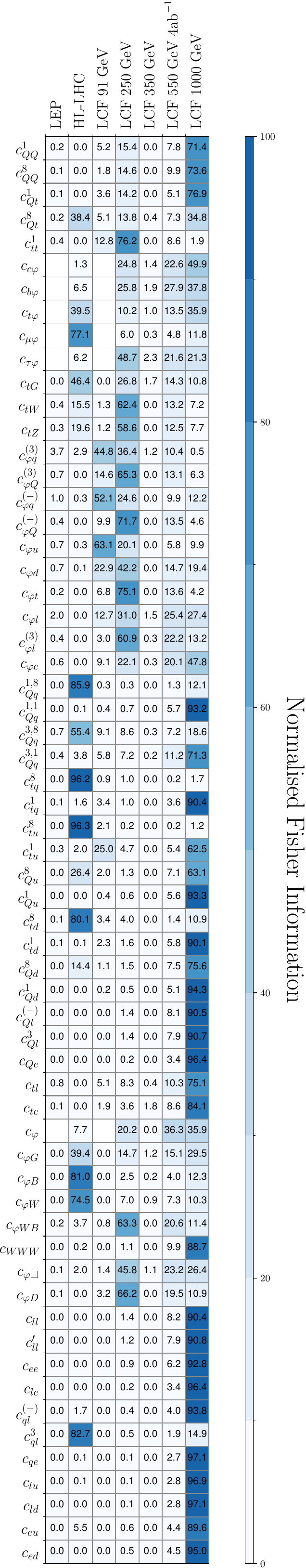}
    \caption{The entries of the Fisher information matrix, evaluated at linear order in the EFT expansion, for the fits based on the LEP3, FCC-ee, LCF-550 and LCF-1000 projections (from left to right). 
    For this analysis, we cluster the input datasets into three groups: LEP, HL-LHC, and either LEP3, FCC-ee, or LCF respectively, in the latter case separating the different $\sqrt{s}$ runs. 
    Theoretical uncertainties are included in the aggressive scenario.
    \label{fig:fisher_heatmap_LCF1000_FCCee}
    }
\end{figure}

To further streamline the interpretation of the results presented in Figs.~\ref{fig:lin_ind_global}--\ref{fig:lin_ind_global_lambda}, we also display in Fig.~\ref{fig:fisher_heatmap_LCF1000_FCCee}
the diagonal entries of the Fisher information matrix~\cite{Ethier:2021bye}, evaluated at linear order in the EFT expansion for the LEP3, FCC-ee and LCF projections, in all cases with LEP and HL-LHC included. 
For the purposes of this Fisher information analysis, we cluster the input datasets into three groups: LEP, HL-LHC, and either LEP3, FCC-ee or LCF, in the latter case separating the different $\sqrt{s}$ runs. 
Each row of the table is normalised to 100.
The larger the entry of these Fisher information matrices, the more dominant a specific dataset is in determining the corresponding operator in the case of individual fits.
For instance, from Fig.~\ref{fig:fisher_heatmap_LCF1000_FCCee} one observes that at the FCC-ee all four-quark operators are constrained predominantly by the Tera-$Z$ observables, while for the LCF the impact of the $Z$-pole run for these operators is mostly subdominant and is instead driven by the energy runs above the $Z$-pole. 
As another example, at the LCF the two- and four-lepton operators are mostly determined by the $\sqrt{s}=1$ TeV run, while at the FCC-ee they are mostly constrained by the runs at $\sqrt{s}=240$ and $365$ GeV.

We now comment on some of the main features observed from Figs.~\ref{fig:lin_ind_global}--\ref{fig:fisher_heatmap_LCF1000_FCCee}.

\paragraph{Individual versus marginalised fits.} Comparing bounds obtained from individual (one-parameter) fits and marginalised fits in Figs.~\ref{fig:lin_ind_global}--\ref{fig:lin_ind_global_lambda}, a first observation is that the bounds obtained in the global marginalised case are in many cases much looser, often by orders of magnitude.
This is the case for most of the four-fermion operators, both those including quarks and those with leptons, and also for some of the purely bosonic and two-fermion operators.
These differences arise because, in the global fit (especially in the linear case), variations in one coefficient may be compensated with correlated variations of other coefficients while still leading to a satisfactory description of the experimental data, hence leading to a degradation of the bounds.

Individual and global marginalised bounds are instead similar for those Wilson coefficients that are weakly correlated in the global fit. 
Operators falling in this category include for example the self-coupling $c_\varphi$ at the HL-LHC and LCF, which can directly be constrained from the di-Higgs production cross-section.  
Similar observations were made when comparing individual and marginalised fits in our previous analysis~\cite{terHoeve:2025omu}.

Any realistic UV completion of the SM model will activate only a subset of dimension-six Wilson coefficients. 
At the same time, only very simple UV extensions would activate only a single SMEFT coefficient at the matching scale $\mu_0$.
Hence the results presented in Figs.~\ref{fig:lin_ind_global}--\ref{fig:lin_ind_global_lambda} need to be understood as bracketing what would be obtained for specific UV scenarios, where the bounds obtained will be more stringent than in the global fit, but looser as compared to the one-parameter analysis. 
This situation will be further discussed in Sect.~\ref{sec:uv_benchmarks}, where we study the impact of future colliders in a number of representative UV benchmark models.

\paragraph{Analysis of individual bounds.}

The spider plot of Fig.~\ref{fig:spider_lin_ind_aggressive} quantifies the expected improvement in the individual bounds for the Wilson coefficients that form our fitting basis in light of the future collider projections with respect to the HL-LHC baseline.
Overall, the FCC-ee displays the best performance, with significant reduction of the bounds on the Wilson coefficients in all categories.
For many directions in the SMEFT parameter space, the improvements of the FCC-ee as compared to the HL-LHC are better by a factor of at least 20.
We note that these improvements also apply to operators, such as the two-light-two-heavy and four-heavy quark types, which are not directly accessed at lepton colliders but to which the fit is sensitive through loop effects via RGE evolution.

One also finds that a small number of operators will not be significantly improved in any of the future colliders considered compared to the HL-LHC, such as the top and muon Yukawa couplings $c_{t\varphi}$ and $c_{\mu\varphi}$, for which a higher-energy proton-proton or muon collider would be needed, or the purely bosonic operators $c_{\varphi W}$ and $c_{\varphi B}$.
In the specific case of the top Yukawa coupling, we remark that our HL-LHC projections are not based on targeted analyses, and that variations of the correlation model for Higgs measurements at the HL-LHC can modify these expectations as illustrated for the kappa framework fits in App.~\ref{app:kappa-additional}.

The LCF reach dominates over the FCC-ee one for some of the operators involving leptons, since these benefit from an increased sensitivity at the higher energy runs at $\sqrt{s}=550$ GeV and $1$ TeV. 
The LCF results are instead less competitive than the FCC-ee ones for the two-fermion-bosonic operators and most of the four-quark operators. These operators are constrained via RGE running to the $Z$-pole observables and are more weakly constrained at the Giga-$Z$ and $\sqrt{s}=250$ GeV runs of the LCF than at the Tera-$Z$ run of the FCC-ee (and to a lesser extent of LEP3).
We note in particular that the Tera-$Z$ run at the FCC-ee is scheduled to have a luminosity almost 70 times larger than the Giga-$Z$ and $\sqrt{s}=250$ GeV runs combined at the LCF.

When considering individual bounds for the Wilson coefficients, LEP3 offers a competitive performance for most operators considered.
The degradation of the bounds as compared to FCC-ee is typically around a factor two.
As in the case of FCC-ee, individual fits at LEP3 constrain two-lepton-two-heavy quark operators even in the absence of the $e^+e^- \to t\bar{t}$ run through RGE evolution to the lower $\sqrt{s}$ observables.

\paragraph{Analysis of global marginalised bounds.}

When comparing the improvement in the bounds to the HL-LHC baseline for the global marginalised fits in Fig.~\ref{fig:spider_lin_global_aggressive}, one observes qualitative differences as compared to the corresponding results at the individual fit level of Fig.~\ref{fig:spider_lin_ind_aggressive}.
First, an inversion of the relative hierarchy in reach for the various colliders: while for individual fits FCC-ee is typically superior with a handful of exceptions, in the global marginalised fits the best performance is achieved by the LCF, especially if the run at $\sqrt{s}=1$ TeV is taken into account.
Second, the improvements with respect to the HL-LHC that could be obtained at the FCC-ee  are now moderate, specifically for the two-light-two-heavy four quark operators. 
Third, LEP3, which was competitive with the other colliders in the individual fit, now leads to the worst constraints for all the operators, with, in many cases only a relatively small improvement compared to the HL-LHC baseline.
Finally, for the LCF, the qualitative pattern of improvement as compared to HL-LHC remains similar in individual and in global marginalised fits, at odds with the FCC-ee case and LEP3. 

The different qualitative behaviour displayed in Fig.~\ref{fig:spider_lin_global_aggressive} as compared to Fig.~\ref{fig:spider_lin_ind_aggressive} may be understood as follows.
For individual fits, FCC-ee excels thanks to its unprecedented precision of the $Z$-pole observables, to which most operators flow via RGE. In the global fit, however, these same $Z$-pole constraints are largely weakened by correlations, whereas the LCF’s larger variety of observables helps break these correlations resulting in tighter marginalised constraints despite its lower overall sensitivity. 
Similarly, LEP3 can also probe with high precision the $Z$-pole observables, leading to strong indirect constraints in the individual fit, but its lack of runs above $\sqrt{s}=230$ GeV prevents competitive bounds in the marginalised fit. 

The results of Figs.~\ref{fig:spider_lin_ind_aggressive} and \ref{fig:spider_lin_global_aggressive} may appear at first inspection to provide a contradictory message.
However, this is not the case: any SMEFT interpretation depends, by construction, on the assumptions made on the operator basis.
Confronting Fig.~\ref{fig:spider_lin_global_aggressive} with Fig.~\ref{fig:spider_lin_ind_aggressive} therefore illustrates the two limiting cases: all operators are assumed to have the same prior relevance, or only one of them is assumed to fully dominate, in both cases at the matching scale $\mu_0$. 

\paragraph{NP mass reach.}

When presented in terms of lower bounds in $\Lambda \equiv c_i^{-1/2}(\mu_0)$, as done in Fig.~\ref{fig:lin_ind_global_lambda}, the results of the individual and global marginalised fits can be interpreted in terms of the characteristic mass scale of UV physics being probed under the assumption that the dimensionless couplings satisfy $g_{\rm UV}\sim 1$ at $\mu_0$ (taken to be the same as the matching scale). 
From the results of individual fits, we find that future lepton colliders can probe values up to $\Lambda$ around $\mathcal{O}\left(10\right)$ TeV, for instance FCC-ee reaches $\Lambda \sim 30$ TeV for two-fermion operators such as $c_{\varphi Q}^{(3)}$ and $c_{\varphi q}^{(3)}$.

Consistently with the trends displayed in the rest of the plots in this subsection, the highest mass reach for individual fits is achieved by the FCC-ee for the two-fermion and most of the four-quark operators, while the LCF typically dominates the reach for the most of the four-fermion operators involving lepton fields, reaching for instance above $\Lambda \sim 100$ TeV for the four-lepton operators and for $c_{Qe}$. 

When considering instead the bounds on the NP mass scale $\Lambda$ obtained from the global fit after marginalisation, the reach on this UV mass scale is markedly reduced, with LCF displaying a superior performance.
For instance, LCF reaches $\Lambda \sim 20$ TeV for the two-lepton-two-light-quark operators, 100 TeV for some of the four-lepton operators, and around 10 TeV for the two-lepton-two-heavy-quark operators.
However, as discussed above, the bounds obtained in the global SMEFT fit are too conservative for any realistic UV model, and for specific UV completions, the reach in the mass scale $\Lambda$ will lie between the bounds of the individual and global fits.

\paragraph{Correlation patterns.}

\begin{figure}[htbp]
    \centering
\includegraphics[width=0.99\linewidth]{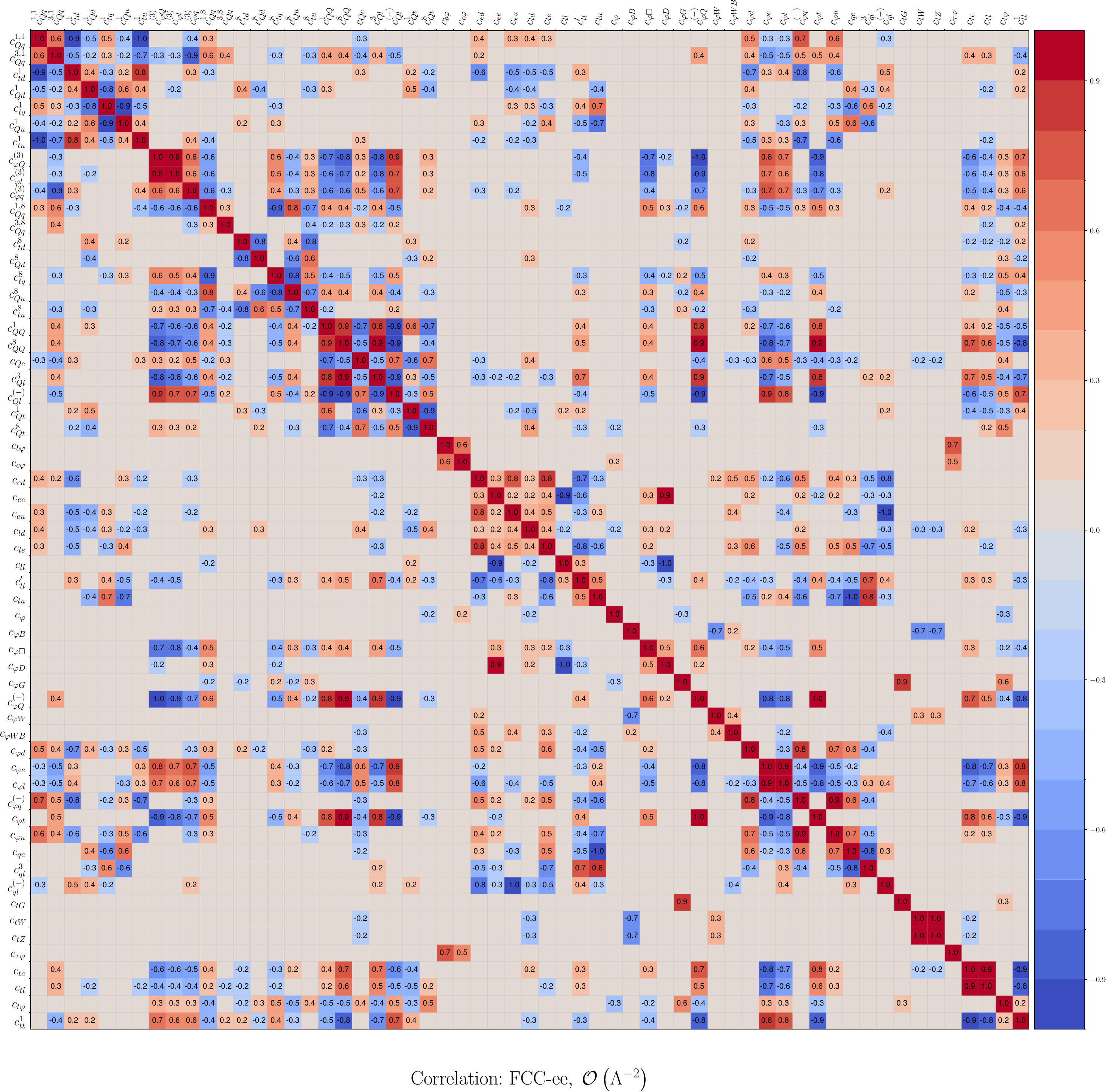}
    \caption{The correlation matrix in the space of Wilson coefficients in a global SMEFT fit carried out at the linear level to LEP, (HL-)LHC, and FCC-ee projections. 
    The numerical value of the correlation is shown only for those entries of the matrix satisfying $|\rho|>0.2$. 
See Fig.~\ref{fig:correlationmap_FCCee_quadratic} for the corresponding correlation map in the quadratic fit.
\label{fig:correlationmap_FCCee_linear}
    }
\end{figure}

To conclude this discussion, we display in Fig.~\ref{fig:correlationmap_FCCee_linear} the correlation matrix for the Wilson coefficients in a global SMEFT fit carried out at the linear level to LEP, (HL-)LHC, and FCC-ee projections. 
The numerical value of the correlation is shown only for those entries of the matrix satisfying $|\rho|>0.2$.
It is clear that the fit parameters are characterised by a complex correlation pattern, although some clear patterns exist. First of all, we note how operators within the same operator class are in general strongly correlated, as can be seen by inspecting for example the two-light-two-heavy operators, the four-heavy or the two-lepton-two-quark operators in Fig.~\ref{fig:correlationmap_FCCee_linear}. Beyond this, we note that cross-correlations between different operator classes also exist, an effect which is enhanced, in particular, through RG mixing effects. Obvious examples here include the four-heavy operators and the operators entering the EWPOs, for which we observe a strong correlation of $|\rho|>0.7$. This is in stark contrast to the milder correlation pattern observed previously in our analysis of Ref.~\cite{Celada:2024mcf} where RG effects were not considered. 
We finally remark that quadratic corrections help reduce significantly the degree of correlation by lifting quasi-flat directions, as verified explicitly in Fig.~\ref{fig:correlationmap_FCCee_quadratic}.

\subsection{Impact of quadratic EFT corrections}
\label{subsec:quadratic_corrections}

The results presented in Sect.~\ref{subsec:results_baseline_linear} are based on EFT calculations truncated at the linear order in the EFT expansion.
Quadratic corrections arising at $\mathcal{O}\lp \Lambda^{-4}\rp$ from the squares of dimension-six operators can be comparable, or in some cases even dominant, to the linear ones for certain observables and values of the Wilson coefficients. 
In order to quantify the stability of our results with respect to the inclusion of $\mathcal{O}\lp \Lambda^{-4}\rp$ corrections, here we present similar results as those of Sect.~\ref{subsec:results_baseline_linear} now keeping the quadratic EFT corrections to the cross sections arising from these squares.
All other ingredients of the global SMEFT fit are kept unchanged. 

\begin{figure}[htbp]
    \centering
\includegraphics[width=\linewidth]{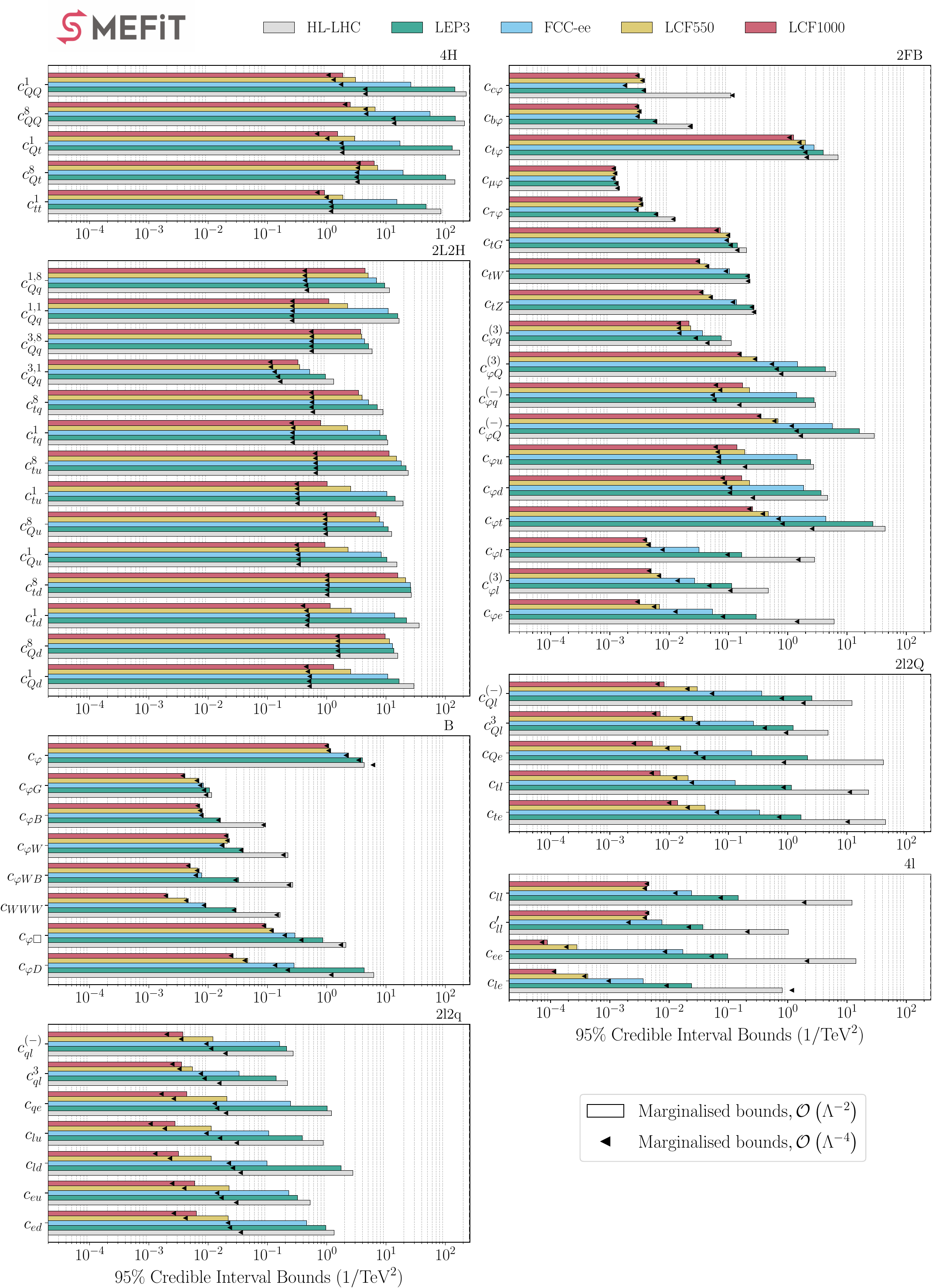}
    \caption{Same as Fig.~\ref{fig:lin_ind_global}, comparing the marginalised bounds obtained in global fits carried out at the linear level in the EFT expansion (filled bars) and those from the quadratic level fits (triangles). 
    Theoretical uncertainties are included in the `aggressive' scenario. 
\label{fig:lin_vs_quad_glob}
}
\end{figure}

\begin{figure}[htbp]
    \centering
\includegraphics[width=\linewidth]{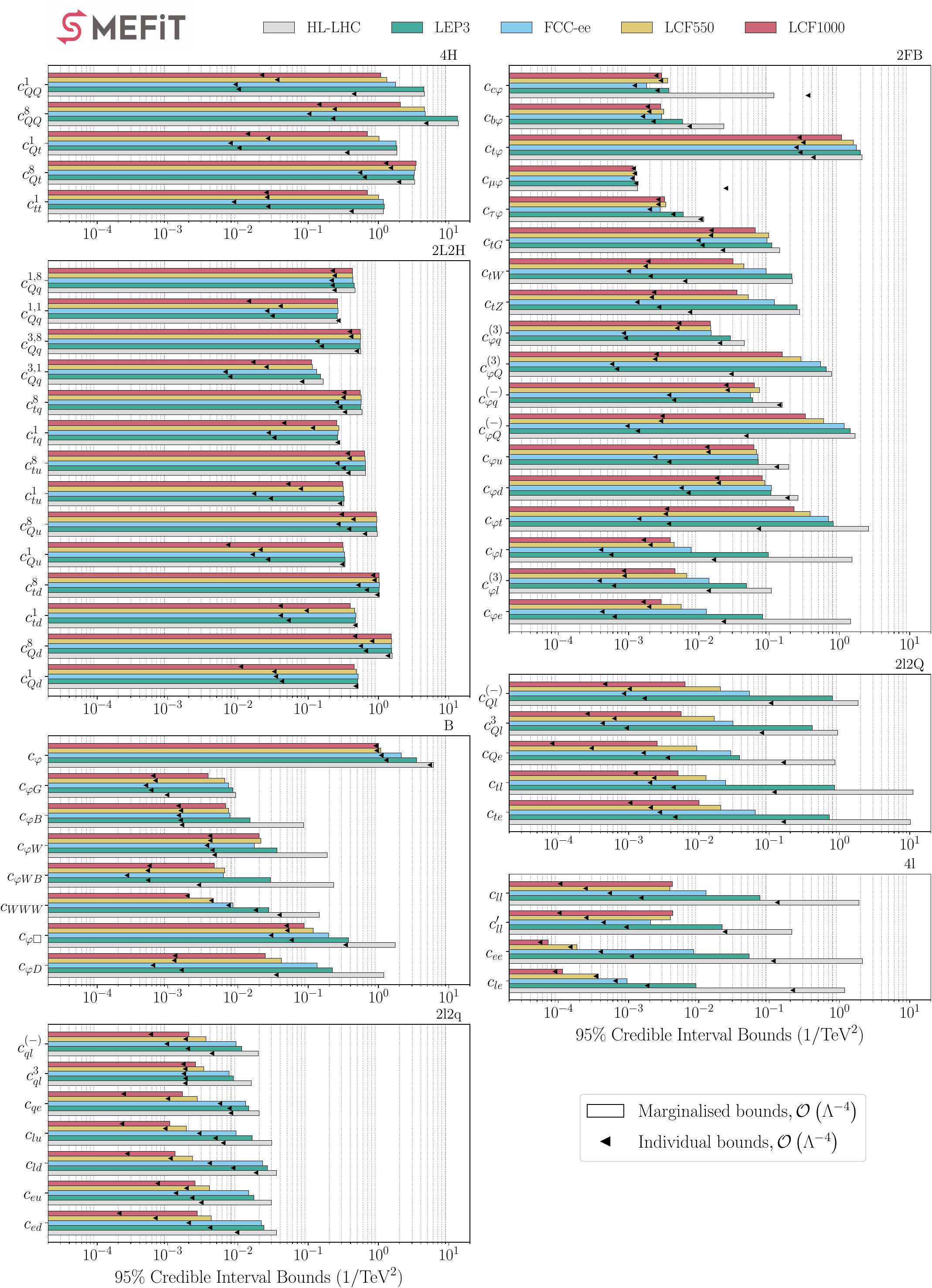}
    \caption{Same as Fig.~\ref{fig:lin_ind_global} for quadratic  fits carried out at $\mathcal{O}\lp \Lambda^{-4}\rp$ in the EFT expansion.
    For each of the colliders considered, the marginalised bounds from the global fit (filled bars) are compared with the individual bounds (triangles).
    Theoretical uncertainties are included in the `aggressive' scenario. 
\label{fig:quad_ind_global}
}
\end{figure}

In order to assess the impact of quadratic corrections, we carried out two comparisons.
First, Fig.~\ref{fig:lin_vs_quad_glob} compares the global marginalised bounds on the Wilson coefficients obtained at the linear and quadratic level for each of the future colliders considered.
Second, Fig.~\ref{fig:quad_ind_global} compares the same marginalised bounds on the quadratic global EFT fit with the analogous results obtained in individual quadratic fits, where only one Wilson coefficient is assumed to be non-zero at the reference scale $\mu_0=10$ TeV.
We now detail some observations that can be derived from Figs.~\ref{fig:lin_vs_quad_glob} and~\ref{fig:quad_ind_global}.

\paragraph{Linear versus quadratic fits.}

We first note that for some operators, once quadratic corrections are included, the HL-LHC projections already saturate the sensitivity, with no or minor subsequent improvements from the lepton colliders.
This is the case especially for the two-light-two-heavy quark operators.

Moreover, for some operators, one finds large differences in the bounds obtained in the linear and the quadratic EFT level, with the latter resulting in more stringent constraints by up to two orders of magnitude, for instance for most of the four-fermion operators. On the other hand, quadratic and linear bounds are similar for certain Wilson coefficients such as the Higgs self-coupling, the Yukawa operators, or the triple gauge coupling $c_{WWW}$. In the case of the LCF projections, differences between linear and quadratic bounds are smaller for operators involving a lepton bilinear, which benefit from the constraints from difermion production processes at high $\sqrt{s}$ values.  

In general, from Fig.~\ref{fig:lin_vs_quad_glob} one concludes that quadratic EFT corrections at the marginalised level can be sizeable for several of the Wilson coefficients included in the global fit, and therefore must be accounted for. They  lead to phenomenologically relevant impact especially within the four-fermion operators and some of the two-fermion ones. 
Quadratic EFT corrections can, however, be safely neglected at future colliders for the Higgs self-coupling, the triple gauge coupling, and for the Yukawa couplings (except for the top Yukawa at the FCC-ee and LEP3).

\paragraph{Individual versus marginalised fits.}

We now consider the results shown in Fig.~\ref{fig:quad_ind_global}, which compares the bounds obtained from individual and marginalised analyses when quadratic EFT corrections are taken into account. The linear EFT fit results discussed in the previous section displayed large variations between the bounds on the Wilson coefficients obtained in the individual and global marginalised fits. 
The same trend is obtained in the case of quadratic fits.

As in the case of the linear fits, also for the quadratic ones the overall pattern of improvement across colliders as compared to the HL-LHC baseline varies dramatically between individual and global fits, as illustrated in particular by the two-fermion-two-bosonic operators.
Concerning the individual quadratic bounds, the best performance is again obtained by the FCC-ee, with LCF being comparable or better for operators involving lepton bilinears. 

The analogous results of the spider plots for the linear fits from Figs.~\ref{fig:spider_lin_ind_aggressive} and~\ref{fig:spider_lin_global_aggressive}, showing the improvement in the bounds of the Wilson coefficients in future lepton colliders as compared to the HL-LHC, are shown for the quadratic fits in App.~\ref{app:extra_smeft}, specifically in Figs.~\ref{fig:spider_quad_ind} and~\ref{fig:spider_quad_glob}. 
These plots illustrate some of the main findings that we just discussed: for individual quadratic fits, FCC-ee shows the best performance except for operators involving lepton bilinears and $c_{WWW}$. 
For global marginalised fits at the quadratic level, LCF leads to the largest reduction, and, finally, the benefits of the Tera-$Z$ run are markedly reduced in the global fit.

\subsection{Impact of theoretical uncertainties}
\label{subsec:theory_errors_impact}

\begin{figure}[htbp]
    \centering
\includegraphics[width=\linewidth]{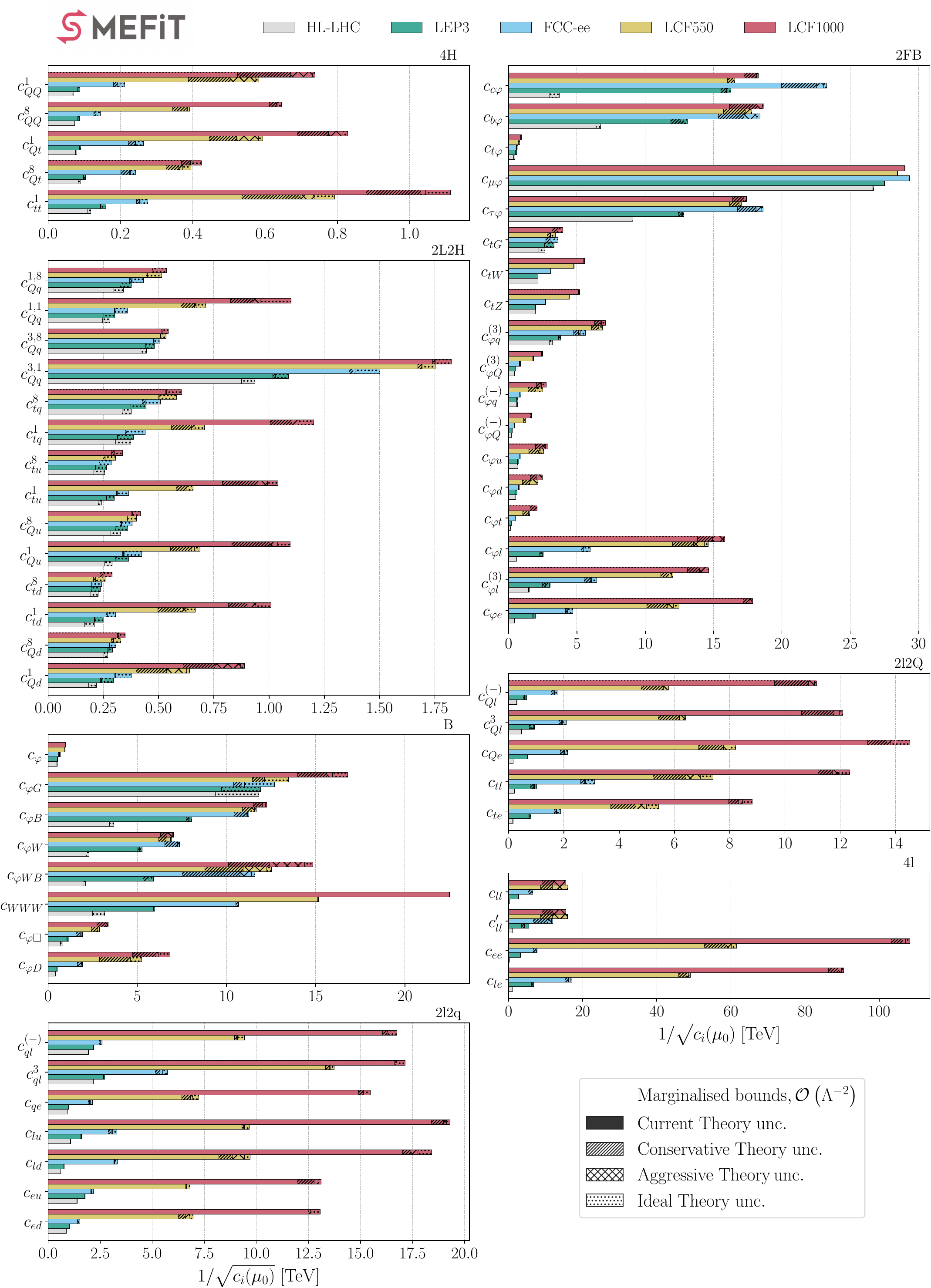}
    \caption{Same as Fig.~\ref{fig:lin_ind_global_lambda} for the marginalised lower bounds on the mass scale $\Lambda=c_i^{-1/2}(\mu_0)$ in the global linear fits at the future colliders considered, comparing the results in the four scenarios for theoretical uncertainties at future lepton colliders: current, conservative, aggressive, and ideal. 
    Note that now the $x$-axis scale is linear and varies for each group of Wilson coefficients. 
    See Fig.~\ref{fig:quad_ind_th_lambda} for the corresponding results in the case of the $\mathcal{O}\lp \Lambda^{-4}\rp$ fits.
\label{fig:lin_glob_th_lambda}
}
\end{figure}

As demonstrated in Refs.~\cite{FCC:2025lpp, Freitas:2019bre}, theoretical uncertainties can significantly degrade the physics reach of future lepton colliders, especially for the high-precision $Z$-pole measurements, but also for Higgs and difermion production.
Here we quantify the role played by theory uncertainties in our SMEFT interpretations of future lepton collider data. Fig.~\ref{fig:lin_glob_th_lambda} presents a similar comparison for the lower bound on the NP mass scale $\Lambda=c_i^{-1/2}(\mu_0)$ as in  Fig.~\ref{fig:lin_ind_global_lambda} for the global marginalised bounds in the linear EFT fit, now for the results in the four scenarios for theoretical uncertainties discussed in Sect.~\ref{subsec:th_errors}: current, conservative, aggressive, and ideal.
Notice that in these comparisons the $x$-axis has a linear scale, and its range has been adjusted to match the typical variations in the different groups of Wilson coefficients. 
See Fig.~\ref{fig:quad_ind_th_lambda} for the corresponding results in the case of the $\mathcal{O}\lp \Lambda^{-4}\rp$ fits. 

From Fig.~\ref{fig:lin_glob_th_lambda} one finds that theoretical uncertainties reduce the reach in the mass scale $\Lambda$ by an amount that depends strongly on the specific operators being probed, the collider, and the scenario for the expected reduction of theoretical uncertainties assumed. 
For instance, for the purely bosonic operator $c_{\varphi W B}$, the projected bounds on $\Lambda$ improve from 7.5 TeV to almost 12 TeV (10 TeV to 15 TeV) at the FCC-ee (LCF550), when moving from the current to the ideal theory scenario.
For other operators, theoretical uncertainties barely have an impact, such as for the Higgs self-coupling $c_{\varphi}$, or the triple gauge coupling $c_{WWW}$. 
In most cases, differences between the `conservative' and `ideal' scenario are found to be at the 10\% to 20\% level for the global marginalised fit results.

\begin{figure}[htbp]
    \centering
\includegraphics[width=\linewidth]{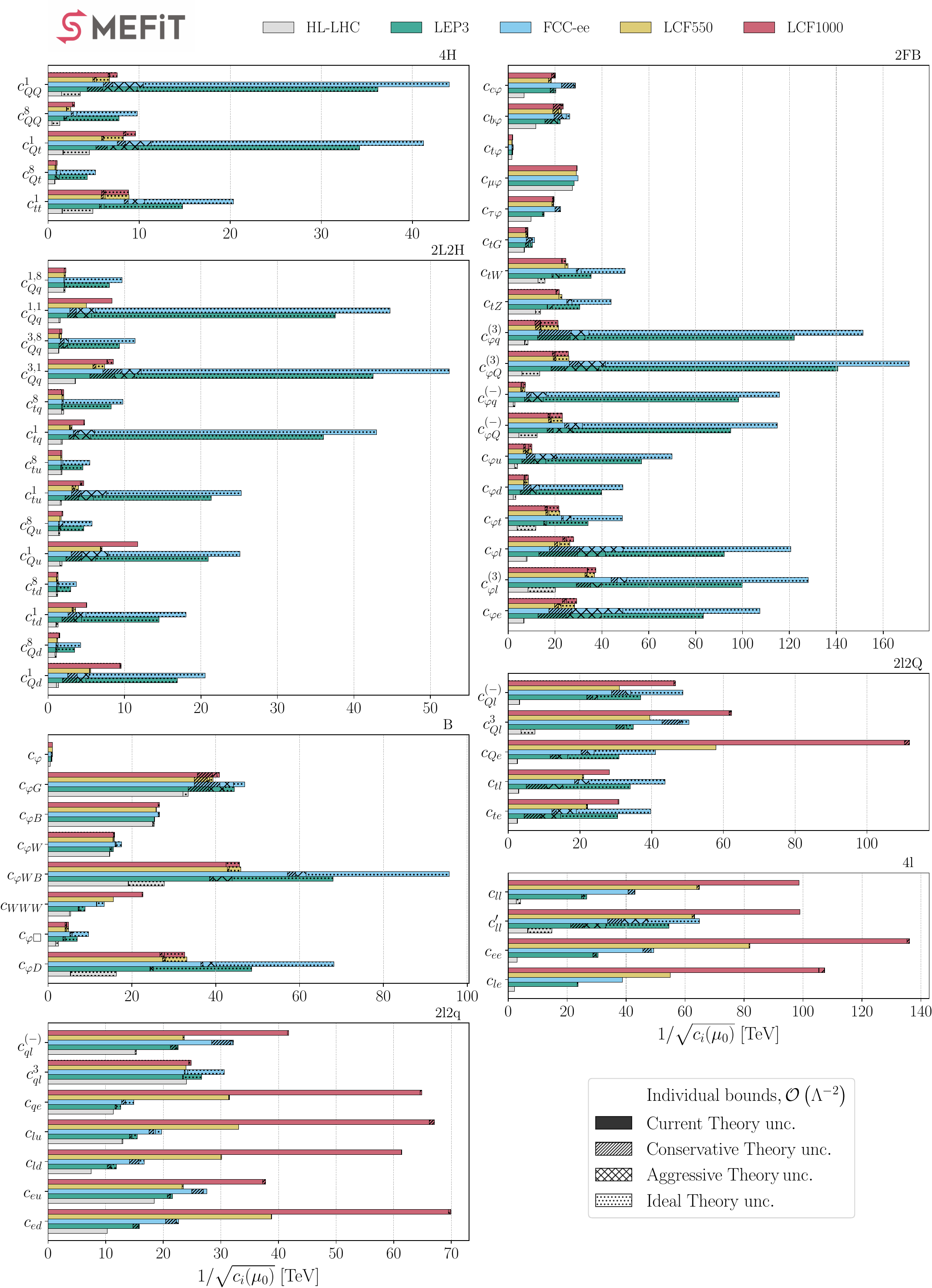}
    \caption{Same as Fig.~\ref{fig:lin_glob_th_lambda} now for individual bounds on the Wilson coefficients. 
\label{fig:lin_ind_th_lambda}
}
\end{figure}

In Fig.~\ref{fig:lin_ind_th_lambda} we present a similar comparison as in Fig.~\ref{fig:lin_glob_th_lambda} for the case of individual fits.
There we find, instead, that the impact of theoretical uncertainties is rather more significant than for the global marginalised analysis.
The observation that stands out the most is how different scenarios for the theory uncertainties impact the individual bounds at FCC-ee and LEP3 for operators that are sensitive to the $Z$-pole observables. For example, the $c_{\varphi q}^{3}$ coefficient shows variations up to a factor 13 at the FCC-ee depending on whether the current or ideal theory scenario is considered. 
By contrast, the effect of the different scenarios is less pronounced at LCF, because FCC-ee is significantly more sensitive to $Z$-pole observables and is therefore more strongly affected by the corresponding theoretical uncertainties.

\subsection{The Higgs self-coupling at future colliders}
\label{subsec:higgs_self_coupling}

The measurement of the Higgs self-interactions is one of the main physics goals of the HL-LHC and of all proposed future particle colliders.
In~\cite{terHoeve:2025omu} we critically assessed the sensitivity of the global SMEFT analysis to deformations of the Higgs self-coupling modifier $\kappa_3$ at the HL-LHC and the FCC-ee.
Our analysis found that improving on the legacy HL-LHC constraints on $\kappa_3$ at the FCC-ee in the global marginalised fit was not possible without the $\sqrt{s}=365$ GeV run;  that individual and marginalised determinations are similar at the HL-LHC while differing by a factor 2 at the FCC-ee; and that quadratic EFT corrections cannot in general be neglected. 
The combination of HL-LHC and FCC-ee data was found to be able to pin down the Higgs self-coupling with $\sim15$\% precision.

Here we extend this analysis to the other lepton colliders discussed in this section (LCF and LEP3) in the framework of the improved global SMEFT fit, with a more extensive operator basis, the most updated projections for future collider observables, and for the conservative and aggressive scenarios for the theoretical uncertainties. 

\begin{figure}[t]
    \centering
\includegraphics[width=0.75\linewidth]{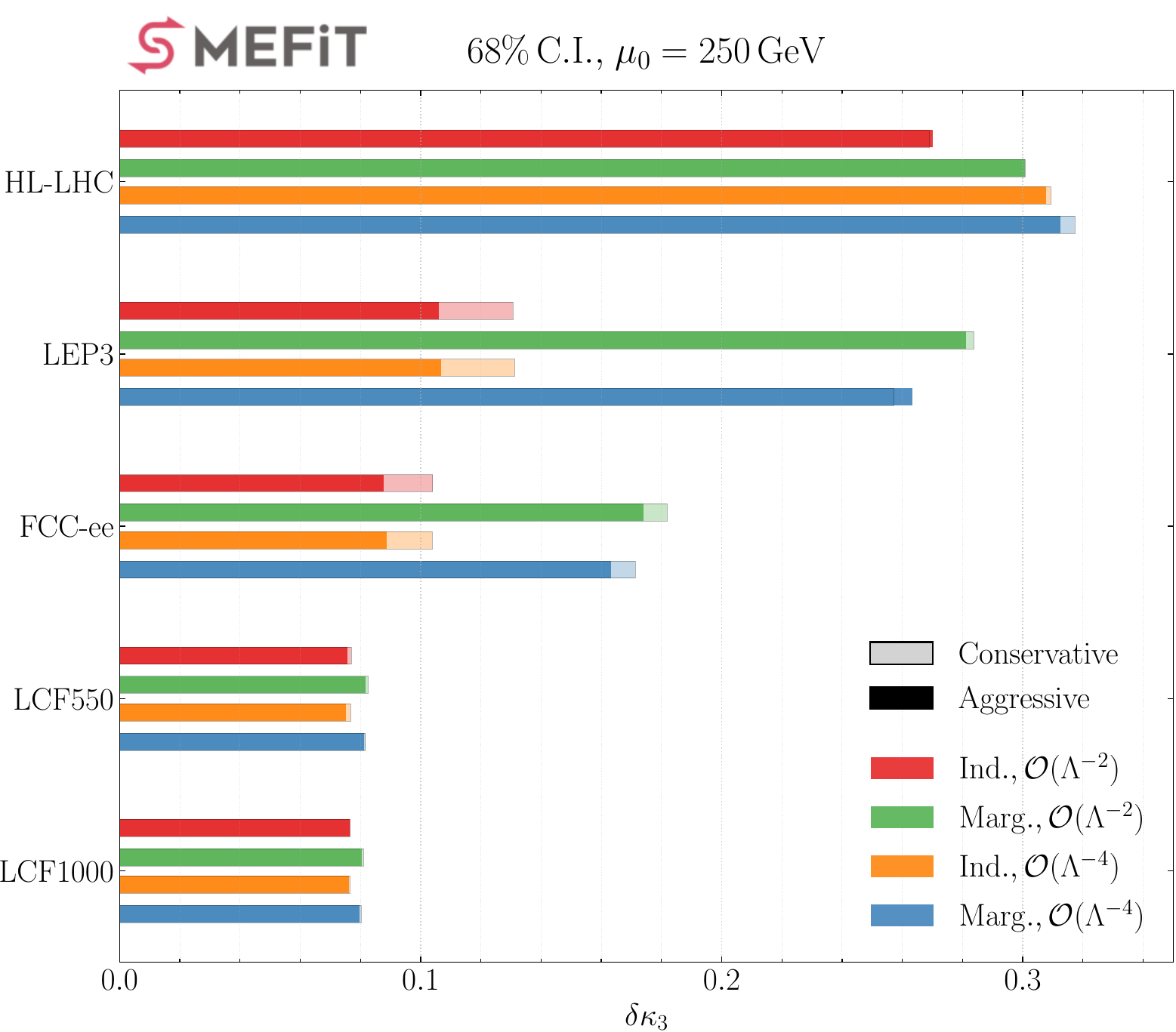}
    \caption{The 68\% C.I. bounds on the Higgs self-coupling modifier $\delta\kappa_3$ obtained in linear and quadratic EFT fits to different future collider projections, for both the conservative and aggressive scenario for theoretical uncertainties. 
    Bounds correspond to a reference scale of $\mu_0=250$ GeV.
    We compare the HL-LHC projections with those from the FCC-ee, LEP3, LCF550 and LCF1000, in all cases with the HL-LHC dataset included.
    See Table~\ref{tab:bounds_trilinear} for the corresponding numerical results.
\label{fig:bound_plot_cphi_hllhc_fcc}}
\end{figure}

Following the approach described in~\cite{terHoeve:2025gey}, Fig.~\ref{fig:bound_plot_cphi_hllhc_fcc} compares the 68\% C.I. bounds on the Higgs self-coupling modifier $\delta\kappa_3$ at the reference scale $\mu_0=250$ GeV for the future collider projections considered in this work. We compare the HL-LHC projections with those from the FCC-ee, LEP3, LCF550, and LCF1000, in all cases with the HL-LHC cross-sections included in the dataset. For each collider, we display the bounds obtained from linear and quadratic EFT fits, as well as individual and global marginalised analyses and distinguish between the conservative and aggressive theory scenarios.
The corresponding numerical bounds are provided in Table \ref{tab:bounds_trilinear}.  

Inspecting the results of Fig.~\ref{fig:bound_plot_cphi_hllhc_fcc} together with the numerical bounds in Table~\ref{tab:bounds_trilinear}, one finds that, first, the HL-LHC and LEP3 give similar results for $\delta\kappa_3$ in the marginalised setup. 
This is consistent with the finding of \cite{terHoeve:2025omu} that demonstrated that the $\sqrt{s}=365$ GeV run was essential to improve on the HL-LHC determination of the Higgs self-coupling. 
We remark that, in comparison to our work in Ref.~\cite{terHoeve:2025omu}, the quadratic marginalised bounds at HL-LHC have changed slightly as a result of the combination of the extended fitting basis, different leptonic flavour assumptions, and updates in the HL-LHC dataset.

At the FCC-ee in the aggressive scenario, we find $|\delta\kappa_3| < 0.18$ and $|\delta\kappa_3| < 0.17$ in the case of linear and quadratic marginalised fits, respectively. As at LEP3, we find that marginalisation weakens the bounds by around a factor 2 up to 3 relative to the corresponding individual bounds, showing how $Zh$ production cannot be studied in isolation, unlike at the HL-LHC where bounds were found to be similar between the individual and marginalised case. 
Moving to the LCF, we find $|\delta\kappa_3| < 0.08$ at the linear marginalised level in the aggressive scenario, an improvement of around a factor two compared to the FCC-ee. 
Marginalised bounds are similar to the corresponding individual bounds, which can be understood by recalling that $c_\varphi$ is directly constrained from di-Higgs production in the high energy runs through direct $\nu \nu HH$ and $ZHH$ measurements. 
This result is consistent with Fig.~\ref{fig:lin_ind_global} where we found similar bounds on $c_\varphi$ in the individual and marginalised setups for LCF and HL-LHC. 

Regarding the impact of different theory scenarios, we find that their impact is generally mild except for the individual fits in the case of LEP3 and FCC-ee, where we find differences up to around 30\%. In the marginalised case, we find a slight degradation of the bounds in the conservative scenario compared to the aggressive one at the few-percent level at FCC-ee, while no significant impact is instead observed in the case of LEP3 and LCF. 

\begin{table}[t]
    \small
    \renewcommand{\arraystretch}{1.4}
    \centering
    
    \begin{tabular}{|l | c|c|c|c|c|}
    \hline
         \multicolumn{2}{|l|}{\multirow{2}{*}{$\delta\kappa_3$~(68\% C.I.)}} & \multicolumn{2}{c|}{Aggressive} & \multicolumn{2}{c|}{Conservative} \\
         \cline{3-6}
          \multicolumn{2}{|l|}{} & Individual & Marginalised & Individual & Marginalised \\
         \hline
\multirow{2}{*}{HL-LHC} & Linear & $0.27$ & $0.30$ & $0.27$ & $0.30$\\
 & Quad. & $0.31$ & $0.31$ & $0.31$ & $0.32$\\
\hline
\multirow{2}{*}{LEP3} & Linear & $0.11$ & $0.28$ & $0.13$ & $0.28$\\
 & Quad. & $0.11$ & $0.26$ & $0.13$ & $0.26$\\
\hline
\multirow{2}{*}{FCC-ee} & Linear & $0.09$ & $0.17$ & $0.10$ & $0.18$\\
 & Quad. & $0.09$ & $0.16$ & $0.10$ & $0.17$\\
\hline
\multirow{2}{*}{LCF550} & Linear & $0.08$ & $0.08$ & $0.08$ & $0.08$\\
 & Quad. & $0.08$ & $0.08$ & $0.08$ & $0.08$\\
\hline
\multirow{2}{*}{LCF1000} & Linear & $0.08$ & $0.08$ & $0.08$ & $0.08$\\
 & Quad. & $0.08$ & $0.08$ & $0.08$ & $0.08$\\
 \hline
    \end{tabular}

   \vspace{0.2cm}
    \caption{The 68\% C.I. on $\delta\kappa_3$ (for $\mu_0=250$ GeV)
    in linear and quadratic SMEFT fits shown in Fig.~\ref{fig:bound_plot_cphi_hllhc_fcc}.
 }
    \label{tab:bounds_trilinear}
\end{table}

\section{Future colliders and benchmark UV models}
\label{sec:uv_benchmarks}

While the SMEFT provides a powerful framework to assess exploration power and detect deviations from the SM in a largely model-agnostic way, a complete understanding of the underlying physics will ultimately require a specific NP model.
In that spirit, in this section we compare different future colliders under the light of representative benchmark UV-complete models matched to the SMEFT.
Specifically, we consider one-particle extensions of the SM, matched to the SMEFT either at tree level or at one-loop, and the Higgs compositeness scenario.
The extension of our analysis to other UV models is straightforward (and mostly automated) following the procedure of~\cite{terHoeve:2023pvs}.
Our projections for the indirect NP reach of the electron-positron colliders considered here (FCC-ee, LEP3, and LCF) can be directly compared with the direct reach of high-energy colliders such as the FCC-hh or the muon collider presented in the ESPPU2026  PBB for benchmarks such as the $W'$ and $Z'$ extensions of the SM or leptoquark scenarios.

In this analysis, we use the same datasets and theoretical predictions from Sect.~\ref{sec:global_smeft_fits}, replacing in the latter case the WCs by their expressions in terms of UV parameters that arise from matching.
This way, we can carry out fits and parameter scans directly in the UV space.
In all cases, results presented in this section are based on the most accurate theory calculations available, in particular including NLO corrections to the SMEFT cross-sections, RGE effects, and quadratic corrections whenever required.
Furthermore, the exact functional dependence of the Wilson coefficients matched to the UV parameters is considered without additional approximations (such as linearisation).

\subsection{Mass reach for one-particle extensions}
\label{sec:mass_reach}

First of all, we consider one-particle extensions of the SM matched to the SMEFT either at tree-level or at one-loop and compare the mass reach of future electron-positron colliders.
The matching relations were obtained using the Granada dictionary~\cite{deBlas:2017xtg}, \texttt{Matchmakereft}~\cite{Carmona:2021xtq} and \texttt{SOLD}~\cite{Guedes:2023azv,Guedes:2024vuf}, and fed into {\sc\small SMEFiT} using \texttt{match2fit}~\cite{terHoeve:2023pvs}.
Compared with previous work~\cite{Celada:2024mcf,terHoeve:2025gey}, we have now added the matching contributions associated to the 2-lepton-2-quark and 4-lepton operators included in the SMEFT predictions, consistently with the settings of Sect.~\ref{sec:global_smeft_fits}.
For each model, we define the $95\%$ C.L. mass reach as the heavy mass $M_{\rm UV}$ for which $\chi^2\left(M_{\rm UV}\right) = \chi^2_{\rm SM} + 3.84$ for fixed values of the UV couplings, as in~\cite{terHoeve:2025gey}. 
We take all dimensionless (dimensionful) UV couplings to be $g_{\rm UV} = 1 $ ($g_{\rm UV} = 1 $~TeV) at the matching scale $\mu=M_{\rm UV}$.

\begin{figure}[t]
    \centering
\includegraphics[width=0.49\linewidth]{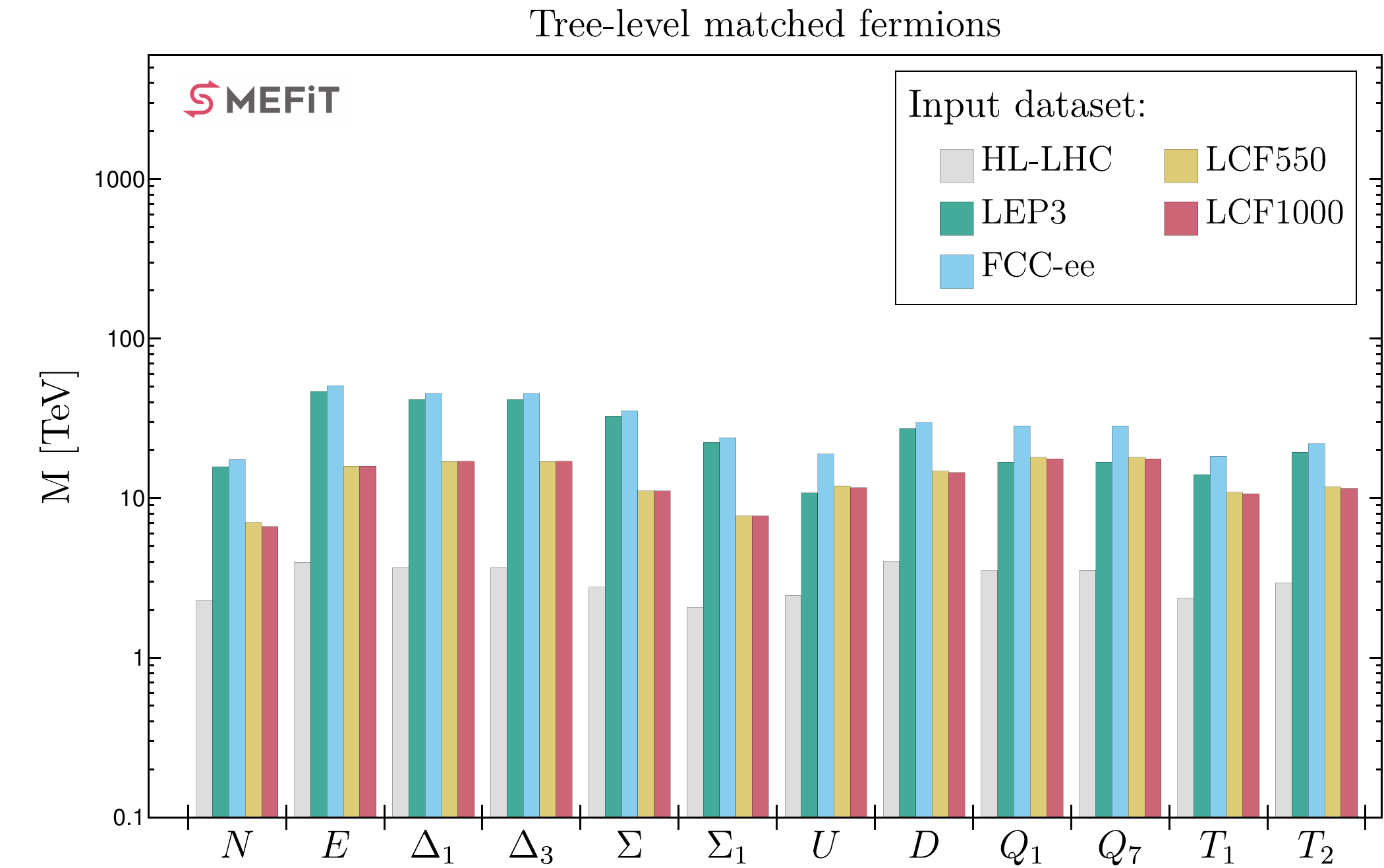}
\includegraphics[width=0.49\linewidth]{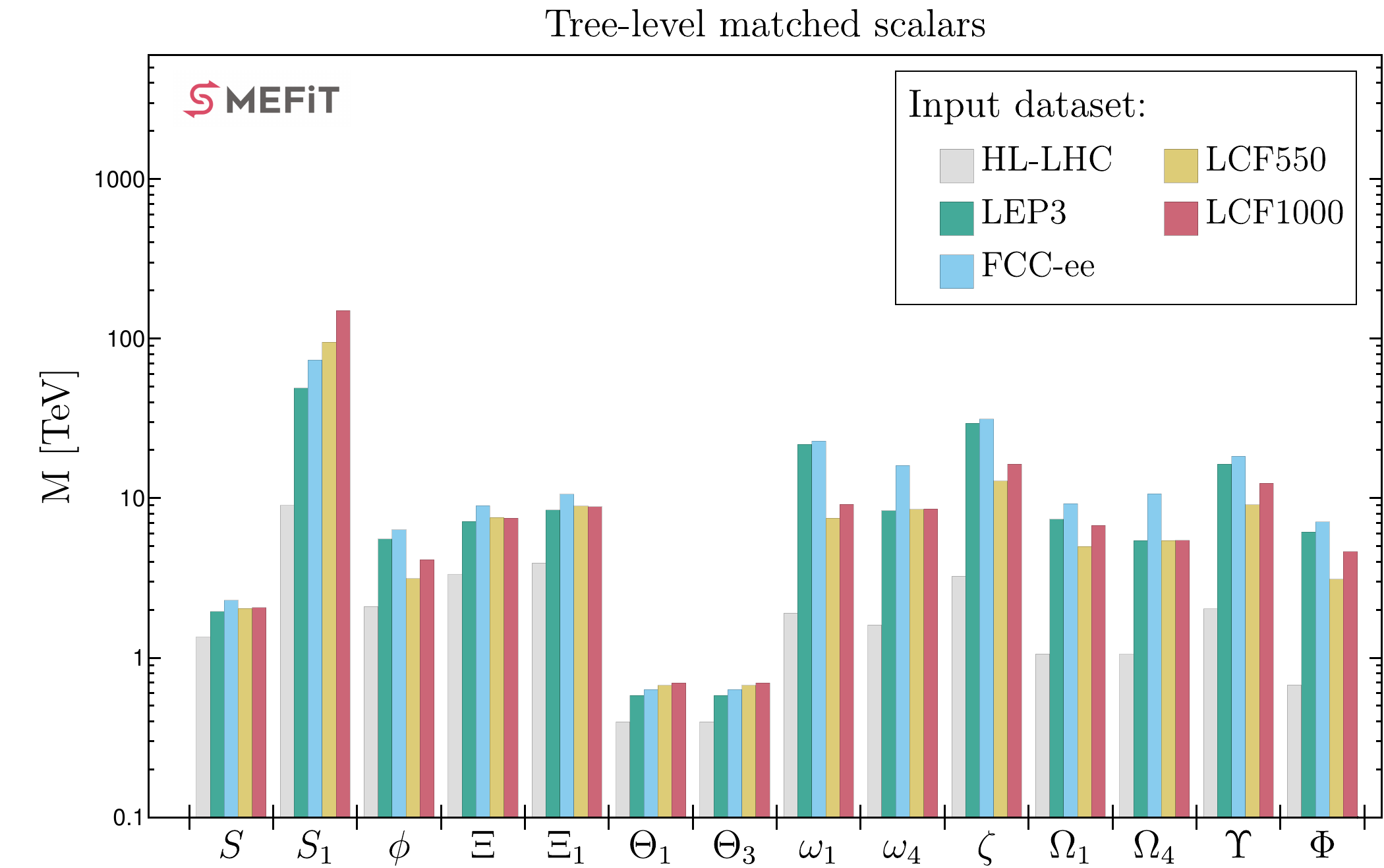}\hfill
\includegraphics[width=0.49\linewidth]{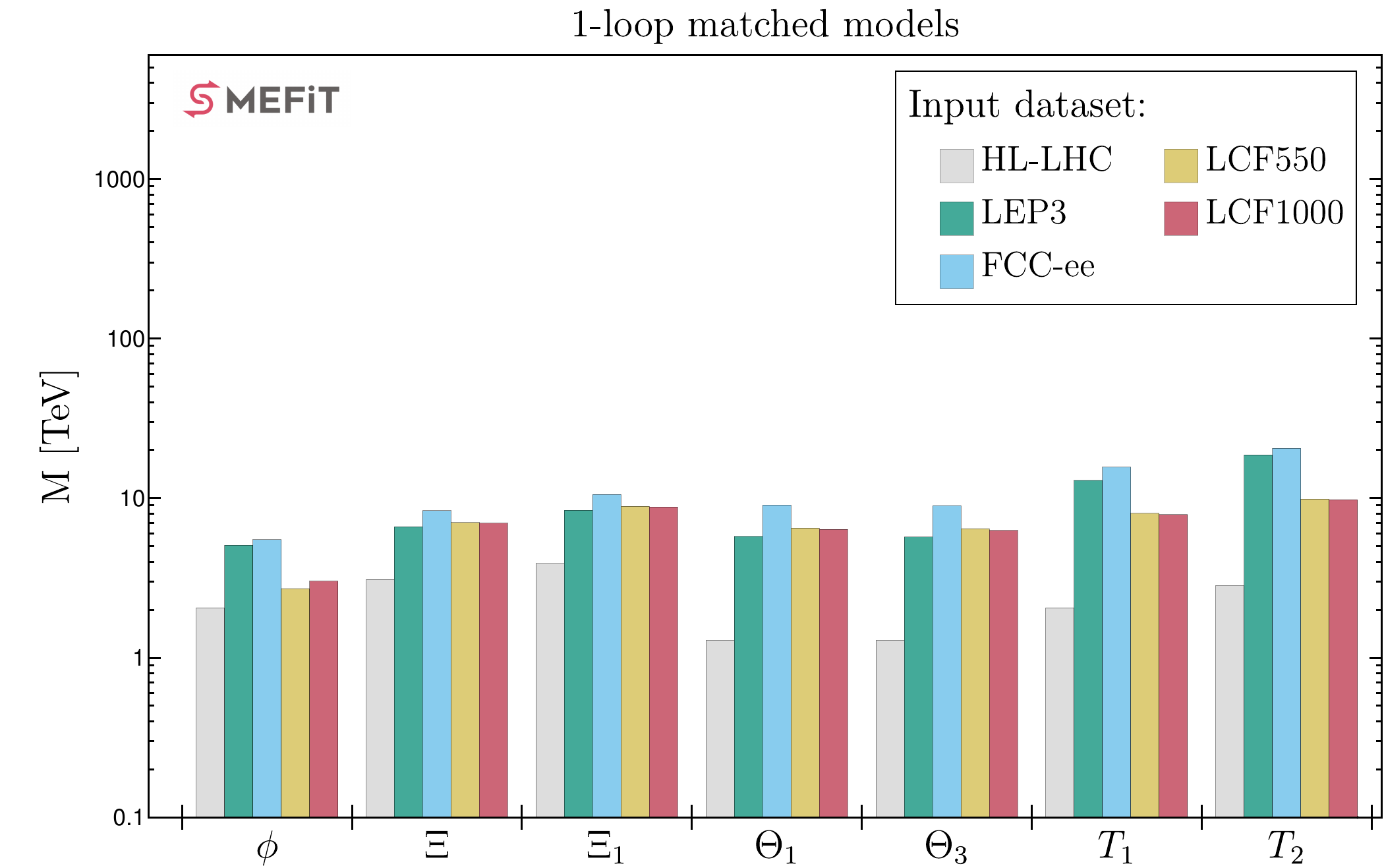}
\includegraphics[width=0.49\linewidth]{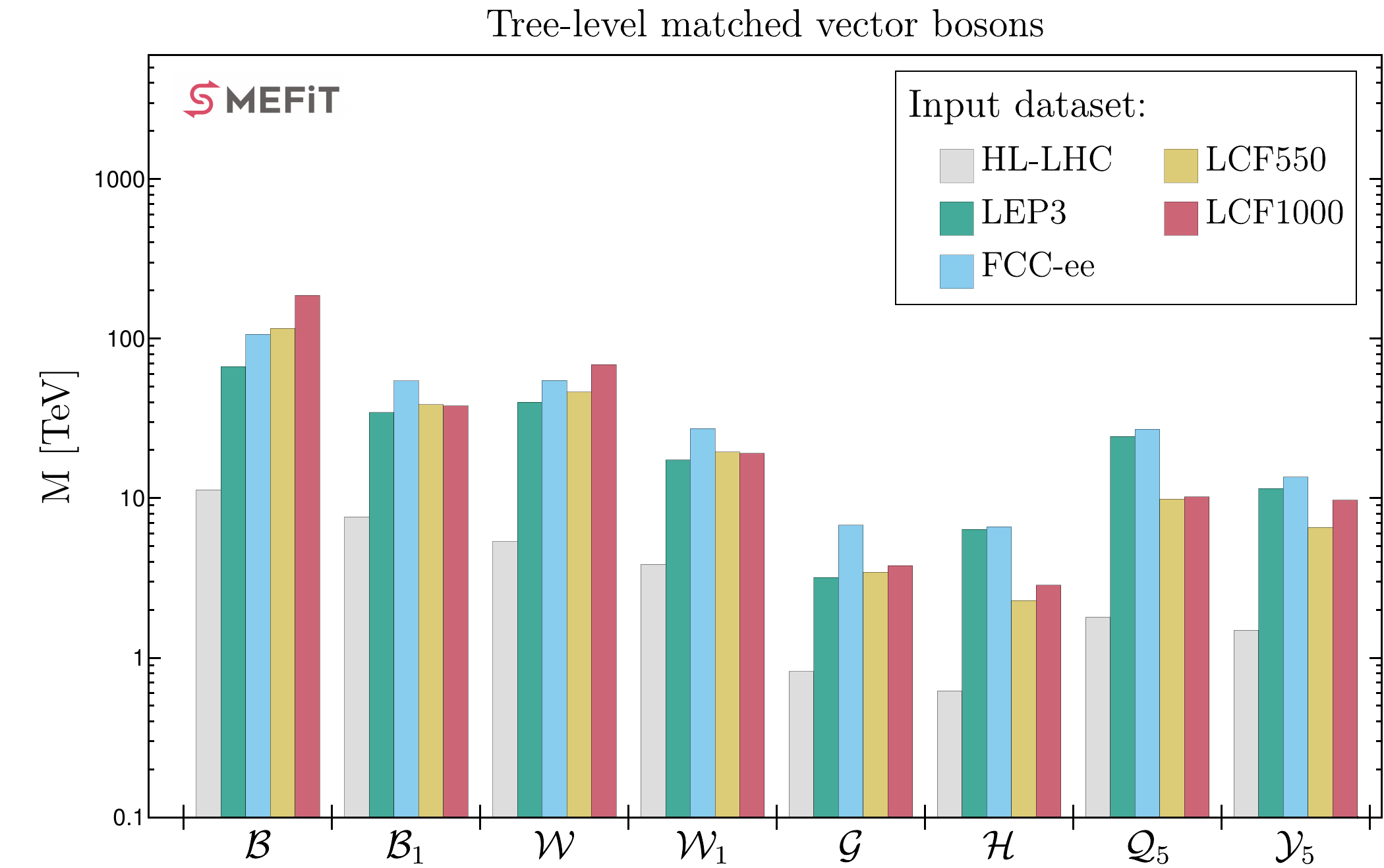}
    \caption{The $95\%$ C.L. reach on the heavy particle mass $M_{\rm UV}$ in representative single-particle extensions of the SM  matched onto the SMEFT.
    This mass reach is determined by a $\chi^2$ scan by imposing the condition $\chi^2(M_{\rm UV})=\chi^2_{\rm SM} + 3.84$.
    All UV couplings $g_{\rm UV}$ are set to be either 1 or $1$~TeV, depending on their dimensionality, at the matching scale $\mu=M_{\rm UV}$. 
   We display the bounds obtained from the HL-LHC and then those upon the addition of the LEP3, FCC-ee, LCF550, or LCF1000 projections.
    We include quadratic EFT dependence in the predictions and  account for theoretical uncertainties in the aggressive scenario. 
    The RGE evolution is performed from $M_{\rm UV}$ down to the scale relevant for each dataset.   
    From the top left in a clockwise direction, we show the results for heavy vector-like fermions, heavy scalars, and heavy vector spin-1 bosons, all matched at tree level, as well as representative heavy scalar and fermion models matched at the one-loop level (bottom-left panel).
    }
\label{fig:mass_reach_future_colliders}
\end{figure}

Results for the mass reach of one-particle extensions of the SM are shown in Fig.~\ref{fig:mass_reach_future_colliders}. 
Beginning our analysis by the tree-level matched fermions (top-left panel), we observe that the HL-LHC reach is in the range $1-5$~TeV for all possible resonances. 
Future lepton colliders will take this reach to the realm of $10$~TeV, with a few exceptions in the case of LCF550 or LCF1000. 
In general, both LEP3 and FCC-ee offer better sensitivity to these models than LCF, since a main driver of the bounds is the effects on Z-pole observables, either via RGE running or tree-level contributions. FCC-ee always offers a better sensitivity than LEP3, with the difference going beyond the few percent in the case of colored fermions that can couple to the top quark, such as $U$, $Q_1$ and $Q_7$, where the $t\bar{t}$-threshold run of FCC-ee can make a difference.

Tree-level matched scalars (top-right panel) offer a less uniform behaviour. 
All colliders offer poor sensitivity, with mass reach of $\mathcal{O}(1)$~TeV, to models that generate at tree-level very few and/or poorly constrained dimension-6 operators, such as the singlet $S$ and the electroweak quadruplets $\Theta_{1}$, $\Theta_{3}$. 
Lepton colliders can probe those models better than HL-LHC but the advantage does not reach the order of magnitude. FCC-ee is the best collider to probe the singlet, while the quadruplets, which only generate $\mcO_\varphi$ at tree level, are better probed at LCF1000, in agreement with our study of the sensitivity to the Higgs self-coupling. 
The charged singlet $S_1$ is the model that can be best probed, since it contributes to the operators $\mcO_{\ell\ell}^{1122}$ and $\mcO_{\ell\ell}^{1221}$, which enter via input shifts and generate energy-growing contact-term contributions to $e^{+}e^{-}\to\mu^{+}\mu^{-}$. 
While the first effect explains the reach of $\sim10$~TeV at HL-LHC, the second explains the gain in reach observed at lepton colliders, which increases as the centre-of-mass energy of the collider grows, reaching $\sim150$~TeV for LCF1000. The remaining colorless scalars, $\phi$, $\Xi$, and $\Xi_1$, yield mass reaches in the range $2-10$~TeV. Any of the considered lepton colliders improves the reach to those models by a factor $\sim2$ and FCC-ee is the one that gives the best reach in all cases, with either variant of LCF matching the reach of LEP3 at best. Finally, for the case of colored scalars, i.e. $\omega_1$, $\omega_4$, $\zeta$, $\Omega_1$, $\Omega_4$, $\Upsilon$, and $\Phi$, we obtained mass reaches around $1-2$~TeV at HL-LHC, while lepton colliders push it towards the $10$~TeV region, reaching $30$~TeV for $\zeta$ at FCC-ee while staying at $\sim 5$~TeV for $\Phi$. The running of tree-level generated four-quark operators into $Z$-pole observables explains the advantage of FCC-ee in exploring these colored scalars, as well as the fact that LCF, in either variant, at best matches LEP3 but never improves on it.

The bottom-right panel shows the result for massive spin-1 vector bosons matched at tree level. 
While HL-LHC offers mass reaches between $1$ and $10$~TeV to these models, lepton colliders comfortably push it towards several $10$s of TeV and even pass the $100$~TeV threshold, with the exception of the color-octet bosons $\mathcal{G}$ and $\mathcal{H}$ where lepton colliders offer equally impressive improvements but the sensitivity remains below $10$~TeV. 
FCC-ee offers the best sensitivity to most of these models with LEP3 as second-best and LCF usually matching LEP3, with two exceptions. 
These are the $\mathcal{B}$ and $\mathcal{W}$ models, which can  be seen as $Z'$ and $W'$ models respectively. In those two models, LCF1000 offers the best sensitivity and LCF550 is close or matches FCC-ee. This is due to their contributions to four-lepton operators that generate energy-growing effects in processes such as $e^+e^-\to e^+e^-/\mu^+\mu^-/\tau^+\tau^-$ and thus profit from the higher energies reached at linear lepton colliders.

Our selection of one-loop matched models (bottom-left panel) includes five colorless scalars, $\phi$, $\Xi$, $\Xi_1$, $\Theta_1$, and $\Theta_3$, and two vector-like fermions, $T_1$ and $T_2$, all of them already included in the figures with tree-level matched models. In general, the inclusion of one-loop matching corrections improves the bounds to all models. This improvement is an order-of-magnitude effect only for the scalar quadruplets $\Theta_{1}$ and $\Theta_3$, where the sensitivity via the tree-level generated $\mcO_\varphi$ is greatly surpassed by the sensitivity via the loop-generated custodial-violating $\mcO_{\varphi D}$, which also explains the FCC-ee dominance in the sensitivity to these models. For all one-loop matched models included here, HL-LHC offers a sensitivity between $1-5$~TeV, while lepton colliders push this to the $5-20$~TeV region. The best sensitivity is obtained at FCC-ee in all cases, with LEP3 as the second best in most cases. 
LCF550 or LCF1000 can match the LEP3 sensitivity for the scalar electroweak triplets and quadruplets, but falls behind when considering the scalar doublet $\phi$ and the two fermions.

The general advantage of FCC-ee in the mass reach to these models, with a few exceptions where LCF becomes the most sensitive collider, is consistent with the result of individual SMEFT fits discussed in Sects.~\ref{subsec:results_baseline_linear}-\ref{subsec:quadratic_corrections}. 
Indeed, all the one-particle models in this study involve between one and at most a handful of UV couplings (see~\cite{terHoeve:2025gey} for details) and all have been set to the same fixed value. In this way, we are effectively exploring one direction in the BSM space with each heavy particle. 
Reproducing a setup more similar to the marginalised global SMEFT fit with a UV model would entail including at least several heavy particles with independent UV couplings at the same time, which we leave for future studies.

\subsection{Testing compositeness at future colliders}
\label{sec:compositeness}

A particularly compelling NP scenario is Composite Higgs (CH)~\cite{Contino:2003ve,Agashe:2004rs,Panico:2015jxa}, which was originally developed to address the Higgs hierarchy problem but has proven to be a framework capable of addressing several issues of the SM. 
There, the Higgs boson arises as a light composite pseudo-Nambu-Goldstone boson from a strongly-interacting heavy sector, while there is an elementary particle corresponding to each of the other SM degrees of freedom.
As discussed in the ESPPU2026 PBB and in related work~\cite{deBlas:2019rxi,deBlas:2025gyz}, the composite Higgs scenario is particularly useful to compare the reach of future particle colliders.
For benchmark purposes, one may consider the simplest composite Higgs scenario, where the effect on low-energy electroweak physics can be described in terms of only two parameters, $g_*$ and $m_*$, which characterise the typical coupling strength and mass of the heavy UV particles.

Here, we assume the Strongly-Interacting Light Higgs (SILH) scenario~\cite{Liu:2016idz}, where the low-energy effects of this new strongly interacting sector, quantified in terms of the UV coupling $g_*$ and mass $m_*$, are parametrised in terms of dimension-six operators in the so-called SILH basis:
\bea
\noindent
&&\frac{c_{6,\varphi,y_{f}}}{\Lambda^2} \propto \frac{g_*^2}{m_*^2}, 
\qquad \frac{c_{W,B}}{\Lambda^2}\propto \frac{1}{m_*^2},
\qquad \frac{c_{2W,2B,2G}}{\Lambda^2} \propto \frac{1}{g_*^2 m_*^2},
\qquad \frac{c_{T}}{\Lambda^2}\propto \frac{y_t^4}{16\pi^2 m_*^2}, \\
&&
\qquad \frac{c_{\varphi W,\varphi B}}{\Lambda^2}\propto \frac{g_{*}^2}{16 \pi^2 m_*^2},\qquad \frac{c_{3W,3G}}{\Lambda^2} \sim \frac{1}{16 \pi^2 m_*^2}, 
\qquad \frac{c_{\gamma, g}}{\Lambda^2}\propto \frac{y_{t}^{2}}{16 \pi^2 m_*^2} \, ,
\eea
where the proportionality factors are model-dependent dimensionless factors typically assumed to be of $\mcO\left(1\right)$.
Without loss of generality, we assume in the following that all such proportionality factors are unity.
Using the known conversion relations between the SILH and the Warsaw basis~\cite{Falkowski:2001958}, we can derive the matching relations between the Wilson coefficients of the Warsaw basis used in {\sc\small SMEFiT} and the UV model parameters $g_*$ and $m_*$, listed for completeness in Table~\ref{tab:matching_CH_SILH}.

\begin{table}[t]
    \centering
\begin{tabularx}{\linewidth}{|X|c|}
\toprule
         Wilson coefficients & SILH matching relation \\
         \midrule
         $c_{\varphi G}$ & $\frac{y_{t}^2\,g_{3}^2}{16\pi^2}\frac{1}{m_{*}^2}$\\
         $c_{\varphi B}$ & $\frac{g_1^2}{64\pi^2}\frac{(4 y_{t}^2-g_*^2)}{m_*^2}$ \\
         $c_{\varphi W}$ & $-\frac{g_2^2}{64\pi^2}\frac{g_*^2}{m_*^2}$ \\
         $c_{\varphi WB}$ & $-\frac{g_1\,g_2}{32\pi^2}\frac{g_*^2}{m_*^2}$\\
         $c_{\varphi\Box}$ & $ \begin{array}{c}
              -\frac{g_*^2}{2\,m_*^2} - \frac{g_1^4\,(1+3\cot^4_\text{w})}{8\,g_*^2\,m_*^2}+\frac{g_1^2}{4}\frac{(1+3\cot^2_\text{w})}{m_*^2} \\
               + \frac{g_1^2\,\left( 1+3\cot_\text{w}^2 \right) }{64\pi^2}\frac{g_*^2}{m_*^2}-\frac{y_t^4}{32\pi^2\,m_{*}^2}
         \end{array} $ \\
         $c_{\varphi D}$ & $ -\frac{g_1^4}{2\,g_*^2\,m_*^2} + \frac{g_1^2}{m_*^2}\left( 1 + \frac{g_*^2}{16\pi^2}\right)-\frac{y_t^4}{8\pi^2\,m_*^2}$ \\
         $c_{WWW}$ & $-\frac{g_2^3}{16\pi^2\,m_*^2}$ \\
         $c_{\varphi q,\,ii}^{(-)} = c_{\varphi Q,\,ii}^{(-)}$ & $ \left( 1-3\cot_{\text{w}}^2 \right) \frac{ g_1^2 \, \left( g_{*}^2 + 16\pi^2 \right) }{ 192 \pi^2  m_*^2 } -
         \left( 1 - 3 \cot_{\text{w}}^4 \right)\frac{g_1^4}{12\,g_*^2\,m_*^2} $ \\
         $c_{\varphi q,\,ii}^{(3)}=c_{\varphi Q,\,ii}^{(3)}$ & $\frac{ g_2^2 }{ 64\pi^2 }\frac{ g_*^2 }{ m_*^2 } -\frac{g_2^4}{4\,g_*^2 m_*^2} + \frac{g_2^2}{4\,m_*^2} $ \\
         $c_{\varphi\ell,ii}^{(1)}$ & $ -\frac{g_1^2}{64\pi^2}\frac{g_*^2}{m_*^2}+\frac{g_1^4}{4\,g_*^2\,m_*^2}-\frac{g_1^2}{4\,m_*^2} $ \\
         $c_{\varphi\ell,ii}^{(3)}$ & $ \frac{g_2^2}{64\pi^2}\frac{g_{*}^2}{m_{*}^2} - \frac{g_2^4}{4}\frac{1}{g_{*}^2\,m_{*}^2} +\frac{g_2^2}{4m_*^2}$ \\ 
         $c_{\varphi u,ii} = c_{\varphi t}$ & $ \frac{g_1^2}{48\pi^2}\frac{g_*^2}{m_*^2} -\frac{g_1^4}{3\,g_*^2\,m_*^2} +\frac{g_1^2}{3 m_{*}^{2}}$ \\
         $c_{\varphi d,ii}$ & $ -\frac{g_1^2}{96\pi^2}\frac{g_*^2}{m_*^2} +\frac{g_1^4}{6\,g_*^2\,m_*^2} -\frac{g_1^2}{6 m_{*}^{2}}$ \\
         $c_{\varphi e}=c_{\varphi \mu}=c_{\varphi \tau}$ & $ -\frac{g_1^2}{32\pi^2}\frac{g_*^2}{m_*^2} + \frac{g_1^4}{2\,g_*^2\,m_*^2} - \frac{g_1^2 }{ 2\,m_{*}^{2} }$ \\
         $c_{\ell\ell,1221}$& $ -\frac{g_2^4}{2\,g_{*}^2m_*^2} $ \\
         $c_{t\varphi}$ & $y_{t}\,\left((1+\frac{g_2^2}{32\pi^2})\frac{g_*^2}{m_*^2}-\frac{g_2^4}{2\,g_*^2\,m_*^2}+\frac{g_2^2}{2\,m_*^2} \right) $ \\
         $c_{b\varphi}$ & $y_{b}\,\left((1+\frac{g_2^2}{32\pi^2})\frac{g_*^2}{m_*^2}-\frac{g_2^4}{2\,g_*^2\,m_*^2}+\frac{g_2^2}{2\,m_*^2} \right) $ \\
         $c_{\tau\varphi}$ & $y_{\tau}\,\left((1+\frac{g_2^2}{32\pi^2})\frac{g_*^2}{m_*^2}-\frac{g_2^4}{2\,g_*^2\,m_*^2}+\frac{g_2^2}{2\,m_*^2} \right) $ \\
         $c_\varphi$ & $\lambda\left( (-1+\frac{g_2^2}{8\pi^2})\frac{g_*^2}{m_*^2} -\frac{g_2^4}{g_*^2\,m_*^2} + \frac{2\,g_2^2}{m_*^2} \right) $ \\
         $c_{\ell\ell,1122} $ & $ \frac{g_2^4-g_1^4}{4\,g_*^2\,m_*^2} $ \\
         \bottomrule
    \end{tabularx}
    \vspace{0.2cm}
    \caption{The matching relations between the Wilson coefficients in the Warsaw basis and the UV parameters  $g_*$ and $m_*$ of the SILH scenario. Here, we have defined $\cot_{\text{w}}=\cot(\theta_W)$ and $\tan_{\text{w}}=\tan(\theta_W)$; $g_1$, $g_2$, and $g_3$ are the U$(1)_Y$, SU$(2)_L$, and SU$(3)_c$ SM gauge couplings respectively; $y_t$, $y_b$, and $y_\tau$ are the SM top, bottom and tau Yukawas, and $\lambda$ is the SM Higgs quartic coupling. 
    }
    \label{tab:matching_CH_SILH}
\end{table}

\begin{figure}[t]
    \centering
    \includegraphics[width=0.90\linewidth]{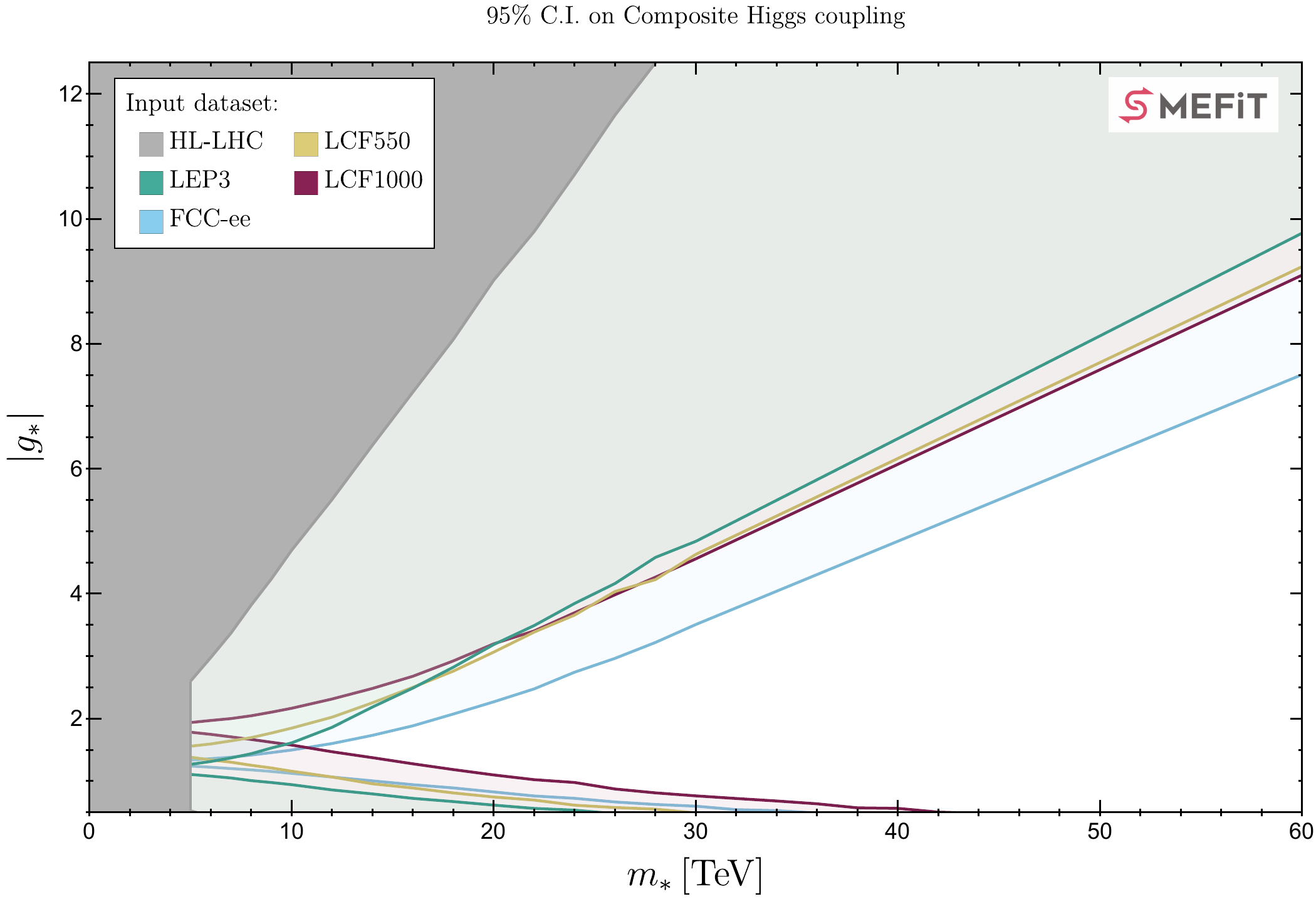}
    \caption{The projected $95\%$ C. I. (H.D.I.) excluded regions (shaded) on the coupling $g_*$ as a function of the mass parameter $m_*$ of the SILH scenario, obtained from the global SMEFT fit with theoretical predictions following the matching relations summarised in  Table~\ref{tab:matching_CH_SILH}.
    The fit is performed at quadratic order in the EFT expansion, includes one-loop RGE effects, and assumes the `aggressive' scenario for theoretical uncertainties.
    As in Fig.~\ref{fig:mass_reach_future_colliders}, we start from the common HL-LHC baseline dataset and then add to it the FCC-ee, LEP3, LCF550, or LCF1000 projections.
    The vertical line at $5$~TeV indicates the minimal mass for which we performed the fit, see text for details.
}
\label{fig:composite_higgs_future_colliders}
\end{figure}

In Fig.~\ref{fig:composite_higgs_future_colliders}, we present the $95\%$ C.I. bounds on the coupling $g_*$ as a function of the mass $m_*$ in the SILH scenario, obtained from a fit to $g_*$ at fixed values of $m_*$ after imposing the matching relations in Table~\ref{tab:matching_CH_SILH}. 
The input datasets are the same as in global SMEFT fits of Sect.~\ref{sec:global_smeft_fits_results}, we assumed the aggressive theory uncertainty scenario, and we performed the RG running from the value of $m_*$ down to the scale of each observable.
As in the case of the analysis in Fig.~\ref{fig:mass_reach_future_colliders}, we start from the common HL-LHC baseline and then add either LEP3, FCC-ee, LCF550 or LCF1000 to it.

The two WCs that drive the bounds are $c_{t\varphi}$ and $c_{\varphi \Box}$, with the former being generated with a coefficient a factor $\sim 2$ bigger than the latter.
However, as seen in Fig.~\ref{fig:quad_ind_global}, all colliders give similar bounds on $c_{t\varphi}$. 
Thus, the bounds on $g_*$ for masses $m_*\gtrsim 20$~TeV emulate the pattern seen for individual bounds on $c_{\varphi \Box}$: any future leptonic collider improves massively on HL-LHC. 
FCC-ee offers the best sensitivity, with LEP3 a $25-30$\% worse. 
The two LCF variants do similarly well and slightly better than LEP3, but $\sim20\%$ worse than FCC-ee. 
At lower masses, LEP3 is not at a disadvantage any more and roughly matches the bounds of LCF.
We note that our bounds are stronger than the ones found in the ESPPU26 PBB~\cite{deBlas:2025gyz} due to differences in the fitting methodology of the UV parameters between the two approaches, in particular our choice of fitting the coupling $g_*$ at fixed values of the mass $m_*$.

Finally, we comment on the lower bounds observed for $|g_*|$ and $m_*$. 
The lower bounds on $|g_*|$ are caused by the non-decoupling effects manifested via contributions $c\propto 1/g_*^2$, in particular to $c_{\ell\ell,1122}$ and $c_{\ell\ell,1221}$. 
The CH scenario has a decoupling limit in only a part of its parameter space, when $f\sim m_*/g_*\to\infty$ such that $\xi\equiv v^2/f^2\to0$ with fixed $v$~\cite{Giudice:2007fh,Panico:2015jxa}.
Thus, at low $m_*$, one could have important non-decoupling effects and the SMEFT description in such region is not recommended, hence making these lower bounds not trustworthy.
Moreover, the narrow wedge of allowed $g_*$ below $m_*\lesssim10$~TeV arises as an interplay between the lower and upper bounds, hence its position changes with the sensitivity to $c_{\ell\ell,1221}$ and $c_{\ell\ell,1122}$.
The lower bound of $m_*\geqslant5$~TeV reflects our choice of a minimal scale to ensure the decoupling regime. 
Our choice has a two-fold justification: first, the presence of datasets in the fit involving momentum transfers in the range $3-4$~TeV (such as top quark pair-production); and second, the HL-LHC mass reach found in Sect.~\ref{sec:mass_reach} for typical resonances in a CHM such as $U$, $D$, $Q_1$, $\mathcal{B}$, and $\mathcal{W}$.
\section{Summary and outlook}
\label{sec:summary}

We have presented a comparative study of the indirect NP reach of future colliders by identifying potential deviations in the interactions of SM particles.
Our results have been obtained in three complementary interpretation frameworks, namely the $\kappa$-framework, the global SMEFT fit, and the effective coupling formalism.
We also considered the matching of the SMEFT to representative UV-complete models, in order to directly derive bounds on the masses and couplings that define the UV model parameter space.
Our study complements the analyses carried out in the ESPPU26 Physics Briefing Book~\cite{deBlas:2025gyz} in a number of aspects, including, for instance,  systematisation of the scans in the UV model space and by quantifying the impact of quadratic corrections in the EFT expansion.

Several interesting findings can be highlighted from our studies.
As already found in previous analyses, the FCC-ee offers unprecedented indirect sensitivity to NP from the interplay between the Tera-$Z$ and the 240 GeV runs, but only if theoretical and experimental systematic errors can be kept under control.
We find LEP3 to provide a competitive programme of precision Higgs and electroweak measurements, but also with major limitations such as the lack of sensitivity to the Higgs self-coupling mostly due to the absence of the $t\bar{t}$ run.
The LCF physics reach is degraded by the lower luminosity of the $Z$-pole run, partially compensated by the higher energy runs which also provide direct access to the Higgs self-coupling and the top Yukawa.

The results of this work could be extended in several complementary directions.
First, the projections used in our analysis assume that the synthetic data follow the SM, while in general model sensitivity is different should the data be described by a BSM scenario. 
It would therefore be interesting to revisit our studies assuming that synthetic data is distributed according to specific BSM benchmarks, and quantify the reach of future colliders to discovery (rather than merely excluding) NP scenarios.
This way, assuming a BSM anomaly is identified at a future electron-positron collider, one can determine the likelihood of the associated particles being directly observed in high-energy colliders such as the FCC-hh or the MuCol.
Likewise, one can identify scenarios leading to  an anomaly at the FCC-ee which may escape direct detection at such colliders.

Secondly, we have carried out SMEFT interpretations only for electron-positron colliders.
Lepton-hadron~\cite{Bissolotti:2023vdw}, proton-proton~\cite{Ethier:2021bye,deBlas:2025xhe}, and muon colliders~\cite{Forslund:2023reu} also offer outstanding prospects to indirectly constrain BSM via effective interactions.
A natural follow up study would hence consider SMEFT projections at the LHeC, MuCol, or FCC-hh, following the same pipeline as done for the lepton colliders.
For instance, the LHeC is especially powerful to constrain two-lepton-two-quark interactions~\cite{Bissolotti:2023vdw}, including those of the first generation for which there is no sensitivity at electron-positron colliders. 
Likewise, high-energy tails in Higgs, top quark pair, Drell-Yan, or diboson production at the FCC-hh are expected to provide indirect NP sensitivity at scales much higher than the kinematic limit ($\sim 50$\,TeV) of direct production.

As a third possible application, recent years have witnessed an explosion of precision higher-order calculations in the SMEFT, from matrix elements and decay rates to RGEs, whose implementation can potentially increase the sensitivity  of future colliders to NP.
As illustrated by the calculation of NLO electroweak corrections to $e^+e^- \to ZH$ in the SMEFT~\cite{Asteriadis:2024xts,Asteriadis:2024xuk}, for a given measurement one can significantly increase the sensitivity to the parameters of interest, in this case the Higgs self-coupling~\cite{terHoeve:2025omu}, through loop corrections.
Extending {\sc\small SMEFiT} with more of these state-of-the-art calculations is hence another avenue to further scrutinize the reach of future facilities to identify NP signatures. 

While the particle physics community moves towards a  momentous decision for the next-generation particle collider, the availability of frameworks that streamline the evaluation of the physics reach of the various proposed options is of utmost importance.
This work contributes to this ongoing effort by releasing a flexible toolbox that facilitates the comparison of future facilities and enables many physics studies related to the possibilities to discover (or exclude) NP scenarios.

\begin{center}
\rule{0.6\linewidth}{0.4pt}
\end{center}

\noindent
The results presented in this work have been obtained with and can be reproduced by means of the open-source {\sc\small SMEFiT} framework:
\begin{center}
\url{https://smefit.science/}
\end{center}
whose installation instructions can be found in
\begin{center}
\url{https://github.com/LHCfitNikhef/smefit}
\end{center}
The experimental data, including projections for future colliders, and the corresponding SM and SMEFT theory calculations used are provided in a separate repository
\begin{center}
\url{https://github.com/LHCfitNikhef/smefit_database}
\end{center}
which also contains example runcards, which the user can adapt to tailor them to their needs.

\subsection*{Acknowledgements}

We warmly thank Jorge de Blas for many fruitful discussions and for assistance with the {\sc\small HepFit} benchmarking.
We are grateful to Christophe Grojean, Matthew McCullough, Hannes Mildner, Victor Miralles, Michele Selvaggi and Giuseppe Ventura for discussions and feedback on some of the topics studied in this work.
T.A. is a Research Fellow of the Fonds de la Recherche Scientifique – FNRS.
S.~T. is supported by a FRIA Grant of the Belgian Fund for Research, F.R.S.-FNRS (Fonds de la Recherche Scientifique-FNRS).
The work of A.~N.~R. is supported by the University of Padua under the 2023 STARS@UniPD Grants Programme (Acronym and title of the project: HiggsPairs -  Precise Theoretical Predictions for Higgs pair production at the LHC), by the Istituto Nazionale di Fisica Nucleare (INFN) through its Iniziativa Specifiche APINE and RD-FCC, and by the Italian MUR via the Departments of Excellence grant
2023- 2027 ``Quantum Frontiers''.
A.~N.~R. and S.~T. are grateful to the CERN Theory Department for its hospitality and support during the completion of this work.
The work of J.t.H. is supported by the UK Science and Technology Facility Council (STFC) consolidated grant ST/X000494/1. 
L.M. acknowledges support from the European Union under the MSCA fellowship (Grant agreement N. 101149078) {\it Advancing global SMEFT fits in the LHC precision era (EFT4ward)}.
The work of E.~C. and E.~V. is supported by the European Research Council (ERC) under the European Union’s Horizon 2020 research and innovation programme (Grant agreement No. 949451).
We acknowledge support from the COMETA COST Action CA22130.

\appendix
\section{SMEFT operator basis}
\label{app:operator_basis}

In this appendix we summarise the definitions of the SMEFT operator basis used in {\sc \small SMEFiT}.
For each operator, we indicate its definition in terms of the SM fields, and the conventions that are used both for the operator and for the coefficient. 
For the theoretical predictions, we adopt the Warsaw basis of dimension-six operators~\cite{Grzadkowski:2010es} assuming as flavour symmetry
\be
U(2)_{q_L}\otimes U(2)_{u_R}\otimes U(3)_{d_R}\otimes (U(1)_{\ell}\otimes U(1)_e)^3 \, ,
\ee
subsequently restricted to $U(2)_{q_L}\otimes U(2)_{u_R}\otimes U(3)_{d_R}\otimes U(3)_{\ell}\otimes U(3)_e$ for the fits presented in Sect.~\ref{sec:global_smeft_fits_results}.
\begin{itemize}
    \item Table~\ref{tab:oper_bos} lists the purely bosonic dimension-six SMEFT operators entering our basis.
        These operators modify the production and decay of Higgs bosons and the interactions of the gluons and electroweak gauge bosons.
        See~\cite{Ethier:2021bye} for more details\footnote{The subtraction of $\frac{v^2}{2}$ in the operators $\Op{\varphi G}$, $\Op{\varphi B}$ and $\Op{\varphi W}$ was not included in the corresponding table of previous {\sc\small SMEFiT} articles due to a typo, but was always accounted for in the predictions.}.

\item Table~\ref{tab:oper_ferm_bos} lists   the operators containing two fermion fields, either quarks or leptons.
  The two-fermion operators are classified in three groups: those involving 3rd generation quarks, those involving 1st and 2nd generation quarks, and those involving two leptons. 
  For the latter (two lepton operators), the flavour index $j$ runs from 1 to 3 following our flavour assumptions. 
  Note that our basis includes the two-fermion operators modifying the Yukawa couplings of the top, bottom, and charm quarks ($c_{t\varphi}$, $c_{b\varphi}$, $c_{c\varphi}$) and of the tau and muon leptons ($c_{\tau\varphi}$, $c_{\mu\varphi}$).
  In Table~\ref{tab:oper_ferm_bos}, the coefficients indicated with (*) do not correspond to physical degrees of freedom
  in the fit, but are rather replaced by  $c_{\varphi q}^{(-)}$, $c_{\varphi Q}^{(-)}$, and
  $c_{tZ}$.

\item Table~\ref{tab:oper_fourfermion} collects the definition of the four-fermion coefficients that enter in the predictions in terms of the coefficients of Warsaw basis operators.
These four fermion operators, involving either four-heavy quarks (4H), two-light-two-heavy quarks (2L2H), two-leptons-two-heavy quarks (2$\ell$2Q) or two-leptons-two-light quarks (2$\ell$2q), are defined as
\begin{align}
	\qq{1}{qq}{ijkl}
	&= (\bar q_i \gamma^\mu q_j)(\bar q_k\gamma_\mu q_l)
	 \nonumber
	,\\
	\qq{3}{qq}{ijkl}
	&= (\bar q_i \gamma^\mu \tau^I q_j)(\bar q_k\gamma_\mu \tau^I q_l)
     \nonumber
	,\\
	\qq{1}{qu}{ijkl}
	&= (\bar q_i \gamma^\mu q_j)(\bar u_k\gamma_\mu u_l)
     \nonumber
	,\\
	\qq{8}{qu}{ijkl}
	&= (\bar q_i \gamma^\mu T^A q_j)(\bar u_k\gamma_\mu T^A u_l)
         \nonumber
	,\\
	\qq{1}{qd}{ijkl}
	&= (\bar q_i \gamma^\mu q_j)(\bar d_k\gamma_\mu d_l)
	,\\
	\qq{8}{qd}{ijkl}
	&= (\bar q_i \gamma^\mu T^A q_j)(\bar d_k\gamma_\mu T^A d_l)
        \label{eq:4f_def}
        \nonumber
	,\\
	\qq{}{uu}{ijkl}
	&=(\bar u_i\gamma^\mu u_j)(\bar u_k\gamma_\mu u_l)
         \nonumber
	,\\
	\qq{1}{ud}{ijkl}
	&=(\bar u_i\gamma^\mu u_j)(\bar d_k\gamma_\mu d_l)
         \nonumber
	,\\
	\qq{8}{ud}{ijkl}
	&=(\bar u_i\gamma^\mu T^A u_j)(\bar d_k\gamma_\mu T^A d_l) \, .
         \nonumber 
\end{align}
%
%
Moreover we include the four-lepton (4$\ell$) operators
\begin{align}
\nonumber
    \qq{}{\ell \ell}{jklm}
    &= \left(\bar \ell_j\gamma_\mu \ell_k\right) \left(\bar \ell_l\gamma^\mu \ell_m\right) 
    ,\\
    \qq{}{\ell e}{jklm}
    &= \left(\bar \ell_j\gamma_\mu \ell_k\right) \left(\bar e_l\gamma^\mu e_m\right)
    ,\\
    \qq{}{e e}{jklm}
    &= \left(\bar e_j\gamma_\mu e_k\right) \left(\bar e_l\gamma^\mu e_m\right)
    \, .
         \nonumber
\end{align}
The flavour index $i$ is either 1 or 2, and $j$ is either 1, 2 or 3.

\end{itemize}

\begin{table}[t] 
  \begin{center}
    \renewcommand{\arraystretch}{1.6}
        \begin{tabular}{lll|lll}
          \toprule
          Operator $\quad$ & Coefficient $\quad$ & Definition& Operator $\quad$ & Coefficient $\quad$ & Definition \\
        \midrule
        \midrule
        $\Op{\varphi G}$ & $c_{\varphi G}$  & $\left( \pdp - \frac{v^2}{2} \right)G^{\mu\nu}_{\sss A}\,
        G_{\mu\nu}^{\sss A}$ 
        & 
        $\Op{\varphi \square}$ & $c_{\varphi \square}$ & $(\pdp)\Box(\pdp)$ \\
        $\Op{\varphi B}$ & $c_{\varphi B}$ & $\left(\pdp - \frac{v^2}{2} \right)B^{\mu\nu}\,B_{\mu\nu}$
        &
        $\Op{\varphi D}$ & $c_{\varphi D}$ & $(\varphi^\dagger D^\mu\varphi)^\dagger(\varphi^\dagger D_\mu\varphi)$ \\ 
        $\Op{\varphi W}$ &$c_{\varphi W}$ & $\left(\pdp - \frac{v^2}{2} \right)W^{\mu\nu}_{\sss I}\,
        W_{\mu\nu}^{\sss I}$ 
        &
        $\mathcal{O}_{W}$&   $c_{WWW}$ & $\epsilon_{IJK}W_{\mu\nu}^I W^{J,\nu\rho} W^{K,\mu}_\rho$ \\ 
        $\Op{\varphi W B}$ &$c_{\varphi W B}$ & $(\varphi^\dagger \tau_{\sss I}\varphi)\,B^{\mu\nu}W_{\mu\nu}^{\sss I}\,$ 
        & 
        $\mathcal{O}_{\varphi}$&   $c_{\varphi}$ & $(\pdp -\frac{v^2}{2})^3$ \\
       \bottomrule
        \end{tabular}
        \vspace{0.2cm}
        \caption{The purely bosonic dimension-six SMEFT operators entering our basis.
        These operators modify the production and decay of Higgs bosons and the interactions of the electroweak gauge bosons.
    For each operator, we indicate its definition in terms of the SM
          fields,
          and the conventions that are used
          both for the operator and for the coefficient. 
          See~\cite{Ethier:2021bye} for more details.
          \label{tab:oper_bos}}
\end{center}
\end{table}

\begin{table}[htbp]
  \begin{center}
    \renewcommand{\arraystretch}{1.45}
    \begin{tabular}{p{1.5cm} p{1.4cm} p{4.4cm} | p{1.5cm} p{1.4cm} p{4.5cm}}
      \toprule
      Operator & Coefficient & $\qquad$ Definition & Operator  & Coefficient & $\qquad$ Definition \\
                \midrule \midrule
      \multicolumn{6}{c}{3rd generation quarks} \\
                \midrule \midrule
    $\Op{\varphi Q}^{(1)}$ & $c_{\varphi Q}^{(1)}$~(*) & $i\big(\varphi^\dagger\lra{D}_\mu\,\varphi\big)
 \big(\bar{Q}\,\gamma^\mu\,Q\big)$ 
 &
 $\Op{tW}$ & $c_{tW}$ & $i\big(\bar{Q}\tau^{\mu\nu}\,\tau_{\sss I}\,t\big)\,
 \tilde{\varphi}\,W^I_{\mu\nu}
 + \text{h.c.}$ \\ 
    $\Op{\varphi Q}^{(3)}$ & $c_{\varphi Q}^{(3)}$  & $i\big(\varphi^\dagger\lra{D}_\mu\,\tau_{\sss I}\varphi\big)
 \big(\bar{Q}\,\gamma^\mu\,\tau^{\sss I}Q\big)$ 
 &
 $\Op{tB}$ & $c_{tB}$~(*) &
 $i\big(\bar{Q}\tau^{\mu\nu}\,t\big)
 \,\tilde{\varphi}\,B_{\mu\nu}
 + \text{h.c.}$\\ 
 & $c_{\varphi Q}^{(-)}$ & $c_{\varphi Q}^{(1)}-c_{\varphi Q}^{(3)}$ 
 & 
 & $c_{tZ}$ & $-s_\theta \, c_{tB}+ c_\theta \,c_{tW}$ \\
    $\Op{\varphi t}$ & $c_{\varphi t}$& $i\big(\varphi^\dagger\,\lra{D}_\mu\,\,\varphi\big)
 \big(\bar{t}\,\gamma^\mu\,t\big)$
 &
  $\Op{t G}$ & $c_{tG}$ & $ig{\sss S}\,\big(\bar{Q}\tau^{\mu\nu}\,T_{\sss A}\,t\big)\,
 \tilde{\varphi}\,G^A_{\mu\nu}
 + \text{h.c.}$ \\ 
     $\Op{t \varphi}$ & $c_{t\varphi}$ & $\left(\pdp\right)
 \bar{Q}\,t\,\tilde{\varphi} + \text{h.c.}$ 
 &
  $\Op{b \varphi}$ & $c_{b\varphi}$ & $\left(\pdp\right)
 \bar{Q}\,b\,\varphi + \text{h.c.}$ \\  
                \midrule \midrule
                \multicolumn{6}{c}{1st, 2nd generation quarks} \\
                \midrule \midrule
    $\Op{\varphi q}^{(1)}$ & $c_{\varphi q}^{(1)}$~(*) & $\sum\limits_{\sss i=1,2} i\big(\varphi^\dagger\lra{D}_\mu\,\varphi\big)
 \big(\bar{q}_i\,\gamma^\mu\,q_i\big)$ 
 &
 ${\Op{\varphi u }}$ &
      ${{c_{\varphi u}}}$ & $\sum\limits_{\sss i=1,2} i\big(\varphi^\dagger\,\lra{D}_\mu\,\,\varphi\big)
 \big(\bar{u}_i\,\gamma^\mu\,u_i\big)$ \\ 
    $\Op{\varphi q}^{(3)}$ & $c_{\varphi q}^{(3)}$ & $\sum\limits_{\sss i=1,2} i\big(\varphi^\dagger\lra{D}_\mu\,\tau_{\sss I}\varphi\big)
 \big(\bar{q}_i\,\gamma^\mu\,\tau^{\sss I}q_i\big)$
 & ${\Op{\varphi d }}$ &
      ${{c_{\varphi d}}}$ & $\sum\limits_{\sss j=1,2,3} i\big(\varphi^\dagger\,\lra{D}_\mu\,\,\varphi\big)
 \big(\bar{d}_j\,\gamma^\mu\,d_j\big)$ \\ 
 & $c_{\varphi q}^{(-)}$ & $c_{\varphi q}^{(1)}-c_{\varphi q}^{(3)}$
 &
 $\Op{c \varphi}$ & $c_{c \varphi}$ & $\left(\pdp\right)
 \bar{q}_2\,c\,\tilde\varphi + \text{h.c.}$ \\ 
                \midrule \midrule
		      \multicolumn{6}{c}{two-leptons} \\
                \midrule \midrule
    $\Op{\varphi \ell_j}$ & $c_{\varphi \ell_j}$ & $ i\big(\varphi^\dagger\lra{D}_\mu\,\varphi\big)
   \big(\bar{\ell}_j\,\gamma^\mu\,\ell_j\big)$ 
   &
   $\Op{\varphi e}$ & $c_{\varphi e}$ & $ i\big(\varphi^\dagger\lra{D}_\mu\,\varphi\big)
 \big(\bar{e}\,\gamma^\mu\,e\big)$  \\
    $\Op{\varphi \ell_j}^{(3)}$ & $c_{\varphi \ell_j}^{(3)}$ & $ i\big(\varphi^\dagger\lra{D}_\mu\,\tau_{\sss I}\varphi\big)
 \big(\bar{\ell}_j\,\gamma^\mu\,\tau^{\sss I}\ell_j\big)$ 
 &
  $\Op{\varphi \mu}$ & $c_{\varphi \mu}$ & $ i\big(\varphi^\dagger\lra{D}_\mu\,\varphi\big)
 \big(\bar{\mu}\,\gamma^\mu\, \mu\big)$ \\  
    $\Op{\mu \varphi}$ & $c_{\mu \varphi}$ & $\left(\pdp\right)
 \bar{\ell_2}\,\mu\,{\varphi} + \text{h.c.}$ 
 & $\Op{\varphi \tau}$ & $c_{\varphi \tau}$ & $ i\big(\varphi^\dagger\lra{D}_\mu\,\varphi\big)
 \big(\bar{\tau}\,\gamma^\mu\,\tau\big)$  \\
  $\Op{\tau \varphi}$ & $c_{\tau \varphi}$ & $\left(\pdp\right)
 \bar{\ell_3}\,\tau\,{\varphi} + \text{h.c.}$ & & & \\
  \bottomrule
\end{tabular}
\vspace{0.2cm}
\caption{Same as Table~\ref{tab:oper_bos}
  for the operators containing two fermion fields, either quarks or leptons.
  Thy are classified in three groups: those involving 3rd generation quarks, those involving 1st and 2nd generation quarks, and those involving two leptons. 
  For the latter (two-lepton operators), the flavour index $j$ runs from 1 to 3 following our flavour assumptions. 
  The coefficients indicated with (*) do not correspond to physical degrees of freedom
  in the fit, but are rather replaced by  $c_{\varphi q}^{(-)}$, $c_{\varphi Q}^{(-)}$, and
  $c_{tZ}$.
\label{tab:oper_ferm_bos}}
\end{center}
\end{table}

\begin{table}[htbp] 
  \begin{center}
    \renewcommand{\arraystretch}{1.53}
        \begin{tabular}{ll| ll}
          \toprule
          DoF $\qquad$ &  Definition (in  Warsaw basis notation) & DoF $\qquad$ &  Definition (in  Warsaw basis notation) \\
          \midrule
          \midrule
          \multicolumn{4}{c}{4-heavy-quark operators} \\
          \midrule
          \midrule
      $c_{QQ}^1$    &   $2\ccc{1}{qq}{3333}-\frac{2}{3}\ccc{3}{qq}{3333}$ 
      &
      $c_{QQ}^8$       &         $8\ccc{3}{qq}{3333}$\\  
     $c_{Qt}^1$         &         $\ccc{1}{qu}{3333}$
     &
     $c_{Qt}^8$         &         $\ccc{8}{qu}{3333}$\\   
     $c_{tt}^1$         &         $\ccc{1}{uu}{3333}$ && \\
            \midrule  
            \midrule
            \multicolumn{4}{c}{2-light-2-heavy quark operators} \\
          \midrule
          \midrule
  $c_{Qq}^{1,8}$       &  	 $\ccc{1}{qq}{i33i}+3\ccc{3}{qq}{i33i}$  
  &
  $c_{Qq}^{1,1}$         &   $\ccc{1}{qq}{ii33}+\frac{1}{6}\ccc{1}{qq}{i33i}+\frac{1}{2}\ccc{3}{qq}{i33i} $   \\    
   $c_{Qq}^{3,8}$         &   $\ccc{1}{qq}{i33i}-\ccc{3}{qq}{i33i} $  
   &
$c_{Qq}^{3,1}$          & 	$\ccc{3}{qq}{ii33}+\frac{1}{6}(\ccc{1}{qq}{i33i}-\ccc{3}{qq}{i33i}) $   \\     
$c_{tq}^{8}$         &  $ \ccc{8}{qu}{ii33}   $ 
  &
$c_{tq}^{1}$       &   $  \ccc{1}{qu}{ii33} $\\   
$c_{tu}^{8}$      &   $2\ccc{}{uu}{i33i}$ 
 &
$c_{tu}^{1}$        &   $ \ccc{}{uu}{ii33} +\frac{1}{3} \ccc{}{uu}{i33i} $ \\   
$c_{Qu}^{8}$         &  $  \ccc{8}{qu}{33ii}$
 &
 $c_{Qu}^{1}$     &  $  \ccc{1}{qu}{33ii}$  \\    
 $c_{td}^{8}$        &   $\ccc{8}{ud}{33jj}$ 
  &
 $c_{td}^{1}$          &  $ \ccc{1}{ud}{33jj}$ \\    
 $c_{Qd}^{8}$        &   $ \ccc{8}{qd}{33jj}$ 
 &
 $c_{Qd}^{1}$         &   $ \ccc{1}{qd}{33jj}$\\
 
             \midrule   
             \midrule  
              \multicolumn{4}{c}{2-heavy-quark-2-lepton operators} \\
          \midrule
          \midrule
  $c_{Q\ell_1}^{(-)}$       &  	 $\ccc{1}{\ell q}{1133}-\ccc{3}{\ell q}{1133}$  
  &
  $c_{Q\ell_1}^{(3)}$         &   $\ccc{3}{\ell q}{1133}$   \\    
  $c_{Q\ell_3}^{(-)}$       &  	 $\ccc{1}{\ell q}{3333}-\ccc{3}{\ell q}{3333}$  
  &
  $c_{Q\ell_3}^{(3)}$         &   $\ccc{3}{\ell q}{3333}$   \\  
  $c_{Q e}$       &  	 $\ccc{}{qe}{3311}$  
  & $c_{Q \tau}$       &  	 $\ccc{}{qe}{3333}$ \\ 
  $c_{t e}$       &  	 $\ccc{}{eu}{1133}$  & $c_{t \tau}$       &  	 $\ccc{}{eu}{3333}$   \\
  $c_{t\ell_j}$         &   $\ccc{}{\ell u}{jj33} $  & \\        
  \midrule     
  \midrule  
  \multicolumn{4}{c}{2-light-quark-2-lepton operators} \\
  \midrule
  \midrule
  $c_{q\ell_1}^{(-)}$       &  	 $\ccc{1}{\ell q}{11ii}-\ccc{3}{\ell q}{11ii}$  
  &
  $c_{q\ell_1}^{(3)}$         &   $\ccc{3}{\ell q}{11ii}$   \\    
  $c_{q e}$       &  	 $\ccc{}{qe}{ii11}$   & & \\
  $c_{\ell_1 u}$         &   $\ccc{}{\ell u}{11ii} $   &  
  $c_{\ell_1 d}$         &   $\ccc{}{\ell d}{11jj} $   \\    
  $c_{e u}$         &   $\ccc{}{e u}{11ii} $   &  
  $c_{e d}$         &   $\ccc{}{e d}{11jj} $   \\ 
  \bottomrule
  \end{tabular}
  \vspace{0.2cm}
  \caption{\small Definition of the four-fermion coefficients that enter in
    the fit in terms of the coefficients of Warsaw basis operators of ~Eq.~(\ref{eq:4f_def}) 
    These coefficients are classified into four-heavy-quark, and two-light-two-heavy-quark, 2-heavy-quark-2-lepton, and 2-light-quark-2-lepton operators.
    The flavour index $i$ is either 1 or 2, 
    and $j$ is either 1, 2 or 3: with the flavour assumptions adopted in this work,  these coefficients will be the same
    regardless of the specific values that $i$ and $j$ take.
\label{tab:oper_fourfermion}}
  \end{center}
\end{table}

In the global fits presented in Sect.~\ref{sec:global_smeft_fits_results} we impose a $U(3)_{\ell}\times U(3)_e$ symmetry in the lepton sector. In total, the number of independent degrees of freedom is $n_{\rm op}=61$.
The flavour universal coefficients definitions are derived by identifying all lepton flavour indices.
The four-lepton operator $\qq{}{\ell \ell}{jklm}$, is split into two invariants, such that~\cite{Brivio:2020onw}
\begin{align}
    c_{\ell \ell}^{jklm} = c_{\ell \ell}\delta^{jk}\delta^{lm} + c_{\ell \ell}^{\prime}\delta^{jm}\delta^{lk} \, .
\end{align}
The coefficients used in the fit are classified into the different groups of operators as specified in Table~\ref{tab:operators_summary}.
For each operator class, we indicate the number of members and the degrees of freedom considered in the SMEFT global fit, see Tables~\ref{tab:oper_bos}--\ref{tab:oper_fourfermion} for the corresponding definitions.
%

\begin{table}[htbp] 
  \begin{center}
    \renewcommand{\arraystretch}{1.60}
        \begin{tabularx}{\textwidth}{|X|c|c|c|}
        \toprule
        Operator class & Subclass & $n_{\rm op}$ & Operators \\
        \midrule
        Purely bosonic (B)  &  & 8 & 
        $c_{\varphi G}$,
        $c_{\varphi B}$,
        $c_{\varphi \square}$, 
        $c_{\varphi D}$,
        $c_{WWW}$,
        $c_{\varphi}$,
        $c_{\varphi W}$,
        $c_{\varphi W B}$\\
        \midrule
\multirow{3}{*}{Two-fermion (2FB)} & 3rd gen quarks & 8 &
$c_{\varphi Q}^{(1)}$, 
$c_{tW}$,
$c_{\varphi Q}^{(3)}$,
$c_{tB}$,
$c_{\varphi t}$,
$c_{tG}$,
$c_{t\varphi}$,
$c_{b\varphi}$\\
\cline{2-4}
& 1st, 2nd gen quarks & 5 &
$c_{\varphi q}^{(1)}$,
$c_{\varphi q}^{(3)}$,
$c_{\varphi d}$,
$c_{\varphi u}$,
$c_{c \varphi}$\\
\cline{2-4}
& two-lepton (2$\ell$) &  5 & 
$c_{\varphi \ell}$,
$c_{\varphi \ell}^{(3)}$,
$c_{\varphi e}$, 
$c_{\mu \varphi}$,
$c_{\tau \varphi}$ \\
\midrule
\multirow{6}{*}{Four-fermion (4F)} & 4H  & 5 & $c_{QQ}^1$,
$c_{QQ}^8$,
$c_{Qt}^1$,
$c_{Qt}^8$,
$c_{tt}^1$ \\
\cline{2-4}
& \multirow{2}{*}{2L2H}  & 
\multirow{2}{*}{14} &
$c_{Qq}^{1,8}$,
$c_{Qq}^{1,1}$,
$c_{Qq}^{3,8}$,
$c_{Qq}^{3,1}$,
$c_{tq}^{8}$,
$c_{tq}^{1}$,
$c_{tu}^{8}$,\\
& & & 
$c_{tu}^{1}$,
$c_{Qu}^{8}$,
$c_{Qu}^{1}$,
$c_{td}^{8}$,
$c_{td}^{1}$,
$c_{Qd}^{8}$,
$c_{Qd}^{1}$\\
\cline{2-4}
& 2$\ell$2Q & 5 & 
$c_{Q\ell}^{(-)}$,
$c_{Q\ell}^{(3)}$,
$c_{Q e}$,
$c_{t\ell}$,
$c_{t e}$ \\
\cline{2-4}
& 2$\ell$2q & 7 & 
$c_{q\ell}^{(-)}$,
$c_{q\ell}^{(3)}$,
$c_{q e}$,
$c_{\ell u}$,
$c_{\ell d}$,
$c_{e u}$,
$c_{e d}$\\
\cline{2-4}
& 4$\ell$ & 4 & 
$c_{\ell\ell}$,
$c_{\ell\ell}^{\prime}$,
$c_{\ell e}$,
$c_{e e}$\\ 
\midrule
        Total  & &  61 & -- \\
        \bottomrule
        \end{tabularx}
        \vspace{0.2cm}
  \caption{\small Overview of Wilson coefficients associated to the dimension-six SMEFT operators that compose our fitting basis.
  For each operator class, we indicate the number of members and the degrees of freedom considered in the SMEFT global fit.
  See Tables~\ref{tab:oper_bos}--\ref{tab:oper_fourfermion} for the corresponding definitions.
Here `H' and `Q' stand for heavy quark fields, $L$ and $q$ for light quark fields, and $\ell$ for leptonic fields.
The index $i$ for the 2-lepton operators runs from 1 to 3 in our flavour assumptions.
\label{tab:operators_summary}}
        \end{center}    
        \end{table}
\section{Derivation of the effective coupling dictionary}
\label{app:effective-derivation}

In this appendix we derive the relations between the effective couplings presented in Sect.~\ref{sec:effective_couplings} and the Warsaw basis operators used in our predictions. 
We follow the dictionary in~\cite{Barklow:2017awn}, adapted to our choices of electroweak input scheme and of SMEFT operator basis.
We start by deriving the definitions of the effective couplings of weak vector bosons to fermions in Eq.~(\ref{eq:eff_couplings_2}). 
Next, we present the Higgs couplings definitions of Eq.~(\ref{eq:eff_couplings_1}), and finally the definitions of the aTGCs of Eq.~(\ref{eq:eff_couplings_3}).

\paragraph{Electroweak couplings.}
Following the notation in~\cite{Barklow:2017awn}, we define $\delta A = \Delta A/A$ in terms of 
the shift $\Delta A$ induced by the SMEFT operators of a given observable $A$ relative to the SM value.\footnote{The relative variation with respect to the measured value or the SM value is actually the same at the first order in $\delta A$, and since we are not interested in accounting for $\Lambda^{-4}$ corrections, how the relative shift is defined is not relevant.}
To connect with the treatment of~\cite{Barklow:2017awn}, we need to relate the different operator basis chosen. 
First, in the Higgs sector, we exchange the operator $\mathcal{O}_{\varphi {\tiny\Dlr}}$ used in~\cite{Barklow:2017awn} by a combination of  $\mathcal{O}_{\varphi\Box}$ and $\mathcal{O}_{\varphi D}$ which are part of the {\sc\small SMEFiT} basis.
Applying the prescription in~\cite{Helset:2020yio} to our $\mathcal{O}_{\varphi\Box}$ definitions we obtain
\begin{equation}
\mathcal{O}_{\varphi D} = (\phi^\dagger D_\mu \phi)^\dagger (\phi^\dagger D_\mu \phi)= \frac{1}{4}\left[\partial_\mu(\varphi^\dagger\varphi)\partial^\mu(\varphi^\dagger\varphi)-(\varphi^\dagger\Dlr^{\,\mu}\varphi)(\varphi^\dagger\Dlr_\mu\varphi)\right] \, ,
\end{equation}
from which we can derive
\begin{align}
c_H &= 2 \left(c_{\varphi\Box}+\frac{1}{4}c_{\varphi D}\right) \voverL\, , \\
c_T &= -\frac{1}{2} c_{\varphi D} \voverL \, ,
\end{align}
connecting the $c_T,c_H$ coefficients used in~\cite{Barklow:2017awn} with those in our fitting basis.

Concerning the operators built upon Higgs fields and  field strength tensors, in the {\sc\small SMEFiT} basis these are defined by subtracting the Higgs vacuum expectation value, namely 
\begin{equation}
    (\varphi^\dagger \varphi) V^a_{\mu\nu}V^{a,\mu\nu}\rightarrow \left(\varphi^\dagger \varphi-\frac{v^2}{2}\right) V^a_{\mu\nu}V^{a,\mu\nu} \, ,
\end{equation}

for $V\in \{G,B,W\}$, while the $\mathcal{O}_{\varphi WB}$ operator is not modified. 
As a consequence of this definition, the vector boson kinetic term normalisation differs from the one used in~\cite{Barklow:2017awn}.
Furthermore, we also take into account the different normalisation of the $\mathcal{O}_{\varphi W B}$ operator, and the fact that in~\cite{Barklow:2017awn} the generator of SU(2) $t^a = \frac{\sigma^a}{2}$ instead of the Pauli matrices $\sigma_a$ appear in the operator definition.
With these considerations, we find the following relations:
\begin{align}
\delta Z_W &= 0  \, ,\\
\delta Z_A &= -2\sw\cw \cpwb \voverL \, , \\
\delta Z_Z &= 2\sw\cw \cpwb \voverL \, , \\
\delta Z_{ZA} &=(\sw^2-\cw^2)\cpwb \voverL \, , \\
\delta Z_H &= -c_H \, ,
\end{align}
where the last one follows from the absence of $\mathcal{O}_{\varphi}$ in the definition of the effective coupling formalism used in this work.

The same operators also affect the Higgs contact interactions with vector bosons, namely operators of the form $HV^{\mu\nu}V_{\mu\nu}$, introducing further modifications between our basis and that used in~\cite{Barklow:2017awn}, which read
\begin{align}
\zeta_W &= 2\cpw\voverL \, ,\\
\zeta_A &= 2\left(\cw^2\cpb-\sw\cw\cpwb+\sw^2\cpw\right)\voverL \, ,\\
\zeta_Z&= 2\left(\sw^2\cpb+\sw\cw \cpwb+\cw^2\cpw \right)\voverL \, ,\\
\zeta_{ZA}&= \left(-2\sw\cw\cpb+(\sw^2-\cw^2) \cpwb+2\sw\cw\cpw \right)\voverL \, .
\end{align}
Finally, the operators modifying the Yukawa interactions, $\mathcal{O}_{q\varphi} $ and $\mathcal{O}_{\ell\varphi}$, are also defined subtracting the vev contribution.
For this reason, in our basis the Yukawa shifts are independent from fermion masses shifts, which are exactly vanishing, $\delta m_f = 0$. 

The effective coupling dictionary can be written in a scheme-independent way, leaving implicit  the mass and EW-coupling shifts~\cite{Barklow:2017awn}.
Here, to simplify the expressions and to facilitate benchmark comparisons, we present the dictionary in the adopted  $\{m_Z,m_W,G_F\}$ electroweak input scheme.
This means that we solve the input scheme equations subject to the conditions
\be
\label{eq:Gfcondition}
\delta G_F=\delta m_W =\delta m_Z=0 \, .
\ee
and additionally we set $\delta m_H = 0$ since in this formalism the Higgs self-coupling is not modified. 
Note that the SMEFT Lagrangian induces a direct shift~\cite{Brivio:2020onw,Ellis:2023zim} in the Fermi constant,
\begin{equation}
 \Delta G_{F} = \frac{1}{\sqrt{2}}  \left(c^{(3)}_{\varphi\ell_1}+c^{(3)}_{\varphi\ell_2}-c_{\ell\ell}\right) \frac{1}{\Lambda^2}\label{eq:GF_shift} \, ,
\end{equation}
and an additional term to Eq.~\eqref{eq:GF_shift} can be included considering the indirect contribution  coming from the Higgs boson vev shift, resulting in
\begin{equation}
\delta G_F =  - 2  \delta v + \sqrt{2} v^2  \Delta G_F \, . 
\end{equation}
In Eq.~\eqref{eq:Gfcondition}, the $\delta A$ are  always understood to  include both the direct and the indirect contributions as done in~\cite{Barklow:2017awn}.

Solving the input parameter relations, Eq.~\eqref{eq:Gfcondition}, we find the following results, 
\begin{align}
\label{eq:deltavev}
\delta v &=  \frac{1}{2}\left(c^{(3)}_{\varphi\ell_1}+c^{(3)}_{\varphi\ell_2}-c_{\ell\ell}\right)\voverL \, , \\
\delta g &= - \delta v \, , \\
\delta g' &= -\delta v +\frac{1}{2\sw^2}c_T - \frac{\cw} {\sw} \cpwb \voverL \, ,
\end{align}
from which we obtain
\begin{equation}
\delta e = \sw^2\delta g + \cw^2 \delta g' + \frac{1}{2} \delta Z_A = - \delta v + \frac{1}{2}\frac{\cw^2}{\sw^2}c_T-\frac{\cw}{\sw}\cpwb\voverL \, .
\end{equation}
The $Z$-boson coupling to fermions shifts can be computed considering both the direct contributions from the SMEFT operators, the indirect ones coming from the EW coupling modifications, and the kinetic term normalisations, resulting in the following effective couplings in terms of the Warsaw basis coefficients:
\begin{align}
\label{eq:eff_couplings_gRf}
\delta g_{fR} &= -\cw^2\delta g + (1+\cw^2) \delta g' +\left( \frac{\delta Z_Z}{2} - \frac{\cw}{\sw} \delta Z_{AZ}\right)+\left[\frac{1}{2 \sw^2Q_f} c_{\varphi f} \right] \voverL\, , \\
\label{eq:eff_couplings_gLf}
\delta g_{fL} &= \frac{1}{\Tf-Q_f \sw^2}\left[\cw^2 (\Tf+Q_f \sw^2) \delta g +\sw^2 \left(\Tf-Q_f(1+\cw^2)\right) \delta g'+\left(\Tf c^{(3)}_{\varphi F} -\frac{1}{2}c_{\varphi F}\right)\voverL\right]+\nonumber\\
+&\frac{Q_f\sw\cw }{\Tf-Q_f \sw^2} \delta Z_{AZ}+\frac{\delta Z_Z}{2} \, ,
\end{align}
where  $F={\ell_i,q_i}$ stands for the SU(2) doublet containing the particles $f={e,\mu,\tau,u,d,s,c,b,t}$.
Furthermore,  instead of including in the analysis the $Z$ boson coupling to neutrinos, we consider the $W\ell_i\nu_i$ coupling modification that can be directly extracted from the $W$ boson leptonic decay,
\begin{equation}
\label{eq:deltagWl}
\delta g_{W\ell_i} = \delta g +c^{(3)} _{\varphi\ell_i} \voverL \, .
\end{equation}

\paragraph{Higgs couplings.}
Moving to the Higgs sector, the Higgs-fermion coupling shifts take the form
\begin{equation}
\label{eq:gHff}
    \delta g_{Hff} = \frac{1}{2}\left(-c_H-2 \frac{c_{f\varphi}}{y_f}\voverL-2\delta v\right).
\end{equation}
The Higgs-gluon coupling incorporates both the contribution coming from the Higgs-Yukawa modifications and the SMEFT contact term coming from the $\mathcal{O}_{\varphi G}$ operator\cite{Brivio:2019myy}, leading to
\begin{equation}
\label{eq:gHgg}
\delta g_{Hgg} = -\frac{c_H}{2}-\delta v -\frac{\alpha_s^2 m_H^3}{72 \pi^3}\voverL\Re\left[\left(\sum_{q=c,b,t} \frac{m_q}{v} I_q\right)\left(\sum_{q=c,b,t} \frac{c_{q\varphi}}{\sqrt{2}} I_q-3\frac{4\pi}{\alpha_s}\frac{c_{\varphi G}}{v}\right)\right]\frac{1}{\Gamma(H\rightarrow gg)_{\rm SM}}
\end{equation}
with 
\begin{equation}
I_q\left(\tau=4\frac{m_q^2}{m_H^2}\right)=\begin{cases} 
        &\frac{3}{2}\frac{\tau}{m_q}\left[1+(1-\tau)\arcsin\left(\frac{1}{\sqrt{\tau}}\right)^2\right],\qquad \text{if } \tau\ge1\\
        &\frac{3}{2}\frac{\tau}{m_q}\left[1-\frac{1}{4}(1-\tau)\left(\log\left(\frac{1+\sqrt{1-\tau}}{1-\sqrt{1-\tau}}\right)-i\pi\right)^2\right], \text{if } \tau<1
\end{cases}
\end{equation}
and $\Gamma(H\rightarrow gg)_{\rm SM}=0.00033$ GeV is taken from \cite{LHCHiggsCrossSectionWorkingGroup:2016ypw}. 

Concerning the effective couplings for the loop induced decays $H\rightarrow \gamma\gamma$ and $H\rightarrow Z\gamma$, following~\cite{Barklow:2017awn} we do not include the contribution coming from the Yukawa-modifying operators, but we keep the indirect  and contact term contributions.
Taking into account \eqref{eq:Gfcondition}, one has
\begin{align}
\label{eq:gHaa}
\delta g_{H\gamma\gamma} &= 263 \zeta_A - \delta Z_A - \frac{c_H}{2} + 2 \delta e - 2 \delta v \, ,\\
\label{eq:gHZa}
\delta g_{HZ\gamma} &= 145 \zeta_{AZ} - \frac{c_H}{2} - (1-3 \sw^2)\delta g + 3 \cw^2 \delta g' + \frac{\delta Z_A}{2} + \frac{\delta Z_Z}{2} - \delta v \, ,
\end{align}

Finally, we need to consider the effective couplings of the Higgs boson to $W$ and $Z$ bosons, namely $\delta g_{HWW}$ and $\delta g_{HZZ}$.
These include the direct modifications to the $HVV$ vertex as well as, given their off-shell character, SMEFT effects on the vector boson decays.
Starting from the $W$ boson case,  its decay into SU(2) doublets is affected by the $\mathcal{O}^{(3)}_{\varphi F}$ operators with $F\in\{\ell_1,\ell_2,\ell_3,q_1,q_2\}$.  These SMEFT contributions can be grouped together into a unique coefficient
\begin{equation}
    C_W =\voverL \sum_F N_{c,F}c^{(3)}_{\varphi F} \Bigg/ \sum_F N_{c,F}= \voverL \frac{c^{(3)}_{\varphi \ell_1}+c^{(3)}_{\varphi \ell_2}+c^{(3)}_{\varphi \ell_3}+6c^{(3)}_{\varphi q}}{9} \, ,
\end{equation}
which  contributes to the $W$ boson width shift,
\begin{equation}
\delta \Gamma_W = 2\delta g + 2C_W\,.
\end{equation}
Since in our basis we do not distinguish the $(u,d)$ and $(c,s)$ doublets the quark contributions are grouped under a unique Wilson coefficient.
For the $Z$-boson couplings, we need to treat separately the left- and right-handed coupling modifications.
We collectively denote by $X$ the SM chiral flavours to which the $Z$ boson can decay\footnote{In the following derivation, the sums over $X$ must be understood as running both over different flavours as well as over different helicities $\{L,R\}$, each one with its own charge $Q_{Z,X}=\left(T^3_X-\sw^2Q_X\right).$}.
In the SMEFT Lagrangian, the modifications of the $Z$-boson couplings to chiral flavours are encapsulated in 
\begin{equation}
  \Delta \mathcal{L}_{\rm SMEFT} \supset \sum_X \frac{g}{c_W} c_{X} Z^\mu \bar X\gamma_\mu X \, ,
\end{equation}
in which the prefactor is chosen to match the SM one and where
\begin{alignat}{2}
c_{Ze_L/\mu_L/\tau_L} &= -\frac{1}{2}\voverL \left(c_{\varphi\ell_{1/2/3}}+c^{(3)}_{\varphi\ell_{1/2/3}}\right),\qquad && c_{Ze_R/\mu_R/\tau_R} = -\frac{1}{2}\voverL \left(c_{\varphi e/\mu/\tau}\right), \\ 
c_{Z\nu_{e,L}/\nu_{\mu,L}/\nu_{\tau,L}} &= -\frac{1}{2}\voverL \left(c_{\varphi\ell_{1/2/3}}-c^{(3)}_{\varphi\ell_{1/2/3}}\right),\qquad &&c_{Z\nu_{e,R}/\nu_{\mu,R}/\nu_{\tau,R}}=0, \\
c_{Zd_L/b_L} &=-\frac{1}{2}\voverL \left(c^{(1)}_{\varphi q/Q}+c^{(3)}_{\varphi q/Q}\right), \qquad &&c_{Zd_R} = -\frac{1}{2}\voverL  c_{\varphi d},\\
c_{Zu_L} &=-\frac{1}{2}\voverL \left(c^{(1)}_{\varphi q}-c^{(3)}_{\varphi q}\right), \qquad &&c_{Zu_R} = -\frac{1}{2}\voverL  c_{\varphi u}.  \\
\end{alignat}
These couplings modify $Z$ boson decays width via a unique modifier defined as
\begin{equation}
C_Z =  \frac{\sum_X c_X Q_{ZX}N_X}{\sum_X Q_{ZX}^2N_X},
\end{equation}
leading to the following shift in the $Z$-boson width
\begin{equation}
    \delta \Gamma_Z = 2\cw^2(1+2Q_r\sw^2)\delta g+2\sw^2(1-2Q_r\cw^2)\delta g'+\delta Z_Z+Q_r\sw\cw\delta Z_{AZ}+2C_Z  \, ,
\end{equation}
where we have defined 
\be
Q_r=\frac{\sum_X Q_{ZX}N_X}{\sum_X Q_{ZX}^2N_X} \, .
\ee
Combining the indirect and the direct contributions from dimension-six SMEFT operators to the effective Higgs couplings of the the $W$ and $Z$ vector bosons, we end up with
\begin{align}
\label{eq:gHww}
\delta g_{HWW} = -\frac{c_H}{2} - 2\delta v -0.375 \zeta_W- 0.44 C_W + 0.53 \delta\Gamma_W \, ,\\
\label{eq:gHZZ}
\delta g_{HZZ} = -\frac{c_H}{2}-\frac{c_T}{2}-\delta v-0.25 \zeta_Z-0.51C_Z+0.59\delta \Gamma_Z \, ,
\end{align}

\paragraph{Triple Gauge Couplings.}
Finally, for completeness we include in our study the anomalous Triple Gauge Couplings (aTGCs), which in the notation used here are given by
\begin{align}
\label{eq:atgc1}
    \delta \lambda_Z &= \frac{3}{2}g \,c_{WWW}\voverL \, ,\\
    \label{eq:atgc2}
    \delta g_{1,Z} &= (1+\sw^2)  \delta g - \sw^2 \delta g' + \frac{\delta Z_Z}{2}  + \frac{\sw}{\cw}\delta Z_{AZ} \, , \\
    \label{eq:atgc3}
    \delta \kappa_\gamma &= \delta e + \frac{\cw}{\sw} \cpwb \voverL \, .
\end{align}
\section{Benchmarking with HEPfit}
\label{app:kappa-validation}

In this appendix we present representative results of benchmark comparisons that we have carried out between the {\sc\small SMEFiT} and {\sc\small HEPfit} frameworks.
Specifically, we compare the outcome of analyses based on the two fitting codes, in turn, for the kappa framework, the effective coupling formalism, and the global SMEFT fit.
These are not, however, tuned benchmarks and some of the input settings may be different in the two sets of analyses.
Hence one should not expect full agreement between the two sets of results.

\subsection{Kappa framework}
\label{subsec:kappa_framework_bench}

\begin{figure}[t]
    \centering
\includegraphics[width=0.49\textwidth]{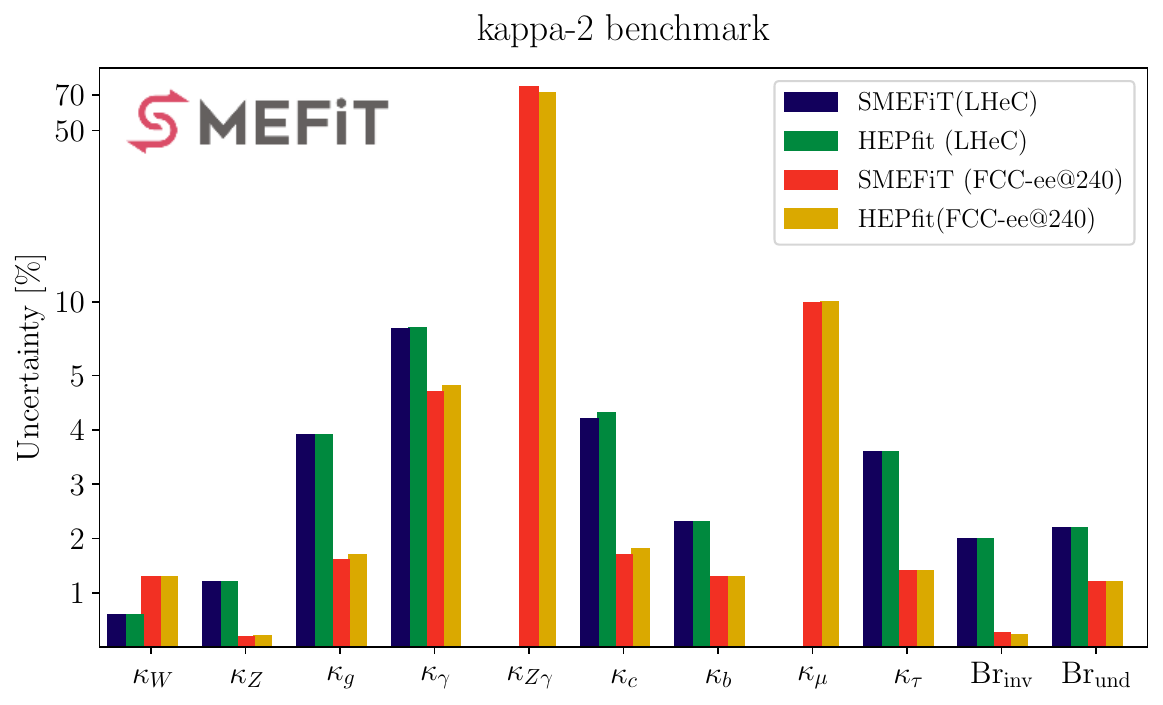}
\includegraphics[width=0.49\textwidth]{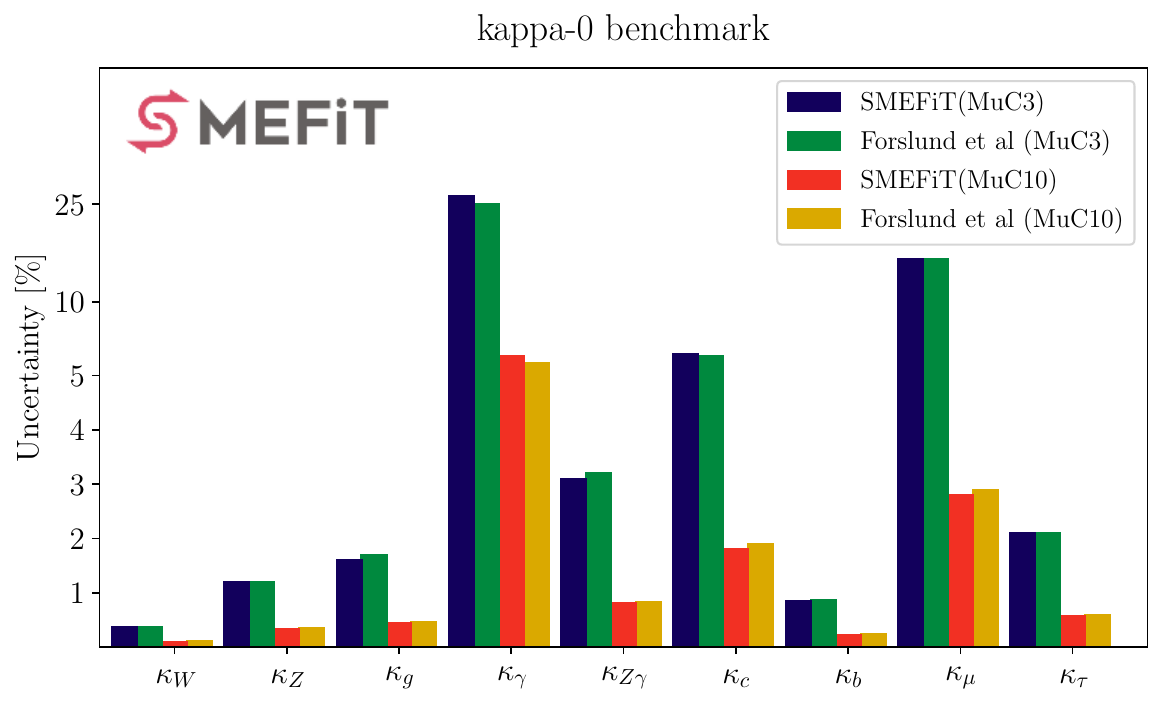}
\vspace{0.2cm}
\caption{Graphical representation of the results of Tables~\ref{tab:kappa-benchmark-1} (left) and~\ref{tab:kappa0_benchmark} (right panel), benchmarking our implementation of the kappa-0 and kappa-2 and frameworks in {\sc\small SMEFiT}. 
Note that the kappa-2 framework is defined in the same way as the kappa-3 fits of Sect.~\ref{subsec:kappa_formalism} but without the HL-LHC dataset as baseline.
\label{fig:kappa-HEPfit}
    }
\end{figure}

First, we present results of benchmark comparisons between {\sc\small SMEFiT} and {\sc\small HEPfit} carried out at the level of the modified couplings framework described in Sect.~\ref{subsec:kappa_formalism}.
Table~\ref{tab:kappa-benchmark-1} displays  the results of kappa-2 fits (defined in the same way as kappa-3 but without the HL-LHC baseline) between {\sc\small SMEFiT} and the {\sc\small HEPfit} study of~\cite{deBlas:2019rxi}. The {\sc\small SMEFiT} setup for the comparison is based on the same experimental inputs as in~\cite{deBlas:2019rxi}.
We consider for illustration two different colliders, the FCC-ee running at $\sqrt{s}=240$ GeV and the LHeC; similar quantitative agreement is obtained for the other colliders.
The uncertainties for the coupling modifiers are reported at the $1\sigma$ level with the exception of the modifiers indicated with $^{(*)}$ whose uncertainties are defined in Eq.~(\ref{eq:err_kv}).
For the branching ratios ${\rm Br}_{\rm inv,und}$ we report the upper  $95\%$ CL bounds.
For the LHeC we impose the constraint $\kappa_V\le 1$ to stabilise the fit.
Fig.~\ref{fig:kappa-HEPfit} displays the graphical representation of these results.
Excellent agreement between the results of the two codes is found.

\begin{table}[t]
\centering
\small
\renewcommand{\arraystretch}{1.3}
\begin{tabularx}{\textwidth}{|X|c|c|c|c|}
\toprule
\multirow{3}{*}{Coupling modifier } & \multicolumn{4}{c|}{$\delta \kappa~(\%)$}\\
 \cline{2-5}
&\multicolumn{2}{c|}{FCC-ee@240}&\multicolumn{2}{c|}{LHeC}\\
\cline{2-3} \cline{4-5}
&HEPfit& {\sc\small SMEFiT} & HEPfit& {\sc\small SMEFiT}\\ \hline
\multicolumn{1}{|l|}{$\kw$}&1.3&1.3&$0.6^*$&$0.6^*$\\ 
\multicolumn{1}{|l|}{$\kz$}&0.21&0.2&$1.2^*$&$1.2^*$\\
\multicolumn{1}{|l|}{$\kg$}&1.7&1.6&3.9&3.9\\ 
\multicolumn{1}{|l|}{$\ka$}&4.8&4.7&7.8&7.7\\ 
\multicolumn{1}{|l|}{$\kza$}&71&75&-&-\\
\multicolumn{1}{|l|}{$\kc$}&1.8&1.7&4.3&4.2\\ 
\multicolumn{1}{|l|}{$\kt$}&-&-&-&-\\ 
\multicolumn{1}{|l|}{$\kb$}&1.3&1.3&2.3&2.3\\ 
\multicolumn{1}{|l|}{$\kmu$}&10&9.9&-&-\\
\multicolumn{1}{|l|}{$\ktau$}&1.4&1.4&3.6&3.6\\ \hline
\multicolumn{1}{|l|}{$\Brinv$}&0.22&0.27&2.2&2\\ 
\multicolumn{1}{|l|}{$\Brund$}&1.2&1.2&2.2&2.2\\
\bottomrule
\end{tabularx}
\vspace{0.2cm}
\caption{Comparison of the results of kappa-2 fits obtained with \smefit\,and \hepfit~\cite{deBlas:2019rxi}, based in both cases on the same experimental inputs as in~\cite{deBlas:2019rxi}.
The kappa-2 framework is defined in the same manner as the kappa-3 one but removing the HL-LHC as baseline.
We consider two different colliders: the FCC-ee running at $\sqrt{s}=240$ GeV and the LHeC; similar agreement is obtained for other colliders.
The percentage uncertainties $\delta\kappa$ for the coupling modifiers are reported at the $1\sigma$ level with the exception of the modifiers indicated with $^*$ whose uncertainties are defined in Eq.~(\ref{eq:err_kv}).
For the branching ratios ${\rm Br}_{\rm inv}$ and${\rm Br}_{\rm  und}$ we report the upper $95\%$ CL bounds.
See Fig.~\ref{fig:kappa-HEPfit} (left panel) for the graphical representation of these results.
}
\label{tab:kappa-benchmark-1}
\end{table}

\begin{table}[t]
\centering
\small
\renewcommand{\arraystretch}{1.3}
\begin{tabularx}{\textwidth}{|X|c|c|c|c|}
\hline
\multirow{3}{*}{Coupling modifier} & 
\multicolumn{4}{c|}{$\delta \kappa~(\%)$} \\
\cline{2-5}
& \multicolumn{2}{c|}{MuCol @ 3 TeV} & \multicolumn{2}{c|}{MuCol @ 10 TeV} \\
 \cline{2-3}  \cline{4-5}
& Forslund {\it et. al.}~\cite{Forslund:2023reu} & {\sc\small SMEFiT}& Forslund {\it et. al.}~\cite{Forslund:2023reu} & {\sc\small SMEFiT} \\ 
\hline
\multicolumn{1}{|l|}{$\kw$} & 0.38& 0.38& 0.11& 0.1\\
\multicolumn{1}{|l|}{$\kz$} & 1.2& 1.2& 0.35& 0.34\\ 
\multicolumn{1}{|l|}{$\kg$} & 1.7& 1.6& 0.46& 0.45\\ 

\multicolumn{1}{|l|}{$\ka$} & 3.2& 3.1& 0.84& 0.81\\
\multicolumn{1}{|l|}{$\kza$} & 25& 27& 5.6& 6\\ 
\multicolumn{1}{|l|}{$\kc$} & 6.0& 6.1& 1.9& 1.8\\
\multicolumn{1}{|l|}{$\kb$} & 0.87& 0.85& 0.24& 0.23\\
\multicolumn{1}{|l|}{$\kmu$} & 15& 15& 2.9& 2.8\\
\multicolumn{1}{|l|}{$\ktau$} & 2.1& 2.1& 0.59& 0.58\\
\bottomrule
\end{tabularx}
\vspace{0.2cm}
\caption{Same as Table~\ref{tab:kappa-benchmark-1} now the kappa-0 framework at the MuCol at both $\sqrt{s}=3$ TeV and 10 TeV. 
The output of {\sc\small SMEFiT} is compared with that of~\cite{Forslund:2023reu} for the same data projections.
See Fig.~\ref{fig:kappa-HEPfit} for the graphical representation.
\label{tab:kappa0_benchmark}
}
\end{table}
The implementation of the kappa framework in {\sc\small SMEFiT} is further validated in the case of the kappa-0 scenario for the muon collider at $\sqrt{s}=3$ TeV and 10 TeV with the results presented in~\cite{Forslund:2023reu}, when the same projections for muon collider measurement are assumed as input. 
As was the case with the kappa-2 comparison with {\sc\small HEPfit}, excellent agreement with the outcome of~\cite{Forslund:2023reu} for the kappa-0 scenario is obtained. 

\subsection{Effective couplings}
\label{app:effective_coupling_benchmark}

\begin{figure}[htp]
    \centering
\includegraphics[width=0.75\textwidth]{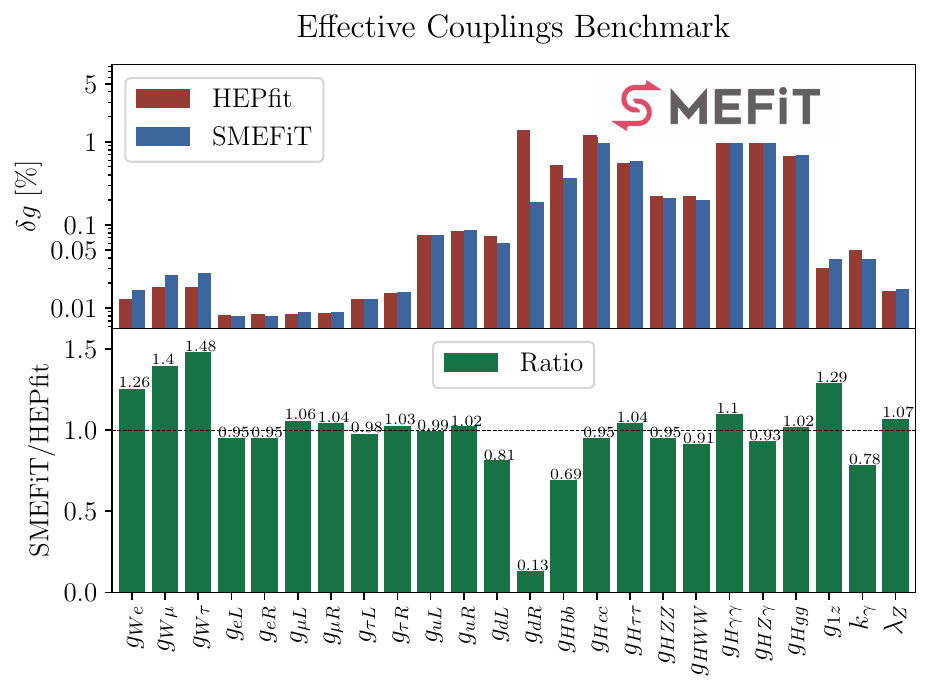}
\caption{Comparison between the results of Higgs and electroweak effective couplings fits to HL-LHC and FCC-ee projections obtained with {\sc\small SMEFiT} and {\sc\small HEPfit}, based in both cases on the Snowmass 2022 settings~\cite{deBlas:2022ofj}.
In particular, the (Snowmass 2022) theoretical uncertainties are taken into account and RGE effects are neglected.
The top panels display the 68\% C.I. relative uncertainties on each effective coupling, while the bottom panels display the ratio between the two sets of results (also indicated numerically).
See text for more details.
}
\label{fig:EffCoupl_comparison}
\end{figure}

Next, we compare the outcomes of {\sc\small SMEFiT} and {\sc\small HEPfit} at the level of the Higgs and electroweak effective couplings defined in Sect.~\ref{sec:effective_couplings}. The results are presented in Fig.~\ref{fig:EffCoupl_comparison}.
The {\sc\small HEPfit} results are taken from the Snowmass analysis of~\cite{deBlas:2022ofj}. For the purpose of this comparison, we adopt a dataset and theoretical settings of the SMEFT predictions which follow as much as possible the setup of~\cite{deBlas:2022ofj}.
Reasonable agreement between the two codes is found, yet some noticeable differences remain.
We can understand some of them as follows.

\begin{itemize}    
    \item  $g_{W\ell}$ couplings: the experimental inputs in the \smefit~ and \hepfit~ analyses being compared here differ because \smefit~ does not include the optimal observables for  $W^+W^-$ production in the hadronic final state at the FCC-ee and neglects the anti-correlations in the different leptonic channels highlighted in~\cite{DeBlas:2019qco}. 
    The missing datasets would further constraint the $\mathcal{O}^{(3)}_{\varphi\ell_1}$ and $\mathcal{O}^{(3)}_{\varphi\ell_2}$ operators entering in the definition of the $g_{W\ell}$ couplings.
    Furthermore, in the SMEFT there are no BSM decays of the $W$-boson, leading to the additional constraint ${\rm Br}(W\rightarrow jj)=1-\sum_i {\rm Br}(W\rightarrow \ell_i \nu_{\ell_{i}})$ connecting the $g_{W\ell}$ effective couplings to the hadronic $W$ decay.
    These considerations explain our worse sensitivity to the $g_{W\ell}$ couplings as compared to {\sc\small HEPfit}, as shown in Fig.~\ref{fig:EffCoupl_comparison}.
    
    \item  $g_{d_R}, g_{d_L}$: our stronger constraints are explained by the different flavour assumptions.
    In {\sc\small SMEFiT}, we do not distinguish the bottom quark from the lighter quarks in the right-singlet operators (that is, we do not have a $\mathcal{O}_{\varphi b}$ operator separate from $\mathcal{O}_{\varphi d}$ in our basis).
    In particular, the bottom-quark asymmetry at the $Z$-pole is in our case directly constraining $g_{d_R}$ and $g_{d_L}$, hence the tighter bound.
    We verified that removing the forward-backward asymmetry $A_{\rm FB}^{b\bar{b}}$ from the fit improves the agreement with \hepfit.
    
   \item   $g_{Hbb}$: In our definition of the Higgs width in the SMEFT, bottom-quark effects are included at NLO, leading to stronger bounds.

   \item  $g_{1Z}$: this effective coupling enters $W^+W^-$ production at the FCC-ee. Since, as explained above, we do not include some datasets relevant for this final state, in particular the optimal observables in $W^+W^-$ production in the hadronic final state, a worse bound is expected.
   
   \item  $k_\gamma$: this effective coupling is strongly dependent on the projected uncertainty for $\alpha_{\rm EW}$. One difference between the \hepfit\, and \smefit\, analyses lies in the choice of electroweak input scheme: \hepfit\, adopts the $\{G_F, m_Z, \alpha_{\rm EW}\}$ scheme, whereas \smefit\, uses the $\{G_F, m_Z, m_W\}$ one (known as the $m_W$ scheme). Hence differences in the $k_\gamma$ bound can be traced back to the different input schemes adopted in the two fits.
\end{itemize} 
All in all, we conclude that the level of agreement between \smefit\, and \hepfit\, is satisfactory for the effective coupling framework and that we can understand the origin of most of the observed differences.

\subsection{Global SMEFT fit}

\begin{figure}[t]
    \centering
\includegraphics[width=0.49\textwidth]{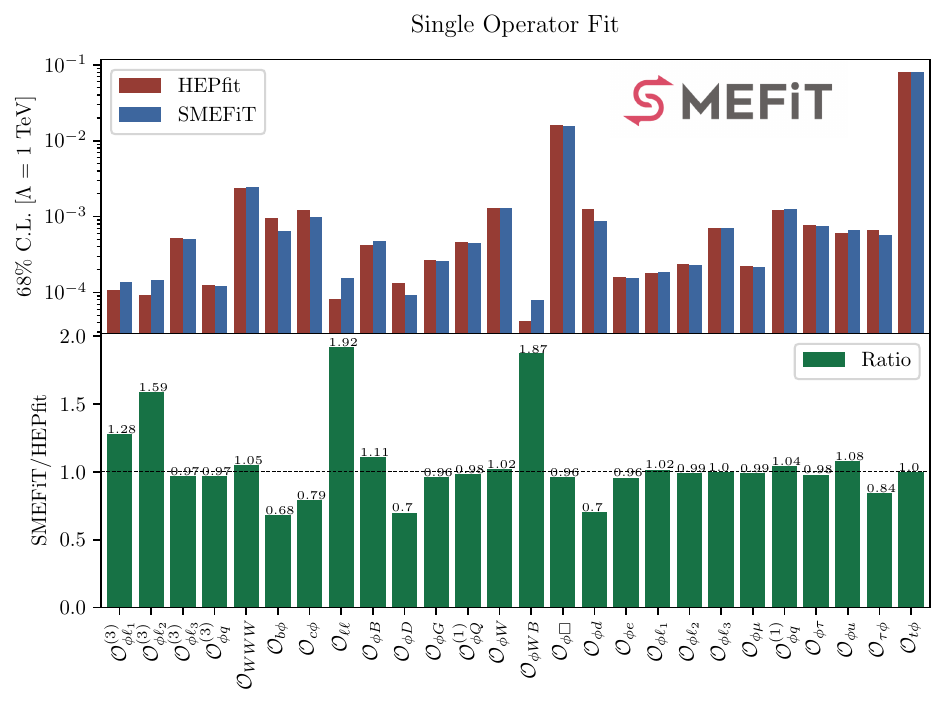}
\includegraphics[width=0.49\textwidth]{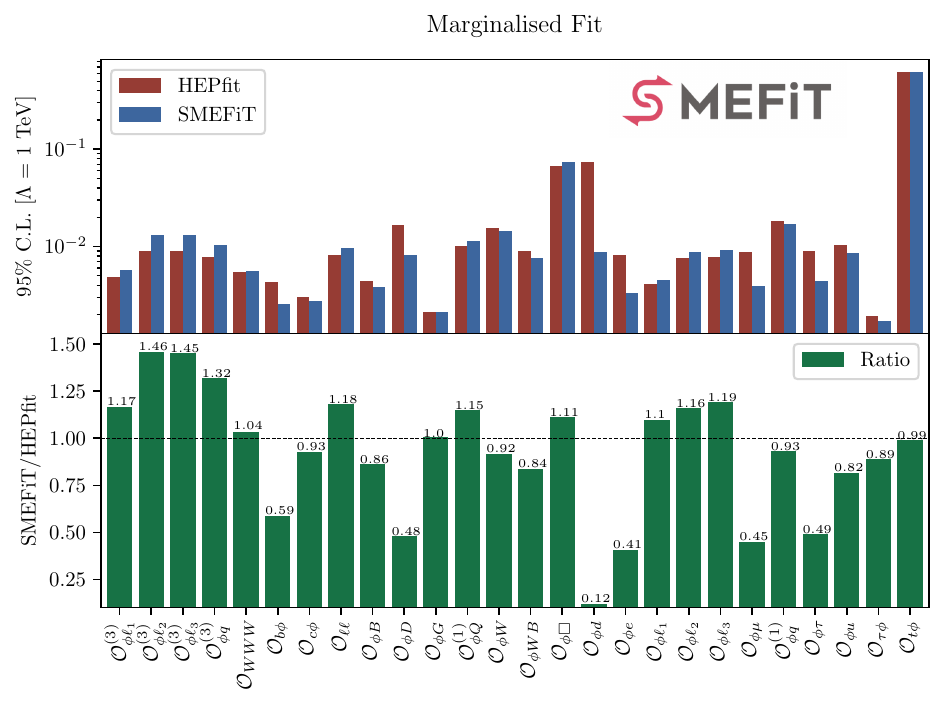}
\caption{Left: comparison between the results of single-parameter  SMEFT fits to HL-LHC and FCC-ee projections obtained with {\sc\small SMEFiT} and {\sc\small HEPfit} based on the Snowmass 2022 settings~\cite{deBlas:2022ofj}.
Fits are carried out at $\mathcal{O}\lp \Lambda^{-2}\rp$ level without accounting for RGE effects and neglecting theory uncertainties.
Right: same, now showing the marginalised bounds from a global fit including all the listed operators.
Here theory errors are instead taken into account in both fits.
The coupling $\mathcal{O}^{(1)}_{\varphi Q}$ stands for the sum of $\mathcal{O}^{(1)}_{\varphi Q}$ and $\mathcal{O}^{(3)}_{\varphi Q}$.
In the two sets of fits, the bottom panels display the ratio between the {\sc\small SMEFiT} and {\sc\small HEPfit} results (also indicated numerically). 
}.
\label{fig:Comparison_SMEFT}
\end{figure}

Fig.~\ref{fig:Comparison_SMEFT} presents a similar comparison as that of Fig.~\ref{fig:EffCoupl_comparison} now at the level of the global SMEFT fit, displaying the comparison between results based on the two frameworks both for one-parameter fits and for a global fit where all operators are simultaneously constrained.
The \hepfit\ results shown in this comparison are taken from~\cite{deBlas:2022ofj}. 
The input dataset consists of HL-LHC and FCC-ee projections (the latter corresponding to the version used in~\cite{deBlas:2022ofj}), and theory uncertainties are accounted for in the global marginalised fit but not in the individual ones.
At the time of Snowmass2022 projections
 \smefit\,  and \hepfit\, fits were realized in two different input schemes ($m_W$ versus $\alpha_{\rm EW}$ respectively).
Since the projected uncertainties provided in~\cite{deBlas:2022ofj} are available only for observables in the $\alpha_{\rm EW}$ scheme, we derive the corresponding errors in the $m_W$ scheme as follows. 
We rescale the  uncertainties in the input parameters in the $m_W$ scheme used in {\sc\small SMEFiT}~\cite{Corbett:2021eux} assuming the same relative improvement as that between the current and future  uncertainties in \cite{deBlas:2022ofj}, while for $\alpha_{\rm EW}$ we propagate the uncertainties for the input parameters ($m_W$ included) as projected to be in FCC-ee. This procedure is not formally rigorous; however, it suffices for the purpose of making an indicative comparison between the two fitting frameworks. We emphasize that in the ESPPU26 \cite{deBlas:2025gyz}, both the quantities and their uncertainties are defined within the $m_W$
 scheme, so this approximate approach was not required and instead we followed the rigorous error propagation method outlined in \ref{subsec:th_errors}.

Similarly to the effective coupling benchmark of Fig.~\ref{fig:EffCoupl_comparison}, also for the global SMEFT fit comparison of Fig.~\ref{fig:Comparison_SMEFT}  there is in general a reasonable agreement between the two codes (specially at the level of one-parameter fits), yet some noticeable differences remain.
Some differences in the single-parameter fits are observed concerning operators entering the input scheme parameter shifts, as expected and documented in~\cite{Brivio:2017bnu}. 
At the level of the global marginalised fit, with the exception of $\mathcal{O}_{\varphi d}$, the results of the two frameworks agree within a 50\% tolerance.
While pinpointing the origin of the observed differences requires a dedicated benchmark comparison, we note that  some of the largest observed discrepancies can be explained by similar considerations as those mentioned in the effective coupling benchmark of App.~\ref{app:effective_coupling_benchmark}.

\section{Comparison with the ESPPU26 PBB results}
\label{app:pbb-bench}

\begin{figure}[t]
    \centering\includegraphics[width=0.9\linewidth]{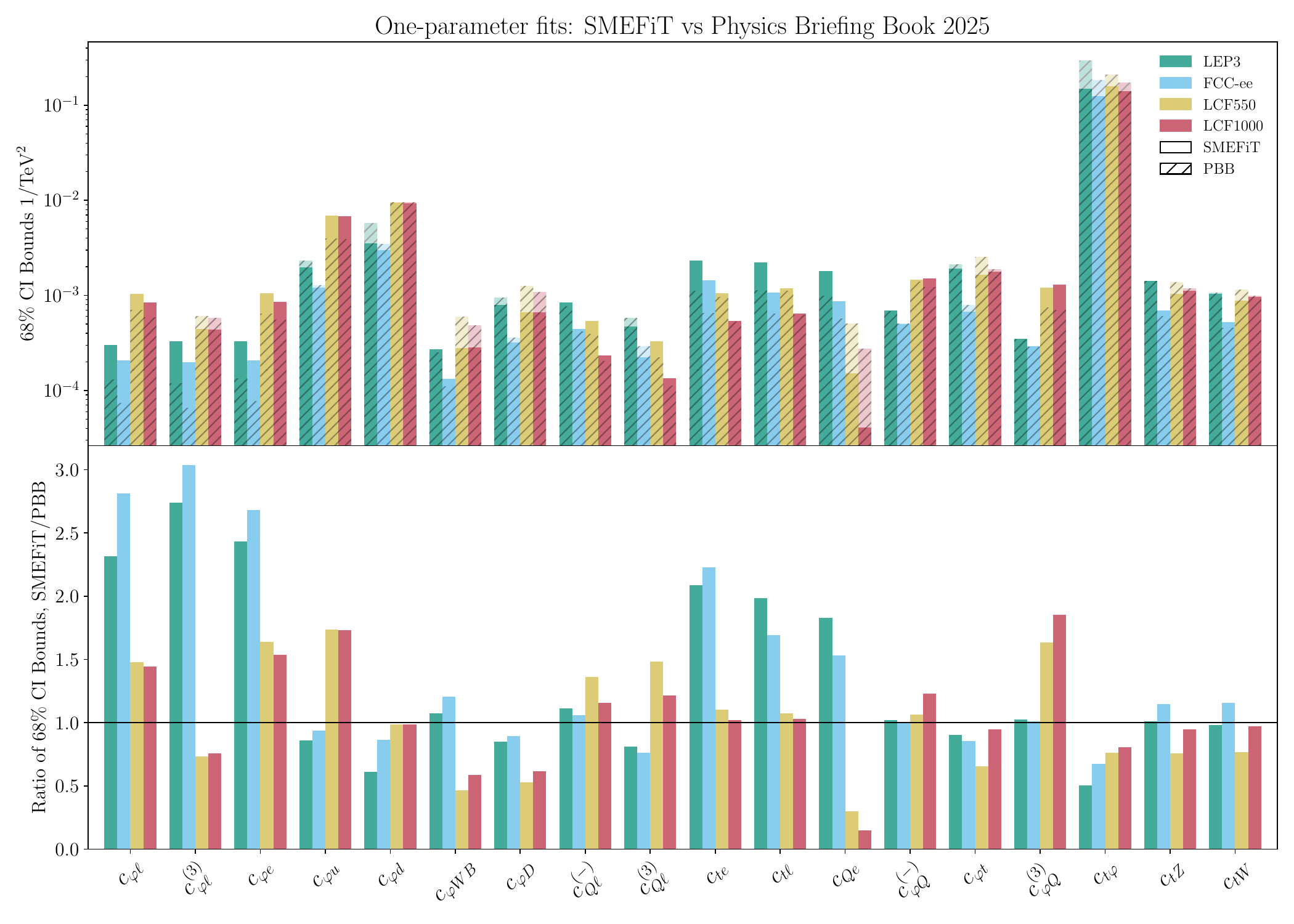}
    \caption{Comparison of representative one-parameter fits to future collider projections (FCC-ee, LEP3, and LCF550/1000) in {\sc\small SMEFiT} with the results presented in the ESPPU2026 PBB~\cite{deBlas:2025gyz}.
    For each operator, we display the 68\% CL bounds (top) and their ratio (bottom panel).
    In both cases, theoretical uncertainties are included in the aggressive scenario and parametric errors are considered.
    In the top panel, the {\sc\small SMEFiT} bounds are indicated with solid bars, while ESPPU2026 PBB one with hatched bars.
\label{fig:smefit_vs_pbb2025}
}
\end{figure}

Fig.~\ref{fig:smefit_vs_pbb2025} displays a comparison of representative one-parameter fits to different future colliders in {\sc\small SMEFiT} with the results presented in the ESPPU2026 PBB~\cite{deBlas:2025gyz}.\footnote{We are grateful to Jorge de Blas for sharing with us the numerical values of the results displayed in the ESPPU2026 PBB.}
For each operator, we display in the top panel the 68\% CL uncertainties obtained with {\sc\small SMEFiT}  and in the ESPPU2026 PBB for the FCC-ee, LEP3, LCF550, and LCF1000 datasets,
and in the bottom panel the corresponding ratios of bounds.
In both cases, EFT calculations at linear order with one-loop RGE effects are considered, theoretical uncertainties are included in the aggressive scenario, and parametric errors are taken into account.
While to the best of our knowledge the settings adopted in the two sets of results are comparable, it should be emphasised that this is not a tuned comparison.

From the comparison in Fig.~\ref{fig:smefit_vs_pbb2025}, general agreement is found between the results obtained within the two frameworks.
For many of the colliders and operators considered, the agreement between the frameworks is at the 10\%--20\% level.
All in all, the differences between {\sc\small SMEFiT} and ESPPU26 PBB are in the range
\begin{equation}
0.5 \lsim \frac{\delta c_{\rm smefit}}{\delta c_{\rm pbb}} \lsim 3 \, , 
\end{equation}
with the only exception being $c_{Qe}$ at the LCF.
For this specific operator, the datasets driving the constraints are the $b \bar b$ production cross section and asymmetry, and the differences at the fit results level originate from a difference in the projected uncertainty.

To fully understand the residual differences between the two sets of results, a dedicated tuned benchmarking comparison would be required, where identical experimental and theoretical inputs are adopted in two frameworks.
We leave this dedicated comparison to future work.
\section{Supplementary results}
\label{app:numerics_and_tables}

In this final appendix, we collect supplementary results (figures and tables) complementing those presented in the main text.
These additional results are presented, in turn,  for the kappa framework (App.~\ref{app:kappa-additional}), the effective couplings analysis (App.~\ref{app:effectivecouplings_supplementary}), and the global SMEFT fit (App.~\ref{app:extra_smeft}) formalism.

\subsection{Kappa framework}
\label{app:kappa-additional}

Table~\ref{tab:future_colliders_kappafits} summarises the information and references of the experimental inputs for the kappa framework analysis presented in this work (which are different from the corresponding inputs to the SMEFT fits listed in Table~\ref{tab:FutureCollider_runs}).
%

\begin{table}[h]
    \centering
    \footnotesize
    \renewcommand{\arraystretch}{1.5}
    \begin{tabularx}{\textwidth}{X|r|c|c|l}
    \toprule
       Collider  & Energy & $\mathcal{L}$ [ab$^{-1}$] & Ref.  & Comments\\ \hline
       \multirow{2}{*}{HL-LHC}  & \multirow{2}{*}{14 TeV} & \multirow{2}{*}{3} & \multirow{2}{*}{\cite{Cepeda:2019klc,TwikiCMS,CMS:2025hfp}} & Correlation matrix and error on signal theory from~\cite{TwikiCMS}.\\
       &&&&
       $\mu\mu$ and $Z\gamma$ signals rescaled to match the projections~\cite{CMS:2025hfp}.\\\hline
       LHeC &  1.3 TeV & 1 & \cite{LHeC:2020van}& Invisible branching ratio from \cite{LHeC:2020van} (Sect.~7.6).\\\hline
       \multirow { 4 }{*}{LCF}  & 250 GeV &2.7&\cite{LinearColliderVision:2025hlt}&-\\
       &350 GeV&0.135&\cite{LinearColliderVision:2025hlt} &-\\
       &500 GeV&6.4&\cite{LinearColliderVision:2025hlt}&Adding $t\bar tH$ inferred from $y_t$ determination\\
       &1 TeV&6.4&\cite{LinearColliderVision:2025hlt}&Adding $t\bar tH$ inferred from $y_t$ determination\\
       \hline
           LEP3 & 230 GeV& 2.304& \cite{LEP3-ESPPU}& Rescaling the input in \cite{Selvaggi:2025kmd} using the Higgs boson number from \cite{LEP3-ESPPU}\\
       \hline
        \multirow {2}{*}{FCC-ee} & 240 GeV &10.8&\cite{Selvaggi:2025kmd}& - \\
       &365 GeV&3.12&\cite{Selvaggi:2025kmd}& -\\ \hline
       FCC-hh & 84 TeV & 30 & \cite{Selvaggi:2025kmd}& -\\\hline
       muC3 & 3 TeV& 3& \cite{Forslund:2022xjq}& -\\\hline
        muC10 & 10 TeV& 10  &\cite{Forslund:2022xjq,InternationalMuonCollider:2025sys,Ruhdorfer:2024dgz} & Including inclusive production    from \cite{Ruhdorfer:2024dgz}    \\ 
        \bottomrule
    \end{tabularx}    
    \caption{References for the experimental inputs entering the kappa framework analysis presented in this work.
    The starting point are the inputs for the ESPPU19~\cite{deBlas:2019rxi}, updated to Snowmass 2022~\cite{deBlas:2022ofj} and subsequently to the ESPPU26~\cite{deBlas:2025gyz} analysis.
    Note that for the kappa framework benchmarks of App.~\ref{subsec:kappa_framework_bench} for consistency we adopt exactly the same experimental inputs as in the corresponding {\sc\small HepFit} analysis, which differ from those in this table.
    }
\label{tab:future_colliders_kappafits}
\end{table}

As discussed in~\cite{Celada:2024mcf}, our HL-LHC projections are mostly based on the direct extrapolation of existing Run II measurements.
Two exceptions are top-quark differential distributions, for which dedicated HL-LHC projections extending to higher $m_{t\bar{t}}$ values are available, and Higgs pair production, for which we use the full projected likelihood~\cite{terHoeve:2025omu}.
This approach to HL-LHC projections has two limitations: it does not account for possible correlations between ATLAS and CMS measurements, and may underestimate the HL-LHC reach in cases for which targeted measurements improve naive Run II extrapolations. 

In this appendix we explore the possible impact of these issues in the context of the kappa framework fits, but the findings are general and may apply also to the SMEFT fits of HL-LHC projections.
We also provide tables collecting the numerical results of the fits displayed in Sect.~\ref{sec:kappa_results}.

For the kappa framework results presented in this work, the HL-LHC projections are included as follows.
First, we do not use as input the separate measurements of production cross-sections and branching ratios~\cite{ATL-PHYS-PUB-2022-037,CMS-PAS-FTR-18-011}, since these assume SM-like Higgs decays and SM-like production respectively.
Instead, we consider as input measurements the signal strengths:
\begin{equation}\label{eq:signal_strength}
    \mu = \frac{\sigma_{\rm exp}\times {\rm Br}_{\rm exp}}{\sigma_{\rm SM}\times {\rm Br}_{\rm SM}}\, .
\end{equation}

Second, we account for the correlations between the Higgs signal strength measurements following~\cite{TwikiCMS}.
Ignoring experimental correlations may lead to underestimating the expected bounds on some coupling modifiers (such as $\kg$, since gluon-fusion production measurements are strongly correlated) and  overestimating them for others (such as $\ktau$, due to the fact that $\tau\tau$ decays in different production channels are anti-correlated).

In Tables~\ref{tab:HL-LHC} and~\ref{tab:HLtheory} we present results for kappa-0 fits at the HL-LHC for different choices of the experimental inputs.
In particular, in Table~\ref{tab:HL-LHC} we first show results based on the CMS inputs alone since they provide the full correlation matrix and the projected uncertainties directly on the signal strength measurements~\cite{TwikiCMS}.
We then repeat the fit using the combined ATLAS+CMS projections for the Higgs signal strengths presented in~\cite{Cepeda:2019klc} for three different scenarios: without correlations and theory uncertainties, without correlations but with theory uncertainties, and with both of them. 
The purpose of this exercise is to assess the improvements obtained considering the combination of the inputs from the two collaborations and the role played by theoretical errors and experimental correlations. 
Since the latter are not provided in~\cite{Cepeda:2019klc}, we assume the CMS correlation model.\footnote{This assumption fails for the $Z\gamma$  channel, for which only ATLAS  provided two different production modes projections whose correlation have not been included.} 
Theory uncertainties arising from the denominator of the signal strength have been included symmetrizing the ones in~\cite{Cepeda:2019klc}. 

\begin{table}[t]
\centering
\small
 \renewcommand{\arraystretch}{1.1}
\begin{tabularx}{\textwidth}{ X|c|c|c|c|c|c|}
\cline{2-7}
& \multicolumn{6}{c|}{HL-LHC}\\
 \cline{2-7}
 &\multicolumn{2}{c|}{CMS} &  \multicolumn{3}{c|}{ATLAS-CMS Combination} & \multirow{2}{*}{ESPPU19}\\
 &{\footnotesize w.o. corr.}&{\footnotesize w. corr.}&  {\footnotesize w.o. theory and corr.} & {\footnotesize w.o. corr.} & {\footnotesize complete  } & \\
\hline
\multicolumn{1}{|l|}{ $\kappa_W$ }&2&1.9&1.5&1.7&1.6&1.7\\ 
\multicolumn{1}{|l|}{ $\kappa_Z$ }&1.9&1.8&1.4&1.5&1.5&1.5\\ 
\multicolumn{1}{|l|}{ $\kappa_g$ }&2.2&2.5&1.6&2&2.3&2.5\\ 
\multicolumn{1}{|l|}{ $\kappa_\gamma$ }&2.5&2.2&1.7&2.2&1.9&1.8\\ 
\multicolumn{1}{|l|}{ $\kappa_{Z\gamma}$ }&-&-&12&12&12&9.8\\ 
\multicolumn{1}{|l|}{ $\kappa_t$ }&2.9&3&2.1&2.5&2.6&3.4\\ 
\multicolumn{1}{|l|}{ $\kappa_b$ }&4.1&4.1&3.2&3.5&3.4&3.7\\ 
\multicolumn{1}{|l|}{ $\kappa_\mu$ }&7&7.1&5.3&5.7&5.6&4.3\\ 
\multicolumn{1}{|l|}{ $\kappa_\tau$ }&2.5&2.3&2&2.2&2.1&1.9\\   
\hline
\end{tabularx}
\vspace{0.2cm}
    \caption{Projected uncertainties (68\% CL) on the coupling modifiers $\kappa_i$ obtained in the kappa-0 framework at the HL-LHC, for different choices of the experimental inputs.
    The last column indicates for reference the results quoted in the ESPP19 report~\cite{EuropeanStrategyforParticlePhysicsPreparatoryGroup:2019qin}.
    See text for more details.
    }
        \label{tab:HL-LHC}
\end{table}

Theory errors on the signal strengths cannot be  obtained by simply combining those from production and decay, since they also enter the experimental acceptances (labelled as $\Delta_{\rm SigAcc}$ in\cite{Cepeda:2019klc}) and can be correlated or even cancel out when taking the ratio in Eq.~\eqref{eq:signal_strength}. 
In order to extract  theory uncertainties, here we evaluate the difference
\begin{equation}\label{eq:thHLLHC}
    \Delta\mu_{\rm th}=\sqrt{\Delta\mu_{\rm CMS, th}^2-\Delta\mu_{\rm CMS, SigAcc}^2} \, ,
\end{equation}
where $\Delta\mu_{\rm CMS, th}$ is the theory error on the signal strengths quoted in~\cite{TwikiCMS} and $\Delta\mu_{\rm CMS, SigAcc}$ is CMS signal modelling theory error on  cross section times branching ratio reported in~\cite{Cepeda:2019klc}.

\begin{table}[t]
\small
 \renewcommand{\arraystretch}{1.1}
\begin{tabular}{c|c|c|c|c|c|c|}
\cline{2-7}
\multirow{2}{*}{}& No theory  & S2 theory err. & S2 theory err. & Eq.~(\ref{eq:thHLLHC}) theory err. & Eq.~(\ref{eq:thHLLHC}) theory err. & \multirow{2}{*}{ESPPU 2019} \\
&errors&No correlations&Correlations&No correlations&Correlations&\\
\hline
\multicolumn{1}{|l|}{ $\kappa_W$ }&1.5&1.7&1.6&1.7&1.6&1.7\\ 
\multicolumn{1}{|l|}{ $\kappa_Z$ }&1.4&1.5&1.5&1.6&1.5&1.5\\ 
\multicolumn{1}{|l|}{ $\kappa_g$ }&1.7&2.0&2.3&2&2.2&2.5\\ 
\multicolumn{1}{|l|}{ $\kappa_\gamma$ }&1.7&2.2&1.9&2.1&1.8&1.8\\ 
\multicolumn{1}{|l|}{ $\kappa_{Z\gamma}$ }&12&12&12&12&12&9.8\\ 
\multicolumn{1}{|l|}{ $\kappa_t$ }&2.1&2.4&2.6&2.5&2.6&3.4\\ 
\multicolumn{1}{|l|}{ $\kappa_b$ }&3.2&3.4&3.4&3.6&3.5&3.7\\ 
\multicolumn{1}{|l|}{ $\kappa_\mu$ }&5.3&5.7&5.6&5.7&5.4&4.3\\ 
\multicolumn{1}{|l|}{ $\kappa_\tau$ }&2&2.2&2.1&2.2&2&1.9\\ 
\hline
\end{tabular}
\vspace{0.2cm}
\caption{Same as Table~\ref{tab:HL-LHC} for different assumptions on theory uncertainties for the HL-LHC projections.}
\label{tab:HLtheory}
\end{table}

Our HL-LHC inputs for the kappa fits are based on Eq.~\eqref{eq:thHLLHC} for the theoretical uncertainties and the CMS model for the signal strength correlation, which lead to results for the coupling modifiers very similar in most cases to those of the last column in Table~\ref{tab:HLtheory}, namely the ESPPU19 reference. 

\paragraph{Kappa-0 numerical results.}

Table~\ref{tab:kappa0} displays the percentage uncertainties for the coupling modifiers, $\delta \kappa_i$, corresponding to the results of the kappa-0 fits shown in
Fig.~\ref{fig:kappa0_histo}. 
Results are provided both with and without accounting for theoretical uncertainties. For the LCF, FCC-ee and FCC-hh colliders, numerical results for a given centre of mass energy $\sqrt{s}$ include also those taken at lower $\sqrt{s}$ values (hence they are cumulative).

\begin{table}[htbp]
\footnotesize
\centering
\begin{tabular} { llc|c|c|c|c|c|c|c|c|c|c|}
\cline{4-13}
& &&$\kw$[\%]  &$\kz$[\%]  &$\kg$[\%]  &$\ka$[\%]  &$\kza$[\%]  &$\kc$[\%]  &$\kt$[\%]  &$\kb$[\%]  &$\kmu$[\%]  &$\ktau$[\%]  \\\hline
\multicolumn{2}{|c|}{\multirow { 2 }{*}{HL-LHC}} &  \multirow { 2 }{*}{\footnotesize exp+th}&\multirow { 2 }{*}{1.6}&\multirow { 2 }{*}{1.5}&\multirow { 2 }{*}{2.1}&\multirow { 2 }{*}{1.9}&\multirow { 2 }{*}{8.4}& \multirow { 2 }{*}{-}&\multirow { 2 }{*}{2.6}&\multirow { 2 }{*}{3.4}&\multirow { 2 }{*}{3.5}&\multirow { 2 }{*}{2}\\ 
\multicolumn{2}{|c|}{}&& & & & & & & & & & \\ \hline 
\multicolumn{2}{|c|}{\multirow { 2 }{*}{LHeC}} &   {\footnotesize exp}&0.74&1.2&3.6&7.5& -&4& -&2.1& -&3.3\\ 
\cline{3-3}\multicolumn{2}{|c|}{} & {\footnotesize exp+th}&0.78&1.2&3.6&7.5& -&4.1& -&2.2& -&3.3\\ \hline 
\multicolumn{1}{|c|}{\multirow { 8}{*}{LCF}}  &\multicolumn{1}{c|}{\multirow { 2 }{*}{250}} &   {\footnotesize exp}&1.1&0.18&1.6&5.2& -&1.7& -&1.2& -&1.3\\ 
\multicolumn{1}{|c|}{} & \multicolumn{1}{c|}{} & {\footnotesize exp+th}&1.2&0.21&1.8&5.2& -&1.8& -&1.2& -&1.3\\ \cline{3-13}
\multicolumn{1}{|c|}{}&\multicolumn{1}{c|}{\multirow { 2 }{*}{350}} &   {\footnotesize exp}&0.8&0.18&1.4&5.1& -&1.5& -&0.91& -&1\\ 
\multicolumn{1}{|c|}{} & \multicolumn{1}{c|}{} & {\footnotesize exp+th}&0.83&0.21&1.5&5.1& -&1.7& -&0.98& -&1\\ \cline{3-13}
\multicolumn{1}{|c|}{}&\multicolumn{1}{c|}{\multirow { 2 }{*}{550}} &   {\footnotesize exp}&0.15&0.13&0.68&2.6& -&0.85&4.4&0.32&15&0.49\\ 
\multicolumn{1}{|c|}{} & \multicolumn{1}{c|}{} & {\footnotesize exp+th}&0.25&0.16&0.79&2.7& -&0.92&4.5&0.47&15&0.55\\ \cline{3-13}
\multicolumn{1}{|c|}{}&\multicolumn{1}{c|}{\multirow { 2 }{*}{1000}} &   {\footnotesize exp}&0.15&0.095&0.52&1.7& -&0.61&1.6&0.26&7.1&0.42\\ 
\multicolumn{1}{|c|}{} & \multicolumn{1}{c|}{} & {\footnotesize exp+th}&0.14&0.096&0.51&1.7& -&0.63&1.6&0.26&7.1&0.42\\ \hline 

\multicolumn{2}{|c|}{\multirow { 2 }{*}{LEP3}} &   {\footnotesize exp}&1.7&0.22&1.9&4.5&14&2.4& -&1.7&13&1.8\\ 
\cline{3-3}\multicolumn{2}{|c|}{} & {\footnotesize exp+th}&1.7&0.24&2&4.5&15&2.5& -&1.7&13&1.8\\ \hline 
\multicolumn{1}{|c|}{\multirow { 6}{*}{FCC}}  &\multicolumn{1}{c|}{\multirow { 2 }{*}{240}} &   {\footnotesize exp}&0.73&0.094&0.82&1.9&6&1.1& -&0.72&5.6&0.78\\ 
\multicolumn{1}{|c|}{} & \multicolumn{1}{c|}{} & {\footnotesize exp+th}&0.78&0.13&1&2.1&5.9&1.3& -&0.83&5.6&0.82\\ \cline{3-13}
\multicolumn{1}{|c|}{}&\multicolumn{1}{c|}{\multirow { 2 }{*}{365}} &   {\footnotesize exp}&0.49&0.083&0.63&1.8&4.9&0.81& -&0.55&5&0.59\\ 
\multicolumn{1}{|c|}{} & \multicolumn{1}{c|}{} & {\footnotesize exp+th}&0.52&0.11&0.77&1.9&4.8&0.89& -&0.66&5.2&0.64\\ \cline{3-13}
\multicolumn{1}{|c|}{}&\multicolumn{1}{c|}{\multirow { 2 }{*}{hh}} &   {\footnotesize exp}&0.43&0.08&0.56&0.32&0.66&0.79&0.89&0.51&0.39&0.55\\ 
\multicolumn{1}{|c|}{} & \multicolumn{1}{c|}{} & {\footnotesize exp+th}&0.45&0.11&0.62&0.34&0.68&0.9&0.93&0.62&0.43&0.61\\ \hline 
\multicolumn{2}{|c|}{\multirow { 2 }{*}{muC 3000}} &   {\footnotesize exp}&0.38&1.2&1.6&3.1&27&6.1& -&0.85&15&2.1\\ 
\cline{3-3}\multicolumn{2}{|c|}{} & {\footnotesize exp+th}&0.42&1.2&1.7&3.2&27&6.1& -&0.98&15&2.1\\ \hline 
\multicolumn{2}{|c|}{\multirow { 2 }{*}{muC 10000}} &   {\footnotesize exp}&0.1&0.19&0.45&0.82&6&1.8& -&0.24&2.8&0.6\\ 
\cline{3-3}\multicolumn{2}{|c|}{} & {\footnotesize exp+th}&0.22&0.21&0.74&0.94&6.1&1.9& -&0.53&2.8&0.62\\ \hline 
\end{tabular} 
\vspace{0.2cm}
\caption{Uncertainties on the $\kappa$-modifiers for the kappa-0 fits shown in Fig.~\ref{fig:kappa0_histo}. 
We provide results without and with inclusion of theoretical uncertainties, see text for more details.
}
\label{tab:kappa0}
\end{table}

%

\paragraph{Kappa-3 numerical results.}
The numerical results for the kappa-3 analyses of Fig.~\ref{fig:kappa3_histo_comparison} are collected in Tables~\ref{tab:kappa3_constrained} and~\ref{tab:kappa3_free} for the cases for which the constraint $\kappa_V\le 1$ is imposed or not, respectively. 
In Table~\ref{tab:kappa3_constrained}, the percentage uncertainty in the coupling modifiers associated to vector bosons $\kappa_W,\kappa_Z$ is evaluated by means of Eq.~(\ref{eq:kvcond}), indicated by a star next to the names of the colliders. 

\begin{table}[!h]
\footnotesize
\centering
\begin{tabular}{ llc|c|c|c|c|c|c|c|c|c|c||c|c|}
\cline{4-15}
&&& \multicolumn{10}{c|}{ [\%] error on the fit ( with $\delta \kappa_V$ as in \eqref{eq:err_kv})} & \multicolumn{2}{c|}{ 95\% U.L.} \\ \cline{4-15}& &&$\kw$&$\kz$&$\kg$&$\ka$&$\kza$&$\kc$&$\kt$&$\kb$&$\kmu$&$\ktau$&${\rm Br}_{\rm inv}$&${\rm Br}_{\rm und}$\\\hline
\multicolumn{2}{|c|}{\multirow { 2 }{*}{HL-LHC*}} &  \multirow { 2 }{*}{\footnotesize exp+th}&\multirow { 2 }{*}{1.8}&\multirow { 2 }{*}{1.8}&\multirow { 2 }{*}{2.1}&\multirow { 2 }{*}{1.6}&\multirow { 2 }{*}{8.2}& \multirow { 2 }{*}{-}&\multirow { 2 }{*}{2.6}&\multirow { 2 }{*}{2.7}&\multirow { 2 }{*}{3.4}&\multirow { 2 }{*}{1.8}&\multirow { 2 }{*}{2.1}&\multirow { 2 }{*}{5.2}\\ 
\multicolumn{2}{|c|}{}&& & & & & & & & & & & & \\ \hline 
\multicolumn{2}{|c|}{\multirow { 2 }{*}{LHeC*}} &   {\footnotesize exp}&0.36&0.72&1.6&1.3&8.2&3.6&2.4&1.1&3.2&1.5&1&1.3\\ 
\cline{3-3}\multicolumn{2}{|c|}{} & {\footnotesize exp+th}&0.4&0.74&1.7&1.4&8.2&3.7&2.5&1.2&3.3&1.5&1.1&1.4\\ \hline 
\multicolumn{1}{|c|}{\multirow { 8}{*}{LCF*}}  &\multicolumn{1}{c|}{\multirow { 2 }{*}{250}} &   {\footnotesize exp}&0.73&0.13&1.1&1.3&8.1&1.4&2.3&0.7&3.2&0.79&0.29&0.65\\ 
\multicolumn{1}{|c|}{} & \multicolumn{1}{c|}{} & {\footnotesize exp+th}&0.72&0.17&1.1&1.3&8.2&1.5&2.4&0.83&3.2&0.8&0.3&0.85\\ \cline{3-15}
\multicolumn{1}{|c|}{}&\multicolumn{1}{c|}{\multirow { 2 }{*}{350}} &   {\footnotesize exp}&0.6&0.13&1.1&1.2&8.1&1.4&2.3&0.65&3.3&0.74&0.28&0.62\\ 
\multicolumn{1}{|c|}{} & \multicolumn{1}{c|}{} & {\footnotesize exp+th}&0.59&0.16&1.1&1.3&8.2&1.5&2.3&0.74&3.1&0.76&0.3&0.78\\ \cline{3-15}
\multicolumn{1}{|c|}{}&\multicolumn{1}{c|}{\multirow { 2 }{*}{550}} &   {\footnotesize exp}&0.13&0.11&0.63&1.2&8.5&0.85&2.1&0.31&3.1&0.46&0.22&0.34\\ 
\multicolumn{1}{|c|}{} & \multicolumn{1}{c|}{} & {\footnotesize exp+th}&0.19&0.13&0.71&1.2&8.4&0.91&2.1&0.43&3.1&0.5&0.25&0.47\\ \cline{3-15}
\multicolumn{1}{|c|}{}&\multicolumn{1}{c|}{\multirow { 2 }{*}{1000}} &   {\footnotesize exp}&0.12&0.079&0.49&1&8.1&0.62&1.3&0.25&2.9&0.4&0.19&0.26\\ 
\multicolumn{1}{|c|}{} & \multicolumn{1}{c|}{} & {\footnotesize exp+th}&0.19&0.1&0.56&1&8.3&0.68&1.3&0.35&2.9&0.42&0.23&0.38\\ \hline 

\multicolumn{2}{|c|}{\multirow { 2 }{*}{LEP3*}} &   {\footnotesize exp}&0.91&0.17&1.1&1.3&7.2&2&2.4&0.85&3.1&0.95&0.12&0.83\\ 
\cline{3-3}\multicolumn{2}{|c|}{} & {\footnotesize exp+th}&0.92&0.2&1.1&1.3&7.1&2.1&2.4&0.97&3.1&0.96&0.12&1\\ \hline 
\multicolumn{1}{|c|}{\multirow { 6}{*}{FCC*}}  &\multicolumn{1}{c|}{\multirow { 2 }{*}{240}} &   {\footnotesize exp}&0.6&0.075&0.58&1.1&4.7&0.91&2.3&0.45&2.7&0.51&0.051&0.37\\ 
\multicolumn{1}{|c|}{} & \multicolumn{1}{c|}{} & {\footnotesize exp+th}&0.59&0.11&0.78&1.1&4.9&1.1&2.3&0.59&2.7&0.56&0.052&0.72\\ \cline{3-15}
\multicolumn{1}{|c|}{}&\multicolumn{1}{c|}{\multirow { 2 }{*}{365}} &   {\footnotesize exp}&0.41&0.066&0.5&1&4.1&0.73&2.3&0.37&2.7&0.43&0.048&0.32\\ 
\multicolumn{1}{|c|}{} & \multicolumn{1}{c|}{} & {\footnotesize exp+th}&0.43&0.092&0.63&1&4.1&0.84&2.4&0.48&2.7&0.48&0.05&0.56\\ \cline{3-15}
\multicolumn{1}{|c|}{}&\multicolumn{1}{c|}{\multirow { 2 }{*}{hh}} &   {\footnotesize exp}&0.37&0.065&0.43&0.3&0.65&0.73&0.79&0.36&0.4&0.42&0.022&0.31\\ 
\multicolumn{1}{|c|}{} & \multicolumn{1}{c|}{} & {\footnotesize exp+th}&0.39&0.09&0.5&0.31&0.65&0.82&0.77&0.46&0.39&0.45&0.023&0.53\\ \hline 
\multicolumn{2}{|c|}{\multirow { 2 }{*}{muC 3000*}} &   {\footnotesize exp}&0.26&0.82&1.1&1.2&7.5&5.9&2.3&0.76&3.1&1.3&0.87&0.98\\ 
\cline{3-3}\multicolumn{2}{|c|}{} & {\footnotesize exp+th}&0.29&0.8&1.2&1.2&7.7&6&2.4&0.84&3.1&1.3&0.95&1.1\\ \hline 
\multicolumn{2}{|c|}{\multirow { 2 }{*}{muC 10000*}} &   {\footnotesize exp}&0.07&0.19&0.41&0.7&4.9&1.7&2.3&0.22&2.1&0.54&0.29&0.29\\ 
\cline{3-3}\multicolumn{2}{|c|}{} & {\footnotesize exp+th}&0.14&0.19&0.64&0.78&4.9&1.9&2.3&0.39&2.1&0.57&0.52&0.53\\ \hline 
\end{tabular} 
\vspace{0.2cm}
\caption{Same as Table~\ref{tab:kappa0}, now for the kappa-3 fits shown in Fig.~\ref{fig:kappa3_histo_comparison} where the constraint $\kappa_V\le 1$ is applied to all colliders.    For the vector boson coupling modifiers $\kappa_W,\kappa_Z$ the percentage uncertainties are defined as in Eq.~\eqref{eq:err_kv}.}
\label{tab:kappa3_constrained}
\end{table}
 
\begin{table}[!h]
\footnotesize
\centering
\begin{tabular}{ llc|c|c|c|c|c|c|c|c|c|c||c|c|}
\cline{4-15}
&&& \multicolumn{10}{c|}{ [\%] error on the fit } & \multicolumn{2}{c|}{ 95\% U.L.} \\ \cline{4-15}& &&$\kw$&$\kz$&$\kg$&$\ka$&$\kza$&$\kc$&$\kt$&$\kb$&$\kmu$&$\ktau$&${\rm Br}_{\rm inv}$&${\rm Br}_{\rm und}$\\\hline
\multicolumn{1}{|c|}{\multirow { 8}{*}{LCF}}  &\multicolumn{1}{c|}{\multirow { 2 }{*}{250}} &   {\footnotesize exp}&0.82&0.23&1.2&1.3&8.3&1.6&2.3&0.96&3.1&0.99&0.32&1.4\\ 
\multicolumn{1}{|c|}{} & \multicolumn{1}{c|}{} & {\footnotesize exp+th}&0.85&0.25&1.2&1.4&8.1&1.7&2.4&1&3.2&1&0.34&1.6\\ \cline{3-15}
\multicolumn{1}{|c|}{}&\multicolumn{1}{c|}{\multirow { 2 }{*}{350}} &   {\footnotesize exp}&0.69&0.22&1.1&1.3&8.1&1.5&2.4&0.85&3.3&0.91&0.33&1.4\\ 
\multicolumn{1}{|c|}{} & \multicolumn{1}{c|}{} & {\footnotesize exp+th}&0.69&0.24&1.1&1.3&8.2&1.6&2.3&0.9&3.2&0.94&0.33&1.5\\ \cline{3-15}
\multicolumn{1}{|c|}{}&\multicolumn{1}{c|}{\multirow { 2 }{*}{550}} &   {\footnotesize exp}&0.22&0.18&0.66&1.2&8.2&0.89&2.1&0.37&3.1&0.52&0.3&1.1\\ 
\multicolumn{1}{|c|}{} & \multicolumn{1}{c|}{} & {\footnotesize exp+th}&0.3&0.19&0.76&1.2&8.2&0.99&2.1&0.5&3.1&0.59&0.31&1.2\\ \cline{3-15}
\multicolumn{1}{|c|}{}&\multicolumn{1}{c|}{\multirow { 2 }{*}{1000}} &   {\footnotesize exp}&0.2&0.15&0.52&1&8.2&0.65&1.3&0.31&2.9&0.44&0.3&0.97\\ 
\multicolumn{1}{|c|}{} & \multicolumn{1}{c|}{} & {\footnotesize exp+th}&0.29&0.17&0.63&1.1&8&0.73&1.3&0.42&2.9&0.5&0.3&1\\ \hline  
\multicolumn{2}{|c|}{\multirow { 2 }{*}{LEP3}} &   {\footnotesize exp}&1&0.27&1.3&1.4&7&2.2&2.4&1.2&3.1&1.2&0.12&1.7\\ 
\cline{3-3}\multicolumn{2}{|c|}{} & {\footnotesize exp+th}&1&0.29&1.3&1.4&7&2.3&2.4&1.2&3.2&1.2&0.12&1.9\\ \hline 
\multicolumn{1}{|c|}{\multirow { 6}{*}{FCC-ee}}  &\multicolumn{1}{c|}{\multirow { 2 }{*}{240}} &   {\footnotesize exp}&0.62&0.12&0.72&1.1&4.8&1&2.3&0.64&2.8&0.69&0.053&0.76\\ 
\multicolumn{1}{|c|}{} & \multicolumn{1}{c|}{} & {\footnotesize exp+th}&0.64&0.14&0.85&1.1&4.8&1.2&2.4&0.73&2.8&0.71&0.054&1\\ \cline{3-15}
\multicolumn{1}{|c|}{}&\multicolumn{1}{c|}{\multirow { 2 }{*}{365}} &   {\footnotesize exp}&0.45&0.1&0.59&1&4&0.8&2.3&0.53&2.6&0.57&0.049&0.66\\ 
\multicolumn{1}{|c|}{} & \multicolumn{1}{c|}{} & {\footnotesize exp+th}&0.48&0.12&0.7&1.1&4&0.9&2.3&0.63&2.7&0.61&0.051&0.85\\ \cline{3-15}
\multicolumn{1}{|c|}{}&\multicolumn{1}{c|}{\multirow { 2 }{*}{hh}} &   {\footnotesize exp}&0.42&0.1&0.54&0.32&0.67&0.8&0.84&0.51&0.4&0.55&0.023&0.65\\ 
\multicolumn{1}{|c|}{} & \multicolumn{1}{c|}{} & {\footnotesize exp+th}&0.42&0.12&0.58&0.33&0.69&0.88&0.84&0.58&0.41&0.57&0.023&0.85\\ \hline 
\multicolumn{2}{|c|}{\multirow { 2 }{*}{muC10}} &   {\footnotesize exp}&0.25&0.21&0.48&0.72&4.9&1.8&2.3&0.33&2.1&0.59&0.24&1.6\\ 
\cline{3-3}\multicolumn{2}{|c|}{} & {\footnotesize exp+th}&0.34&0.23&0.74&0.83&4.9&1.9&2.3&0.59&2.2&0.65&0.25&1.8\\ \hline 
\end{tabular} 
\vspace{0.2cm}
\caption{Same as Table~\ref{tab:kappa3_constrained} now without imposing any condition on $\kappa_V$.
Note that in this case only colliders that can measure $\Gamma_H$ in a model-independent way can enter the analysis.
}
\label{tab:kappa3_free}
\end{table}

\paragraph{Universal kappa modifier numerical results.}
Finally, we present here the numerical results for the coupling modifier fits associated to the kappa-universal fit of Fig.~\ref{fig:kappa-universal} and the fermion-vector kappa modifiers fits of Figs.~\ref{fig:kappa-FvsV_histo} and~\ref{fig:kfvskv_contours}. As before, when the constraint Eq.~(\ref{eq:kvcond}) is imposed to stabilise the fit, we  indicate it with a (*) next to the name of the corresponding collider.

\begin{table}[t]
\centering
\small
 \renewcommand{\arraystretch}{1.1}
\begin{tabular}{c|cc|cc|cc|}
\cline{2-7}
&\multicolumn{2}{c|}{$\kappa_{\rm univ}$[\%]}&\multicolumn{2}{c|}{$\Brinv$[\%]}&\multicolumn{2}{c|}{$\Brund$[\%]}\\ 
& Exp & Exp+Th& Exp & Exp+Th& Exp & Exp+Th\\\hline
\multicolumn{1}{|l|}{HL-LHC*}& \multicolumn{2}{c|}{0.55}& \multicolumn{2}{c|}{1.6}& \multicolumn{2}{c|}{2.5}\\ 
\multicolumn{1}{|l|}{LHeC*}&0.22&0.27&0.88&1&0.98&1.2\\ 
\multicolumn{1}{|l|}{LCF*}&0.051&0.08&0.18&0.24&0.23&0.41\\ 
\multicolumn{1}{|l|}{CEPC*}&0.038&0.071&0.056&0.06&0.18&0.43\\ 
\multicolumn{1}{|l|}{LEP3*}&0.13&0.17&0.12&0.12&0.67&0.94\\ 
\multicolumn{1}{|l|}{FCC-ee*}&0.053&0.08&0.048&0.05&0.26&0.51\\ 
\multicolumn{1}{|l|}{FCC-ee+hh*}&0.051&0.077&0.022&0.022&0.25&0.45\\ 
\multicolumn{1}{|l|}{muC 3000*}&0.2&0.22&0.82&0.9&0.9&1\\ 
\multicolumn{1}{|l|}{muC 10000*}&0.055&0.094&0.27&0.48&0.27&0.49\\
\hline\end{tabular} 
\vspace{0.2cm}
\caption{Results of the kappa-framework fits with a universal modifier and $\kappa_{\rm univ} \le 1$.}
\label{tab:kappa_univ_constrained}
\end{table}

\begin{table}[htbp]
\centering
\small
 \renewcommand{\arraystretch}{1.1}
\begin{tabular}{c|cc|cc|cc|}
\cline{2-7}
&\multicolumn{2}{c|}{$\kappa_{\rm univ}$[\%]}&\multicolumn{2}{c|}{$\Brinv$[\%]}&\multicolumn{2}{c|}{$\Brund$[\%]}\\ 
& Exp & Exp+Th& Exp & Exp+Th& Exp & Exp+Th\\\hline
\multicolumn{1}{|l|}{LCF}&0.14&0.16&0.3&0.3&0.9&1\\ 
\multicolumn{1}{|l|}{CEPC}&0.08&0.1&0.061&0.063&0.51&0.67\\ 
\multicolumn{1}{|l|}{LEP3}&0.25&0.27&0.12&0.12&1.6&1.7\\ 
\multicolumn{1}{|l|}{FCC-ee}&0.092&0.11&0.05&0.05&0.59&0.74\\ 
\multicolumn{1}{|l|}{FCC-ee+hh}&0.095&0.11&0.023&0.023&0.59&0.73\\ 
\multicolumn{1}{|l|}{muC 10000}&0.039&0.07&0.16&0.21&0.27&0.54\\ 
\hline \end{tabular} 
\vspace{0.2cm}
\caption{Same as Table~\ref{tab:kappa_univ_constrained} without the constraint on $\kappa_{\rm univ}.$} 
\label{tab:kappa_univ_free}
\end{table}

\begin{table}[htbp]
\centering
\small
 \renewcommand{\arraystretch}{1.1}
\begin{tabular}{c|cc|cc|cc|cc|}
\cline{2-9}
&\multicolumn{2}{c|}{$\kappa_{F}$[\%]}&\multicolumn{2}{c|}{$\kappa_{V}$[\%]}&\multicolumn{2}{c|}{$\Brinv$[\%]}&\multicolumn{2}{c|}{$\Brund$[\%]}\\ 
& Exp & Exp+Th& Exp & Exp+Th& Exp & Exp+Th& Exp & Exp+Th\\\hline
\multicolumn{1}{|l|}{HL-LHC*}& \multicolumn{2}{c|}{1.2}& \multicolumn{2}{c|}{0.57}& \multicolumn{2}{c|}{1.6}& \multicolumn{2}{c|}{2.5}\\ 
\multicolumn{1}{|l|}{LHeC*}&0.86&0.94&0.25&0.28&0.98&1.1&1.2&1.2\\ 
\multicolumn{1}{|l|}{LCF*}&0.23&0.31&0.053&0.08&0.2&0.23&0.26&0.4\\ 
\multicolumn{1}{|l|}{CEPC*}&0.2&0.32&0.047&0.071&0.058&0.06&0.23&0.44\\ 
\multicolumn{1}{|l|}{LEP3*}&0.59&0.69&0.15&0.18&0.11&0.12&0.8&1\\ 
\multicolumn{1}{|l|}{FCC-ee*}&0.27&0.35&0.063&0.08&0.048&0.049&0.33&0.5\\ 
\multicolumn{1}{|l|}{FCC-ee+hh*}&0.2&0.23&0.057&0.077&0.022&0.023&0.3&0.47\\ 
\multicolumn{1}{|l|}{muC 3000*}&0.66&0.68&0.2&0.23&0.79&0.89&0.86&1\\ 
\multicolumn{1}{|l|}{muC 10000*}&0.21&0.33&0.055&0.1&0.26&0.51&0.27&0.54\\ 
\hline 
\end{tabular} 
\vspace{0.2cm}
\caption{Same as Table~\ref{tab:kappa_univ_constrained} for the universal vector and fermion coupling modifiers, with $\kappa_V\le 1$.}
\label{tab:k_VvsF_constrained}
\end{table}

\begin{table}[!h]
\centering
\small
 \renewcommand{\arraystretch}{1.1}
\begin{tabular}{c|cc|cc|cc|cc|}
\cline{2-9}
&\multicolumn{2}{c|}{$\kappa_{F}$[\%]}&\multicolumn{2}{c|}{$\kappa_{V}$[\%]}&\multicolumn{2}{c|}{$\Brinv$[\%]}&\multicolumn{2}{c|}{$\Brund$[\%]}\\ 
& Exp & Exp+Th& Exp & Exp+Th& Exp & Exp+Th& Exp & Exp+Th\\\hline
\multicolumn{1}{|l|}{LCF}&0.27&0.32&0.15&0.16&0.29&0.3&0.89&0.96\\ 
\multicolumn{1}{|l|}{CEPC}&0.21&0.33&0.084&0.1&0.062&0.061&0.54&0.68\\ 
\multicolumn{1}{|l|}{LEP3}&0.67&0.72&0.27&0.27&0.12&0.12&1.7&1.7\\ 
\multicolumn{1}{|l|}{FCC-ee}&0.3&0.37&0.1&0.11&0.049&0.05&0.64&0.79\\ 
\multicolumn{1}{|l|}{FCC-ee+hh}&0.22&0.25&0.1&0.11&0.022&0.022&0.62&0.72\\ 
\multicolumn{1}{|l|}{muC 10000}&0.26&0.37&0.16&0.18&0.23&0.24&0.99&1.1\\ 
\hline \end{tabular} 
\vspace{0.2cm}
\caption{Same as Table~\ref{tab:k_VvsF_constrained} without the constraint on $\kappa_V$.}
\label{tab:k_VvsF_free}
\end{table}

\subsection{Effective couplings}
\label{app:effectivecouplings_supplementary}

\begin{table}[htbp]
\scriptsize
\centering
 \renewcommand{\arraystretch}{1.08}
\begin{tabular}{|c|c|c|c|c|c|c|c|}
\toprule
 &  & HL-LHC & LEP3 & FCC-ee & LCF550 & LCF1000 \\
\midrule
\multirow[c]{4}{*}{$\delta g_{eL}[\%]$} & No theory unc. & 0.13 & 0.0013 & 0.00076 & 0.0078 & 0.0081 \\
 & Aggr. theory unc. & 0.13 & 0.0042 & 0.0017 & 0.0082 & 0.0088 \\
 & Cons. theory unc. & 0.13 & 0.0098 & 0.008 & 0.01 & 0.011 \\
 & Curr. theory unc. & 0.13 & 0.044 & 0.035 & 0.016 & 0.016 \\
\cline{1-7}
\multirow[c]{4}{*}{$\delta g_{\mu L}[\%]$} & No theory unc. & 0.31 & 0.0029 & 0.0014 & 0.014 & 0.014 \\
 & Aggr. theory unc. & 0.31 & 0.0049 & 0.0022 & 0.016 & 0.016 \\
 & Cons. theory unc. & 0.31 & 0.011 & 0.0097 & 0.019 & 0.019 \\
 & Curr. theory unc. & 0.31 & 0.11 & 0.11 & 0.051 & 0.048 \\
\cline{1-7}
\multirow[c]{4}{*}{$\delta g_{\tau L}[\%]$} & No theory unc. & 0.22 & 0.0029 & 0.0015 & 0.02 & 0.02 \\
 & Aggr. theory unc. & 0.22 & 0.0049 & 0.0022 & 0.021 & 0.021 \\
 & Cons. theory unc. & 0.22 & 0.011 & 0.0093 & 0.023 & 0.023 \\
 & Curr. theory unc. & 0.22 & 0.096 & 0.095 & 0.049 & 0.049 \\
\cline{1-7}
\multirow[c]{4}{*}{$\delta g_{u L}[\%]$} & No theory unc. & 0.77 & 0.0037 & 0.0018 & 0.066 & 0.06 \\
 & Aggr. theory unc. & 0.77 & 0.011 & 0.0098 & 0.067 & 0.061 \\
 & Cons. theory unc. & 0.77 & 0.044 & 0.035 & 0.074 & 0.068 \\
 & Curr. theory unc. & 0.77 & 0.11 & 0.077 & 0.094 & 0.088 \\
\cline{1-7}
\multirow[c]{4}{*}{$\delta g_{d L}[\%]$} & No theory unc. & 0.52 & 0.0051 & 0.0026 & 0.054 & 0.05 \\
 & Aggr. theory unc. & 0.52 & 0.0095 & 0.0082 & 0.056 & 0.052 \\
 & Cons. theory unc. & 0.52 & 0.039 & 0.031 & 0.059 & 0.056 \\
 & Curr. theory unc. & 0.52 & 0.083 & 0.06 & 0.071 & 0.065 \\
\bottomrule
\multicolumn{7}{c}{}\\[-0.8em]
\midrule
\multirow[c]{4}{*}{$\delta g_{eR}[\%]$} & No theory unc. & 0.16 & 0.0018 & 0.001 & 0.0086 & 0.0089 \\
 & Aggr. theory unc. & 0.16 & 0.0059 & 0.0024 & 0.0098 & 0.01 \\
 & Cons. theory unc. & 0.16 & 0.015 & 0.012 & 0.013 & 0.013 \\
 & Curr. theory unc. & 0.16 & 0.064 & 0.045 & 0.02 & 0.019 \\
\cline{1-7}
\multirow[c]{4}{*}{$\delta g_{\mu R}[\%]$} & No theory unc. & 0.46 & 0.0042 & 0.0021 & 0.015 & 0.015 \\
 & Aggr. theory unc. & 0.46 & 0.0069 & 0.0031 & 0.016 & 0.016 \\
 & Cons. theory unc. & 0.46 & 0.016 & 0.014 & 0.021 & 0.021 \\
 & Curr. theory unc. & 0.46 & 0.16 & 0.15 & 0.064 & 0.063 \\
\cline{1-7}
\multirow[c]{4}{*}{$\delta g_{\tau R}[\%]$} & No theory unc. & 0.3 & 0.0043 & 0.0023 & 0.02 & 0.02 \\
 & Aggr. theory unc. & 0.3 & 0.007 & 0.0031 & 0.021 & 0.021 \\
 & Cons. theory unc. & 0.3 & 0.016 & 0.014 & 0.026 & 0.026 \\
 & Curr. theory unc. & 0.3 & 0.14 & 0.13 & 0.068 & 0.067 \\
\cline{1-7}
\multirow[c]{4}{*}{$\delta g_{u R}[\%]$} & No theory unc. & 1.8 & 0.02 & 0.0097 & 0.083 & 0.076 \\
 & Aggr. theory unc. & 1.8 & 0.023 & 0.014 & 0.086 & 0.076 \\
 & Cons. theory unc. & 1.8 & 0.12 & 0.12 & 0.12 & 0.11 \\
 & Curr. theory unc. & 1.8 & 0.62 & 0.44 & 0.21 & 0.19 \\
\cline{1-7}
\multirow[c]{4}{*}{$\delta g_{d R}[\%]$} & No theory unc. & 7.7 & 0.17 & 0.083 & 0.2 & 0.19 \\
 & Aggr. theory unc. & 7.7 & 0.17 & 0.083 & 0.2 & 0.2 \\
 & Cons. theory unc. & 7.7 & 0.72 & 0.6 & 0.25 & 0.24 \\
 & Curr. theory unc. & 7.7 & 2.3 & 1.4 & 0.32 & 0.3 \\
\bottomrule
 \multicolumn{7}{c}{}\\[-0.8em]
\midrule
\multirow[c]{4}{*}{$\delta g_{We}[\%]$} & No theory unc. & 0.18 & 0.0038 & 0.0019 & 0.0033 & 0.0037 \\
 & Aggr. theory unc. & 0.18 & 0.0039 & 0.0019 & 0.0033 & 0.0037 \\
 & Cons. theory unc. & 0.18 & 0.0078 & 0.0054 & 0.0062 & 0.0063 \\
 & Curr. theory unc. & 0.18 & 0.022 & 0.017 & 0.011 & 0.01 \\
\cline{1-7}
\multirow[c]{4}{*}{$\delta g_{W\mu}[\%]$} & No theory unc. & 0.19 & 0.005 & 0.0023 & 0.0034 & 0.0037 \\
 & Aggr. theory unc. & 0.19 & 0.0051 & 0.0024 & 0.0034 & 0.0037 \\
 & Cons. theory unc. & 0.19 & 0.0081 & 0.0055 & 0.0063 & 0.0063 \\
 & Curr. theory unc. & 0.19 & 0.022 & 0.017 & 0.011 & 0.011 \\
\cline{1-7}
\multirow[c]{4}{*}{$\delta g_{W\tau}[\%]$} & No theory unc. & 0.39 & 0.0046 & 0.0028 & 0.015 & 0.017 \\
 & Aggr. theory unc. & 0.39 & 0.0057 & 0.0034 & 0.016 & 0.017 \\
 & Cons. theory unc. & 0.39 & 0.011 & 0.009 & 0.017 & 0.019 \\
 & Curr. theory unc. & 0.39 & 0.057 & 0.055 & 0.037 & 0.038 \\
\cline{1-7}
\multirow[c]{4}{*}{$\delta g_{1z}[\%]$} & No theory unc. & 0.16 & 0.027 & 0.012 & 0.0093 & 0.0091 \\
 & Aggr. theory unc. & 0.16 & 0.033 & 0.013 & 0.0096 & 0.0098 \\
 & Cons. theory unc. & 0.16 & 0.036 & 0.017 & 0.012 & 0.013 \\
 & Curr. theory unc. & 0.16 & 0.05 & 0.031 & 0.019 & 0.018 \\
\cline{1-7}
\multirow[c]{4}{*}{$\delta k_{\gamma}[\%]$} & No theory unc. & 0.64 & 0.031 & 0.014 & 0.014 & 0.01 \\
 & Aggr. theory unc. & 0.64 & 0.033 & 0.015 & 0.015 & 0.011 \\
 & Cons. theory unc. & 0.64 & 0.037 & 0.02 & 0.019 & 0.016 \\
 & Curr. theory unc. & 0.64 & 0.07 & 0.045 & 0.03 & 0.025 \\
\cline{1-7}
\multirow[c]{4}{*}{$\delta \lambda_Z[\%]$} & No theory unc. & 0.23 & 0.035 & 0.012 & 0.0058 & 0.0027 \\
 & Aggr. theory unc. & 0.23 & 0.036 & 0.012 & 0.0058 & 0.0027 \\
 & Cons. theory unc. & 0.23 & 0.036 & 0.012 & 0.0059 & 0.0027 \\
 & Curr. theory unc. & 0.23 & 0.037 & 0.012 & 0.0059 & 0.0028 \\
\bottomrule
\end{tabular}
\vspace{0.2cm}
\caption{Numerical results for the effective coupling fits shown in Fig. \ref{fig:eff_couplings_results_EW}.
From top to bottom, we show the 68\% C.I. bounds on the electroweak left-handed  and right-handed couplings of the $Z$-boson and the effective couplings of the $W$-boson both to fermions and to other gauge bosons. 
Results are presented for the four scenarios considered in this work for theoretical uncertainties at future colliders.
\label{tab:app_effective_couplings_1}
}
\end{table}
\begin{table}[t]\footnotesize
\centering
 \renewcommand{\arraystretch}{1.02}
\begin{tabular}{|cc|c|c|c|c|c|c|}
\toprule
 &  & HL & LEP3 & FCC-ee & LCF550 & LCF1000 \\
\midrule
\multirow[c]{4}{*}{$\delta g_{H\mu\mu}[\%]$} & No theory unc. & 3.7 & 2.8 & 2.4 & 2.6 & 2.6 \\
 & Aggr. theory unc. & 3.7 & 2.8 & 2.4 & 2.6 & 2.6 \\
 & Cons. theory unc. & 3.7 & 2.8 & 2.4 & 2.6 & 2.6 \\
 & Curr. theory unc. & 3.7 & 2.9 & 2.5 & 2.7 & 2.6 \\
\cline{1-7}
\multirow[c]{4}{*}{$\delta g_{H\tau\tau}[\%]$} & No theory unc. & 3.2 & 0.76 & 0.37 & 0.44 & 0.42 \\
 & Aggr. theory unc. & 3.2 & 0.78 & 0.37 & 0.44 & 0.43 \\
 & Cons. theory unc. & 3.2 & 0.78 & 0.38 & 0.45 & 0.44 \\
 & Curr. theory unc. & 3.2 & 0.83 & 0.44 & 0.5 & 0.48 \\
\cline{1-7}
\multirow[c]{4}{*}{$\delta g_{Hcc}[\%]$} & No theory unc. & 28 & 0.84 & 0.42 & 0.73 & 0.62 \\
 & Aggr. theory unc. & 28 & 0.84 & 0.42 & 0.73 & 0.62 \\
 & Cons. theory unc. & 28 & 0.84 & 0.43 & 0.78 & 0.64 \\
 & Curr. theory unc. & 28 & 0.9 & 0.55 & 0.86 & 0.74 \\
\cline{1-7}
\multirow[c]{4}{*}{$\delta g_{Hbb}[\%]$} & No theory unc. & 3 & 0.44 & 0.22 & 0.23 & 0.21 \\
 & Aggr. theory unc. & 3 & 0.44 & 0.22 & 0.23 & 0.21 \\
 & Cons. theory unc. & 3 & 0.45 & 0.22 & 0.25 & 0.23 \\
 & Curr. theory unc. & 3 & 0.48 & 0.26 & 0.29 & 0.27 \\
\cline{1-7}
\multirow[c]{4}{*}{$\delta g_{Htt}[\%]$} & No theory unc. & 3.8 & 1.6 & 1.6 & 1.5 & 1.2 \\
 & Aggr. theory unc. & 3.8 & 1.8 & 1.8 & 1.6 & 1.2 \\
 & Cons. theory unc. & 3.8 & 1.8 & 1.8 & 1.7 & 1.2 \\
 & Curr. theory unc. & 3.8 & 1.9 & 1.8 & 1.7 & 1.3 \\
\bottomrule
\multicolumn{7}{c}{}\\[-0.6em]
\toprule
\multirow[c]{4}{*}{$\delta g_{Hgg}[\%]$} & No theory unc. & 4.7 & 0.63 & 0.37 & 0.49 & 0.43 \\
 & Aggr. theory unc. & 4.7 & 0.66 & 0.39 & 0.5 & 0.45 \\
 & Cons. theory unc. & 4.7 & 0.71 & 0.43 & 0.53 & 0.47 \\
 & Curr. theory unc. & 4.7 & 0.9 & 0.65 & 0.65 & 0.62 \\
\cline{1-7}
\multirow[c]{4}{*}{$\delta g_{H\gamma\gamma}[\%]$} & No theory unc. & 3.8 & 1 & 0.89 & 0.97 & 0.93 \\
 & Aggr. theory unc. & 3.8 & 1.1 & 0.91 & 1 & 0.97 \\
 & Cons. theory unc. & 3.8 & 1.1 & 0.91 & 1 & 0.98 \\
 & Curr. theory unc. & 3.8 & 1.1 & 0.96 & 1.1 & 1 \\
\cline{1-7}
\multirow[c]{4}{*}{$\delta g_{HZ\gamma}[\%]$} & No theory unc. & 7.7 & 5.5 & 3.3 & 3.8 & 3.9 \\
 & Aggr. theory unc. & 7.7 & 5.6 & 3.3 & 3.8 & 4.1 \\
 & Cons. theory unc. & 7.7 & 5.6 & 3.4 & 4 & 4.1 \\
 & Curr. theory unc. & 7.7 & 5.7 & 3.6 & 4.5 & 4.4 \\
\cline{1-7}
\multirow[c]{4}{*}{$\delta g_{HWW}[\%]$} & No theory unc. & 2.6 & 0.18 & 0.098 & 0.12 & 0.1 \\
 & Aggr. theory unc. & 2.6 & 0.18 & 0.1 & 0.12 & 0.11 \\
 & Cons. theory unc. & 2.6 & 0.19 & 0.11 & 0.14 & 0.12 \\
 & Curr. theory unc. & 2.6 & 0.25 & 0.19 & 0.19 & 0.18 \\
\cline{1-7}
\multirow[c]{4}{*}{$\delta g_{HZZ}[\%]$} & No theory unc. & 2.5 & 0.19 & 0.1 & 0.12 & 0.11 \\
 & Aggr. theory unc. & 2.5 & 0.19 & 0.11 & 0.13 & 0.11 \\
 & Cons. theory unc. & 2.5 & 0.19 & 0.12 & 0.14 & 0.13 \\
 & Curr. theory unc. & 2.5 & 0.25 & 0.19 & 0.19 & 0.18 \\
\bottomrule

\end{tabular}
\vspace{0.2cm}
\caption{Same as Table~\ref{tab:app_effective_couplings_1} for the Higgs boson coupling modifiers, showing the numerical results of Fig.~\ref{fig:eff_couplings_results_Higgs}.
\label{tab:app_effective_couplings_2}
}
\end{table}

We report here the numerical results for the effective coupling fits corresponding to Figs.~\ref{fig:eff_couplings_results_EW}--\ref{fig:eff_couplings_results_Higgs} in Sect.~\ref{sec:effective_couplings_results}.
Specifically, Table~\ref{tab:app_effective_couplings_1}
displays the numerical results for the effective coupling fits shown in Fig.~\ref{fig:eff_couplings_results_EW}, where from top to bottom, we show the
68\% C.I. bounds on the electroweak left-handed and right-handed couplings of the $Z$-boson and the effective couplings
of the $W$-boson both to fermions and to other gauge bosons. 
Results are presented for the four scenarios considered
in this work for theoretical uncertainties at future colliders.
Then Table~\ref{tab:app_effective_couplings_2}
 presents the same comparison now for the Higgs boson coupling modifiers, showing the numerical results corresponding to Fig.~\ref{fig:eff_couplings_results_Higgs}.

\subsection{Global SMEFT fit}
\label{app:extra_smeft}

Here we collect additional results and comparisons relevant for the global SMEFT fit studies complementing those already presented in Sect.~\ref{sec:global_smeft_fits_results}.

\begin{figure}[htbp]
    \centering
\includegraphics[width=0.99\linewidth]{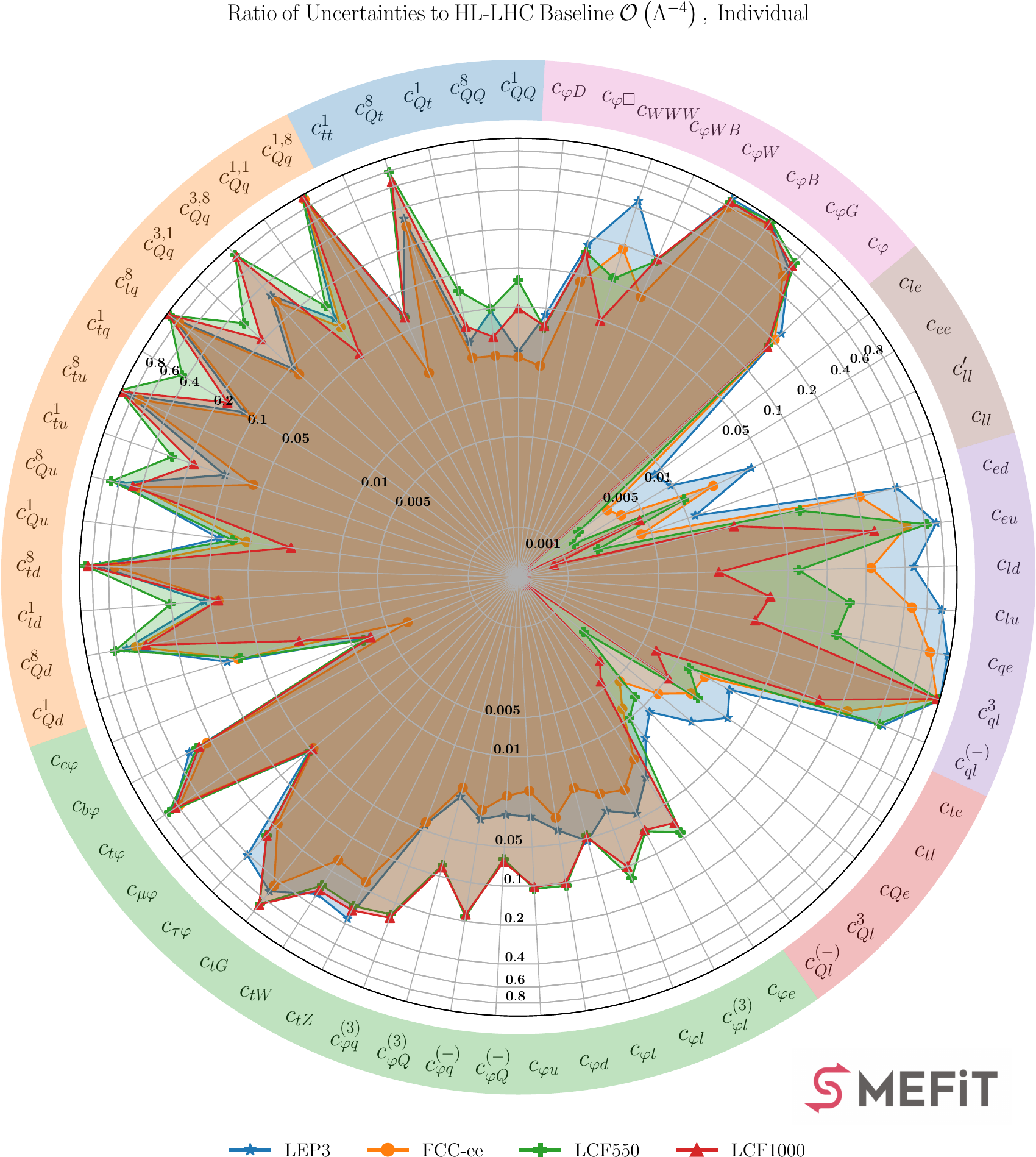}
    \caption{Same as Fig.~\ref{fig:spider_lin_ind_aggressive} for individual fits carried out at the quadratic level in the EFT expansion.
\label{fig:spider_quad_ind}
}
\end{figure}

\begin{figure}[htbp]
    \centering
\includegraphics[width=\linewidth]{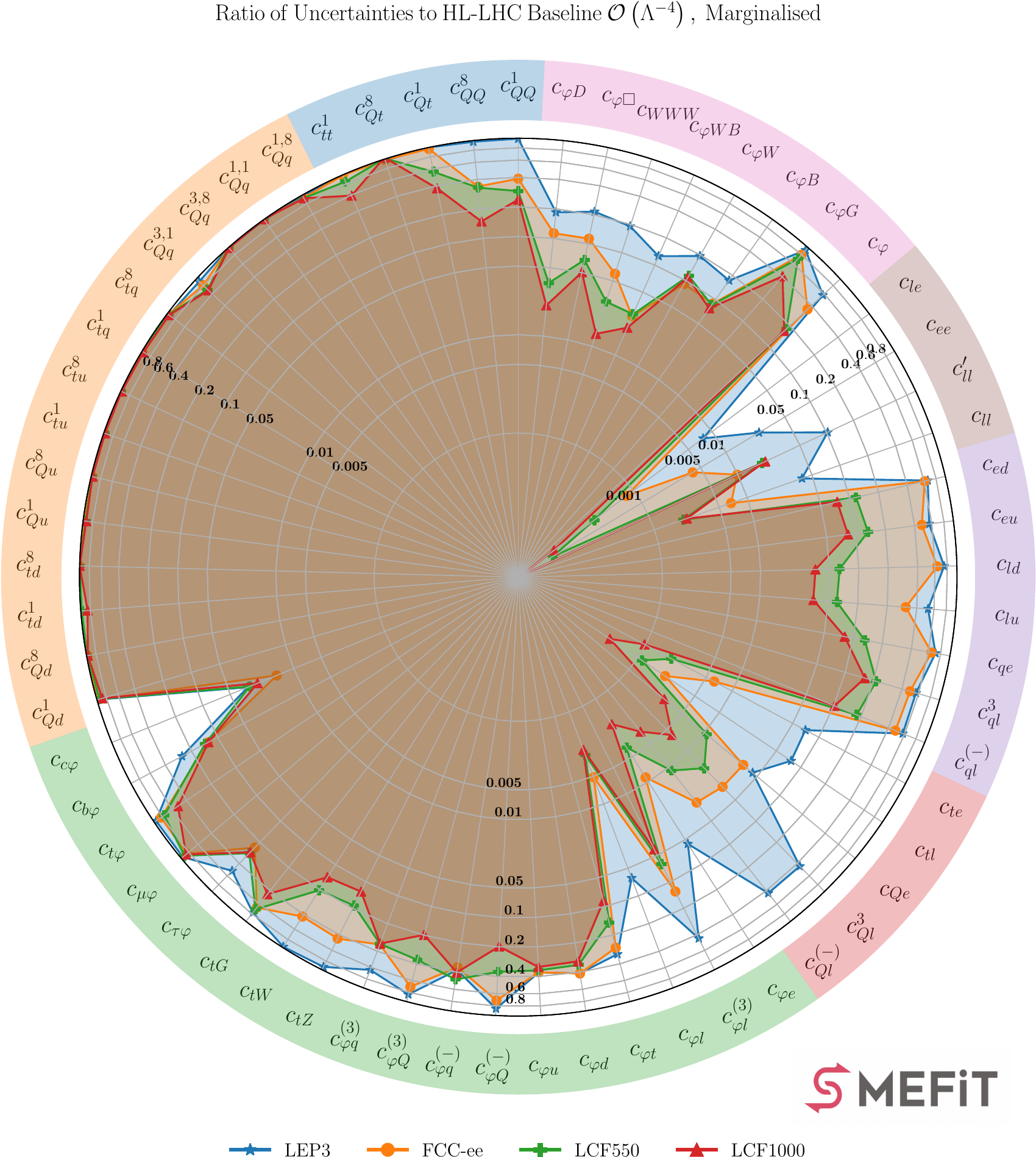}
    \caption{Same as Fig.~\ref{fig:spider_lin_global_aggressive} for marginalised global fits carried out at the quadratic level in the EFT expansion. 
\label{fig:spider_quad_glob}
}
\end{figure}

\begin{figure}[htbp]
    \centering
\includegraphics[width=\linewidth]{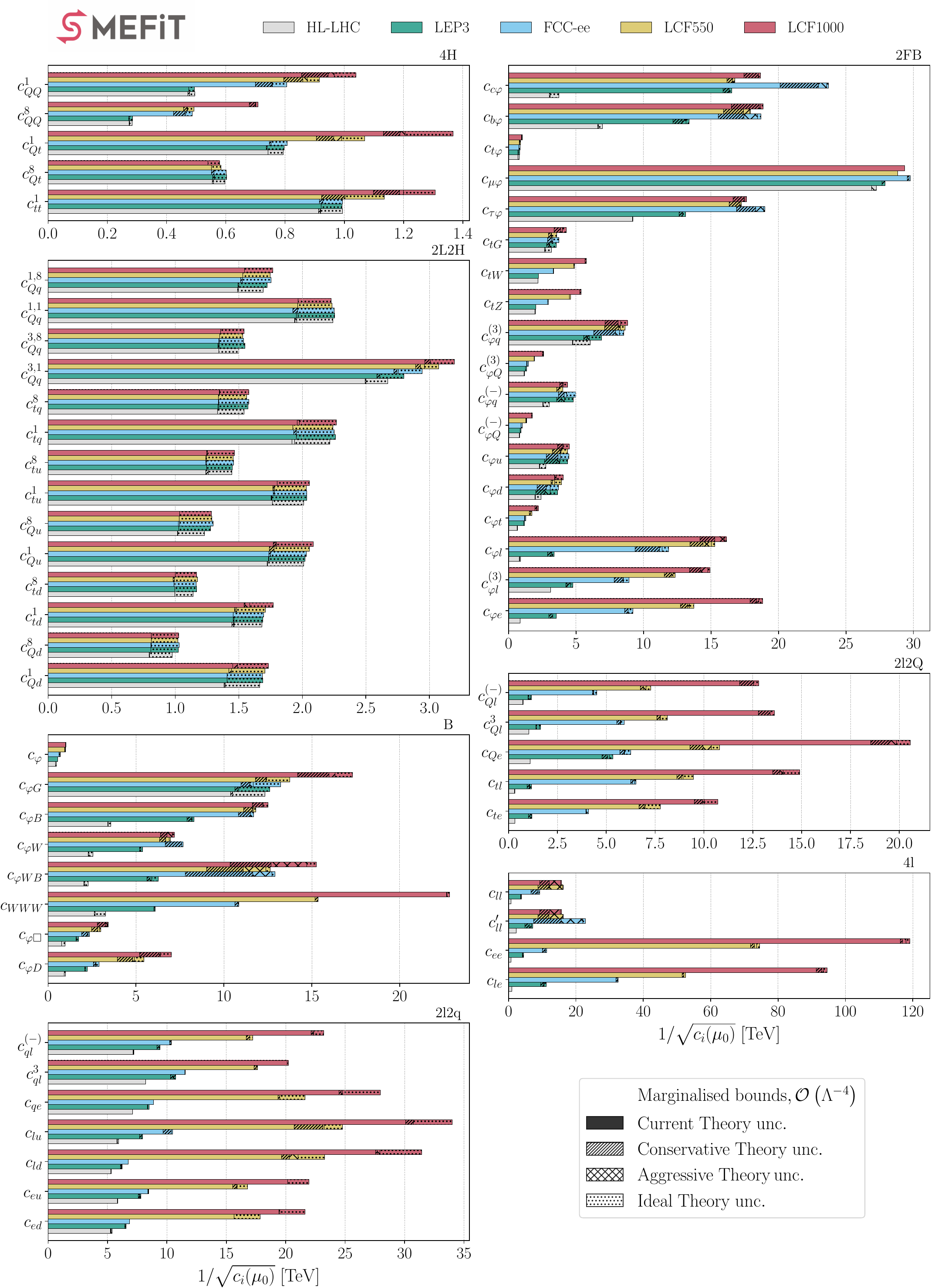}
    \caption{Same as Fig.~\ref{fig:lin_glob_th_lambda}, now assessing the impact of theoretical uncertainties in the global marginalised bounds for the reach in mass scale $\Lambda$ obtained in quadratic fits. 
\label{fig:quad_ind_th_lambda}
}
\end{figure}

\begin{figure}[htbp]
    \centering
\includegraphics[width=\linewidth]{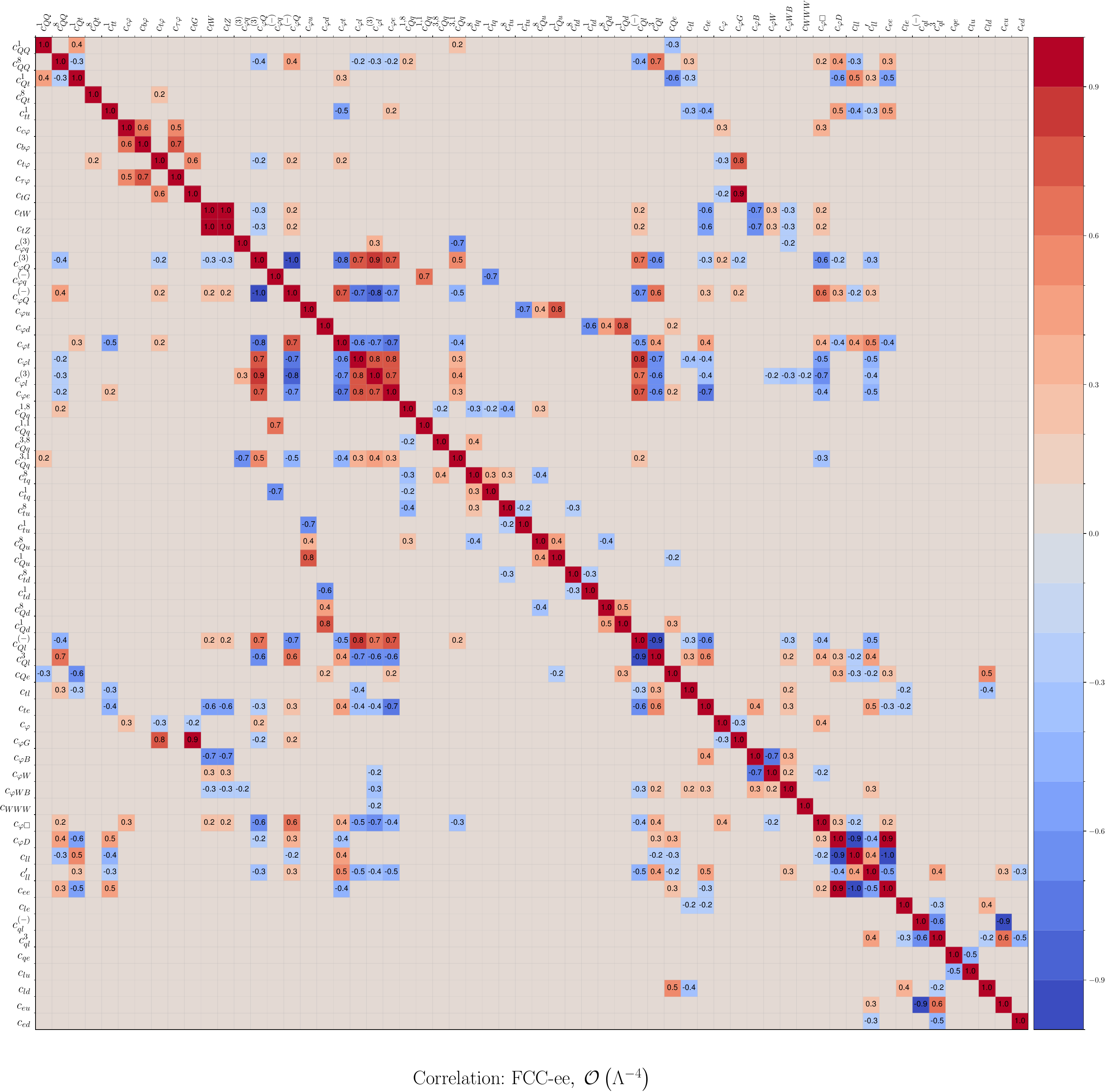}
    \caption{Same as Fig.~\ref{fig:correlationmap_FCCee_linear}, now for the quadratic fit. 
\label{fig:correlationmap_FCCee_quadratic}
}
\end{figure}

\begin{itemize}

\item Figs.~\ref{fig:spider_quad_ind} and~\ref{fig:spider_quad_glob} pthe resent analogous results of Figs.~\ref{fig:spider_lin_ind_aggressive}and~\ref{fig:spider_lin_global_aggressive}, namely the improvement in the bounds on the Wilson coefficients as compared to the HL-LHC baseline for individual and global marginalised fits respectively, now in the case of EFT carried out at the quadratic level. 

\item Fig.~\ref{fig:quad_ind_th_lambda}  presents a similar comparison as in Fig.~\ref{fig:lin_ind_global_lambda}, which assessed the impact of theoretical uncertainties in the global marginalised bounds for the reach in mass scale $\Lambda$, now for fits carried out at quadratic order in the EFT expansion.
From this comparison, one finds that
for quadratic fits the impact of theoretical uncertainties is more substantial than in the global marginalised fit, especially in the case of the FCC-ee. 
For instance, for some of the two-fermion operators, such as $c_{\varphi Q}^{(3)}$, there is up to a factor 5 difference between the lower bounds in $\Lambda$ obtained between the `current' and `ideal' scenarios for theoretical uncertainties.

\item Fig.~\ref{fig:correlationmap_FCCee_quadratic} presents the same correlation map in the space of Wilson coefficients as in Fig.~\ref{fig:correlationmap_FCCee_linear}, but now when quadratic EFT corrections are also considered.
Comparison of the correlation maps in the linear and the quadratic fits indicate that in the latter correlations are in general milder due to the effect of quadratic EFT corrections, which may break poorly constrained directions in the parameter space.

\clearpage

\end{itemize}

\clearpage
\providecommand{\href}[2]{#2}\begingroup\raggedright\endgroup

\end{document}